\documentclass[11pt,a4paper]{book}

\usepackage[latin1]{inputenc} 
\usepackage{graphics} 
\usepackage[pdftex]{graphicx}
\usepackage{epstopdf}
\usepackage{fancyhdr}
\usepackage[dvips]{epsfig} 
\usepackage{epic} 
\usepackage{eepic} 
\usepackage{latexsym,chapterbib}
\usepackage{color}
\usepackage{amsfonts,amssymb,amsmath}
\usepackage{graphicx}
\usepackage{mathbbol,mathrsfs}
\usepackage{wrapfig}
\usepackage{appendix}
\usepackage{makeidx}
\usepackage{psfrag}
\usepackage{float}
\usepackage{textcomp} 
\usepackage[final]{pdfpages}
\usepackage[retainorgcmds]{IEEEtrantools}
\newcommand{\azul}{\color[rgb]{0,0,1}}

\newcommand{\inilista}[1]{\begin{list}{\bf#1.\theenumi}{\usecounter{enumi}}}
\newcommand{\finlista}[1]{\setcounter{#1}{\value{enumi}}\end{list}}
\newcommand{\parrafo}[1]{\par\vspace*{3mm}}

\makeindex

\usepackage{cite}
\usepackage{captionGABRIEL}
\newcommand*\captionwidthset[1]{
\setlength\captionwidth{#1}
\takecaptionwidthtrue}
\captionwidthset{11.5cm}

\allowdisplaybreaks[1]

\begin{document}
\thispagestyle{empty}
\begin{center}
\textbf{\LARGE A  Quantum-mechanical Approach}\\
\vspace*{3mm}
\textbf{\LARGE for Constrained Macromolecular Chains}\\
\vspace*{10mm}
{\bf\large Gabriel F. Calvo$^{1,{\azul*}}$ and Ram\'{o}n F. Alvarez-Estrada$^{2,\azul{\dagger}}$}\\
 \vspace*{10mm}
 {\em $^{1}$Departamento de Matem\'{a}ticas, ETS de Ingenieros de Caminos, Canales y Puertos and  
IMACI-Instituto de Matem\'{a}tica Aplicada a la Ciencia y la Ingenier\'{\i}a,
 Universidad de Castilla-La Mancha, 13071, Ciudad Real, Spain}\\[0.3cm]
 {\em $^{2}$Departamento de F\'{\i}sica Te\'{o}rica I, Facultad de Ciencias F\'{\i}sicas, 
Universidad Complutense, 28040 Madrid, Spain}\\[0.4cm]
 {\azul $*$ \em\bf gabriel.fernandez@uclm.es} and {\azul$\dagger$ \em\bf rfa@fis.es}\\[0.2cm]
(\today)\\
\end{center}
\vfill

\newpage
\thispagestyle{empty}



\newpage
\frontmatter

\tableofcontents
\mainmatter
\chapter{Introduction }
\label{intro}
Many approaches to three-dimensional constrained macromolecular chains at thermal equilibrium, at about room temperatures, are based upon constrained Classical  Hamiltonian Dynamics (cCHDa). Quantum-mechanical approaches (QMa) have also been treated by different researchers for decades. QMa address a fundamental issue (constraints versus the uncertainty principle) and are  versatile:  they  also yield classical descriptions (which may not  coincide  with those from cCHDa, although  they  may  agree for certain relevant quantities). Open issues include whether QMa have enough practical consequences which differ from and/or improve  those from cCHDa.  We shall treat cCHDa  briefly and deal with QMa,  by outlining old approaches and focusing on recent ones. In QMa, we  start with Hamiltonians for $N(\gg 1)$ non-relativistic quantum particles, interacting among themselves through potentials which include strong vibrational ones (constraining bond lengths and bond angles)    and other weaker interactions. We get (by means of variational calculations) effective three-dimensional constrained quantum partition functions at equilibrium ($Z_{Q}$) and Hamiltonians ($H_{Q}$) for single-stranded (ss) macromolecules (freely-jointed,  freely-rotating, open or closed) and for double-stranded (ds) open macromolecules. Due to crucial cancellations, we can neatly separate the constrained degrees of freedom (by getting the large constant vibrational zero-point energies associated to them) from  the  slow unconstrained angular  variables (accounted for by $Z_{Q}$ and $H_{Q}$). In the classical limit,  we obtain classical partition functions $Z_{C}$ from $Z_{Q}$. The $Z_{C}$'s are  respectively  different from the classical partition functions found starting from cCHDa for similar  chains. Thus, they differ in determinants of the sort referred to in Ref.\cite{SSTtiEche}: QMa determinants are simpler than cCHDa ones. For ss macromolecules, we compare several quantities  (bond-bond correlations, squared end-to-end distances, etc) from QMa with the standard Gaussian model in Polymer Science: the comparisons display  good consistencies (which are also met with cCHDa). For double-stranded DNA (dsDNA) macromolecules,  the  $Z_{C}$'s from QMa have structures  which generalize those  obtained by other researchers, and enable to study  thermal denaturation. 
\par

There are many excellent general references on macromolecular chains (or polymers), from different standpoints.      A 
(relatively 
small) set of them is: \cite{Leh,Volk,Flo,Gros,McQ,Doi,Freed,Proh,Frank,Elias,Gotoh,deGen,desClois,Pol2}. 
\par
The contents of this tutorial review complements the subjects presented in the following contributions:  \cite{SSTtiEche,SSTtiSkRe,SSTtiHaCi,SSTtiMaHa,SSTtiElHe,SSTtiHunRe}.  
\par 
Numerical computations for macromolecules are  extremely difficult to carry out, because of the enormously large number of 
degrees of freedom involved. Fortunately, a subset of those degrees of freedom turn out to be, very approximately, constrained 
  (taking on essentially constant values), 
namely,  bond lengths,   
    bond angles between successive bond vectors and, depending on the chain, other coordinates as well.
The interest of accomplishing an adequate treatment of constraints in large molecular chains  is easy to understand: it yields a reduction in the number of effective degrees of freedom to be treated and, hence, to  a simplification in the 
simulations. The following pattern will be met in all macromolecules to be treated here. If the macromolecule is formed by $N(\gg 1)$ atomic constituents, one would have to deal, a priori, with $aN$ degrees of freedom, $a$ being  some positive integer. By treating suitably various constraints, at the end, the number  of  
effective  degrees of freedom which remain unconstrained and, hence, have to be considered in the numerical computations  
will be $bN$, with $b$ a positive integer such that $b<a$.   
\par  
The fact that the constituents of a macromolecule are atoms (described through Quantum Mechanics) raises some issues  of principle, even 
if, at the end, one employs Classical Statistical Mechanics for practical purposes. Could one start the analysis of a constrained macromolecule safely from  
Classical (Statistical) Mechanics  or, rather, should one  begin with Quantum (Statistical) Mechanics and, at a later 
stage, proceed to its classical limit? The following argument will indicate that the answer to that question may not be 
straightforward.   If, in a system of classical particles (say, strictly in the framework of Classical Mechanics), some coordinates (denoted by $q_{con}$) are fixed (constrained), while the others (named as $q$) are not but evolve in time, then, 
as remarked by Brillouin \cite{Brillou}, the momenta canonically conjugate of $q_{con}$ will not vanish and will vary with 
time in general, in a way completely determined by both $q$ and the momenta canonically conjugate to the latter. 
    On the other hand, if, in a system of microscopic particles described trough Quantum Mechanics (for which, in principle, 
Brillouin's remark  does not apply),  certain coordinates $q_{con}$ take on essentially fixed values (with very small, or almost vanishing, uncertainty), then, according to the quantum-mechanical uncertainty principle, their canonically conjugate momenta take on any possible value and have an almost infinite uncertainty! 
\par
We shall   remind     some  works   devoted to analyze constrained motions of simpler microscopic 
systems from    first principles: the 
quantum-mechanical motion (via a Schr\"{o}dinger equation) of a three-dimensional 
microscopic particle, which is constrained (by the action of some  adequate potential) to move, in a suitable limit, 
along a given 
curve 
or on a surface
  \cite{JenKo,DaCos1,DaCos3}.  The analysis   was generalized in \cite{DaCos2} 
to the quantum motion of a system described by a  larger set 
of coordinates ($q_{j}$ and $q_{con,i}$, to be unconstrained and constrained, respectively) \cite{DaCos2}: 
under the condition  that no 
crossed term  containing second partial derivatives (like  
$[\partial/\partial q_{j}][\partial/\partial q_{con,i}]$)  
appears in the kinetic energy part of the quantum 
hamiltonian,   then   the quantum constrained system is described by 
a unique Schr\"{o}dinger equation (which generalizes those for 
the surface and curve cases). Unfortunately, the kinetic 
energy operators typically associated to even the simplest 
macromolecular chains (with $N\geq 3$)  will  contain such crossed terms [see 
Eqs. (\ref{eq:Ha}) and (\ref{eq:T1})]. Then,  the above 
condition is not fulfilled and there is no guarantee that 
the procedure in \cite{DaCos2} would yield a unique quantum hamiltonian. See also comments in \cite{AE}. 

 The above discussions contribute to confirm the inherent difficulties 
involved in the study of constraints in  quantum systems. 
Is it possible to formulate a complete quantum theory of constraints for macromolecules which be free of difficulties?

\begin{figure}[t]
\begin{center}
{
\includegraphics[scale=0.34]{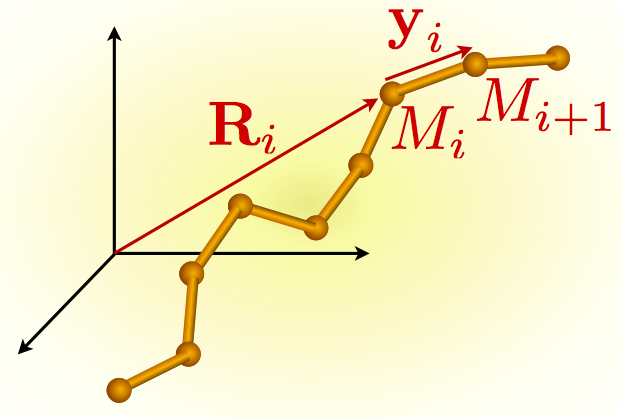}}
\end{center}
\vspace*{-3mm}
\caption{\small Single-stranded (ss) macromolecular chain.}
\vspace*{-4mm}
\label{fig:RFAE.GFC.1}
\end{figure}

 The following general aspects which will apply for all single-stranded (ss) macromolecular chains to be treated here. In three-dimensional space, we shall consider one single ss macromolecular chain formed by  $N$ non-relativistic atoms (treated either as classical or as quantum particles), with masses  $M_{i}(>0)$. Let  ${\bf R_{i}}$ be the position vector of the $i$-th atom, with $i=1,\ldots,N$, and let  $M_{tot}$ be the total mass of the whole ss macromolecule (see figure \ref{fig:RFAE.GFC.1}). Also, let  the center-of-mass (CM) position vector and the relative ones for the successive atoms (the bond vectors) be denoted by ${\bf R_{CM}}$,
  and ${\bf y}_{i}$ (where $1 \leq i \leq N-1$), respectively. One has:
\begin{eqnarray}
{\bf R}_{CM} &=& \frac{\sum_{i=1}^{N} M_{i} {\bf R}_{i} }{M_{tot}}, \, M_{tot} = \sum_{i=1}^{N} M_{i}, \label{eq:Rcm}\\
{\bf y}_{i} &=& {\bf R}_{i+1} - {\bf R}_{i}\; .      \label{eq:y}
\end{eqnarray}
Using spherical coordinates, we write:  
\begin{eqnarray}
{\bf y}_{i} & =  &y_{i} {\bf u}_{i}\; , \label{eq:yi} \\
{\bf u}_{i} & = &\left( \cos \varphi_{i} \sin \theta_{i}, \sin \varphi_{i} \sin \theta_{i}, \cos \theta_{i} \right) \;,\label{eq:ui} \\
{\bf u}_{\theta_{i}} & = &\left( \cos \varphi_{i} \cos \theta_{i}, \sin \varphi_{i} \cos \theta_{i}, -\sin \theta_{i} \right) \; , \label{eq:thetai} \\
{\bf u}_{\varphi_{i}} & = &\left( -\sin \varphi_{i}, \cos \varphi_{i}, 0 \right)\; . \label{eq:phii}
\end{eqnarray}
Notice that, for fixed $i$,   the   vectors ${\bf u}_{i}, {\bf u}_{\varphi_{i}}, {\bf u}_{\theta_{i}}$ constitute an orthonormal set.
\par  
   
The total number of atoms in any single macromolecule to be studied here  (namely, $N$ in the ss-case) is supposed to be 
enormously  large.  This justifies the applicability of Statistical Mechanics (and, eventually, of Thermodynamics) to those 
chains. The possibility of applying  the latter  frameworks   to molecular chains which are not large has
recently  attracted  attention (in particular, related to single-molecule stretching experiments) \cite{Rubi,Ritort}; we shall not analyze this subject  here. 
 \par
We shall always deal with large three-dimensional macromolecules specifically  in thermodynamical equilibrium, at  absolute temperature $T$ in an interval of  
 physical interest,  which includes room temperature ($\simeq 300$ K). We shall anticipate two ubiquitous  physical facts. At $T$  not far from  $300$ K, there are many  allowed 
states for the macromolecule and, for each allowed state,  all interatomic distances $y_{i}$, $i=1,\ldots,N-1$, turn out 
to be approximately constant and $\simeq d_{i}>0$ (the bond lengths), namely, approximately constrained.  Also, 
the energies associated  to  individual unconstrained angular degrees of freedom of microscopic constituents  in the 
macromolecule  are, typically, smaller than $k_{B}T$ ($k_{B}$ being Boltzmann' s constant).

We shall  restrict in Chapters \ref{sec:RFAE.GFC.2}  to \ref{sec:RFAE.GFC.5} through  to what characterizes the freely-jointed 
(or unhindered) chains.  In them,   all $y_{i}=d_{i}$, $i=1,\ldots,N-1$,  while all  
angular variables  will remain  unconstrained (except for a closed-ring constraint in \ref{sec:RFAE.GFC.5}). 
  Chapter 2 outlines   approaches 
to constraints in freely-jointed ss macromolecules, through Classical  Hamiltonian Dynamics (cCHDa), displays certain controversies arising 
there and  turn to  quantum approaches for them.  
Chapter \ref{sec:RFAE.GFC.3} justifies   the quantum-mechanical physical assumptions and the variational approach to constrained macromolecules 
  to be employed later. Chapter  \ref{sec:RFAE.GFC.4} 
   treats   freely-jointed macromolecules  quantum-mechanically.  
    Chapters \ref{sec:RFAE.GFC.5} and  \ref{sec:RFAE.GFC.6} 
 summarize   generalizations for  single-stranded macromolecules with constrained bond lengths and with further constraints  
(closed-ring and angular ones).  
   Chapter \ref{sec:RFAE.GFC.7} is devoted to   double-stranded (ds) open chains with constraints. 
Chapter \ref{sec:RFAE.GFC.8}   deals 
    with applications   to   dsDNA. 

 \chapter{Macromolecules with rigid and stiff constraints }
\label{sec:RFAE.GFC.2}  
\section{Classical models, classical partition functions and paradoxes}
\label{subsec:RFAE.GFC.2.1}
  
Let all particles be regarded as classical. A dot above a time-dependent function will denote its first time derivative. The classical kinetic energy  
$T_{o} =\frac{1}{2}\sum_{i=1}^{N}M_{i}\dot{{\bf R}}_{i}^{2}$ of the ss macromolecule, by employing   Eqs. (\ref{eq:Rcm}) and   (\ref{eq:y}), also reads:
\begin{eqnarray}
T_{o}=\frac{1}{2}M_{tot} \dot{{\bf R}}_{CM}^{2}+\frac{1}{2}\sum_{i,j=1}^{N-1}
\dot{{\bf y}}_{i}(B^{-1})_{ij}\dot{{\bf y}}_{j}\;  ,
\label{eq:21} 
\end{eqnarray}
where the total mass $M_{tot}$ was given in (\ref{eq:Rcm}).  Regarding the $(N-1)\times(N-1)$ matrix $B^{-1}$, with elements $(B^{-1})_{ij}$, it   suffices   to know that: i) $B^{-1}$ is symmetric ($(B^{-1})_{ij}=(B^{-1})_{ji}$); 
ii) $B^{-1}$ has positive eigenvalues ($T_{o}$ being a positive definite quadratic form); iii) the inverse matrix $B$, with elements $B_{ij}$, also has positive eigenvalues and it is a tridiagonal symmetric matrix (for  $i=2,\ldots,N-2$, $B_{ij}=0$, unless $i=j-1,j,j+1$, and $B_{13}=\cdots=B_{1N-1}=0$, $B_{N-11}=\cdots=B_{N-1N-3}=0$).  
Explicit computations for low values of $N$ allow to confirm  i)-iii) easily.  
  See \cite{HuMcC} and 
references therein about how to find analytically the inverse of a general tridiagonal matrix.  
\par
 We shall start with the constrained Classical  Hamiltonian Dynamics approach (cCHDa), in which all $y_{i}=d_{i}(>0 )$, 
$i=1,\ldots,N-1$, hold   (holonomic or classical rigid  constraints  ). In so doing, there is not a unique procedure 
(alternative procedures  being  equivalent to one another). One, due to Kramers \cite{Kram} and   
 used   in \cite{KirkRise,Hass,Hass2},   employs   only independent 
unconstrained variables. We shall disregard the overall motion of the CM.  In Kramers'  rigid model   the independent unconstrained variables are all angles 
$\theta_{i}$ and $\varphi_{i}$, $i=1,\ldots,N-1$, to be denoted collectively as $q_{i}$, $i=1,\ldots,2(N-1)$. 
As $\dot{{\bf y}}_{i}=d_{i}(\dot{\theta_{i}}{\bf u}_{\theta_{i}}+ \dot{\varphi_{i}}{\bf u}_{\varphi_{i}}  )$, with 
$i=1,\ldots,N-1$, Eq. (\ref{eq:21}) becomes (with the CM disregarded): 
$T_{o, con}=   2^{-1}\sum_{i,j=1}^{2(N-1)}G_{ij}\dot{q_{i}}\dot{q_{j}}$.
 The $[2(N-1)]\times[2(N-1)]$ matrix $G$, with elements $G_{ij}$, is obtained   upon replacing the above 
$\dot{{\bf y}}_{i}$ into Eq. (\ref{eq:21}). Notice that $G_{ij}$ depends on $q_{i}$, with $i=1,\ldots,2(N-1)$.   
There may also be some potential energy $V_{con}=V_{con}(q_{1},\ldots,q_{2(N-1)})$, so that the   lagrangian (with the CM disregarded) 
is $L_{con}=T_{o,con}-V_{con}$. Throughout this work, the subindex "c" in a dynamical variable will always remind that the latter is classical 
(namely,  not a quantum operator  ). The classical  momentum canonically conjugate to $\theta_{i}$ and $\varphi_{i}$ are 
$\pi_{\theta;i,c}$ and $\pi_{\varphi;i,c}$, $i=1,\ldots,N-1$, to be denoted, collectively, as $\pi_{an,i,c}$. One  
  has:  
$\pi_{an,i,c}=\partial L_{con}/\partial {\dot{q_{i}}}= \sum_{j=1}^{2(N-1)}G_{ij} \dot{q_{j}}$.  
 The  associated  classical hamiltonian is: 
\begin{eqnarray}&&
 H_{con } = \sum_{i,j=1}^{2(N-1)}\dot{q_{i}} \pi_{an,i,c}   
 - L_{con}=\frac{1}{2}\sum_{i,j=1}^{2(N-1)}(G^{-1})_{ij} \pi_{an,i,c}\pi_{an,j,c}+V_{con} \;  .
\label{eq:25}
\end{eqnarray}
The   matrix $G^{-1}$, with elements $(G^{-1})_{ij}$, is the inverse of $G$. The very large chain in 
thermodynamical equilibrium at temperature $T$   is now described  by Classical Statistical Mechanics, thereby  
disregarding     quantum effects.   
 We shall disregard the    CM motion and  concentrate on   all 
$q_{i}$ and $\pi_{an,i,c}$. The classical equilibrium partition function for the macromolecule   is: 
 \begin{eqnarray}
  Z_{con} &= & \int \frac{\left[\rm d\mbox{\boldmath{$u$}}\right] \left[\rm d\mbox{\boldmath{$\pi$}}_{an,c}\right]}{(2\pi \hbar)^{2(N-1)}}   
 \exp\left[-\frac{H_{con}}{k_{B}T}\right] \;  ,
\label{eq:26}\\
 \left[\rm d\mbox{\boldmath{$u$}}\right]
 &=&\prod_{i=1}^{N-1}d\theta_{i} d\varphi_{i} \; , \quad 
  \left[\rm d\mbox{\boldmath{$\pi$}}_{an,c}\right]=\prod_{i=1}^{2(N-1)}d\pi_{an,i,c}  \;      .\label{eq:27}
\end{eqnarray}
  By performing  all Gaussian integrations on $\pi_{an,i,c}$, 
 one finds: 
\begin{eqnarray}
 Z_{con} =(2\pi k_{B}T)^{N-1}\int \frac{\left[\rm d\mbox{\boldmath{$u$}}\right]}{(2\pi \hbar)^{2(N-1)}}\left[\det G\right]
^{1/2}\exp\left[-\frac{V_{con}}{k_{B}T}\right] \;  .
\label{eq:28}
\end{eqnarray}
 $\det G$  being  the determinant of the matrix $G$.  
One   generic difficulty of this classical model   is that  $\det G$ depends on the angles. See Fixman 
\cite{Fix}  for a  study of $\det G$.  Other cCHDas  to macromolecules ( equivalent to 
 Kramers'  one) work with all variables (all  ${\bf y}_{i}$ and their canonically conjugate momenta) and suitable 
Lagrange multipliers, see \cite{AE,EdGood}. Another tutorial review in this volume \cite{SSTtiMaHa} discusses  generic  
$N$-dimensional Lagrangian systems with $D$ independent holonomic constraints, including the case in which  the $N-D$ 
unconstrained coordinates  cannot be constructed explicitly, not even locally. The latter formulation (the ambient space one) 
and  the associated 
Langevin and Fokker-Planck equations  (to treat   dynamics) are reviewed in \cite{SSTtiMaHa}. 
\par 
  
 The classical and unavoidable   time variation of the momenta canonically conjugate to the $y_{i}$'s  \cite{Brillou} 
turns out to be one of the great difficulties met in simulations employing the molecular dynamics method 
for the study of macromolecules,  when constraints are taken into account  \cite{Ryck,Ciccotti}. Nevertheless, considerable progress along this research 
line has been facilitated mostly due to efficient  algorithms allowing to integrate numerically the $3N$ Cartesian equations 
of a system of  $N$ classical point particles subject to holonomic constraints \cite{Shake}. 
As a matter of fact, the holonomic constraints considered in \cite{Shake} included not only  bond lengths  but angular
 variables (bond angles) as well. The computational  algorithms in \cite{Shake} (implementing molecular dynamics simulations 
and referred to, at present, as ``the Shake ones'') have been subsequently followed and developed further by a good 
number of authors. See    
\cite{SSTtiElHe}. 
\par
By taking  \cite{EdGood} as starting point and through  a detailed analysis, Mazars \cite{Mazars1,Mazars2} has obtained that the critical exponents for Kramers' model are the same as those of the Gaussian model  \cite{Doi,deGen}. See also \cite{Mazars3}. For other studies, aimed at comparing Kramers' model with the Gaussian one, see  \cite{AE} and references therein. 
\par 
We now turn to  another classical  model considered by Fraenkel \cite{Fraenk} (and pursued  by  
\cite{FixKo,FixEv,Tit1,Tit2,Rou,Zimm,Lod,Hinc,Bird,Ed,Freed1}), which is also based upon Classical Mechanics and employs 
harmonic springs (hs).   Fraenkel's model, which 
is not equivalent to Kramers' one, is 
characterized by the fact that the constraints $y_{i}=d_{i}$, with $i=1,\ldots,N-1$ are not imposed from the very beginning,  
but at a later stage, by letting the harmonic springs to become very stiff:  a classical flexible model with infinite 
stiffness (also known as stiff model).   Flory has  employed   similar  ideas  in his classic 
 work about statistics  of macromolecules  \cite{Flo}. The lagrangian is now $L_{hs}=T_{o}-V_{0}$, where  $T_{o}$ is given 
in Eq. (\ref{eq:21})) and the potential $V_{0}$   now depends  on all $y_{i}$, $\theta_{i}$ and $\varphi_{i}$, 
$i=1,\ldots,N-1$. Upon   introducing the canonically conjugate momenta  
 $\mbox{\boldmath{$\pi$}}_{i,c}=\partial L_{hs}/\partial  \dot{{\bf y}}_{i}= 
\sum_{j=1}^{N-1} (B^{-1})_{ij} \dot{{\bf y}}_{j}$ and disregarding the contribution of the CM, the actual hamiltonian reads: 
 \begin{eqnarray}&&
H_{hs,in} = H_{in}+V_{0},\,H_{in}=\frac{1}{2}\sum_{i,j=1}^{N-1} \mbox{\boldmath{$\pi$}}_{i,c} B_{ij} 
\mbox{\boldmath{$\pi$}}_{j,c}  \;  \label{eq:30} 
 \end{eqnarray}
 $B_{ij}$ are the elements of the matrix $B$ [the inverse of $B^{-1}$ in (\ref{eq:21})]. 
 The  partition function for the chain  at thermal equilibrium  is:  
\begin{eqnarray}&&
Z_{hs}=\int \frac{\left[{\rm d}{\bf y} \right] \left[{\rm d}\mbox{\boldmath{$\pi$}}_{c}\right]}
{(2 \pi\hbar)^{3(N-1)}} \exp \left[-\frac{H_{hs,in}}{k_{B}T}\right] \! ,\label{eq:33}\\&&
\left[\rm d{\bf y}\right]=\prod_{i=1}^{N-1}{\rm d}^{3}{\bf y}_{i}  ,  \quad
\left[{\rm d}\mbox{\boldmath{$\pi$}}_{c}\right]= \prod_{i=1}^{N-1}{\rm d}^{3}\mbox{\boldmath{$\pi$}}_{i,c} \;  .
\label{eq:32}
\end{eqnarray}
 Upon performing the Gaussian integrations  over all momenta, one gets: 
\begin{eqnarray}
 Z_{hs} =\frac{(2\pi k_{B}T)^{3(N-1)/2}}{\left[\det B\right]^{1/2}}\int \frac{ \left[\rm d{\bf y} \right] }
{(2\pi \hbar)^{3(N-1)}}  
\exp\left[-\frac{V_{0}}{k_{B}T}\right] \;  .
\label{eq:34}
\end{eqnarray}
  $\det B$ (the determinant of the matrix $B$) is  a constant.  We shall suppose that $V_{0}=V_{con}+U$, 
where $V_{con}$ is   similar to the one in Eq. (\ref{eq:25}) and in Eq. (\ref{eq:28}), while $U$ is the following 
harmonic-oscillator-like (or harmonic spring) potential: 
 \begin{eqnarray}
U=\frac{1}{2}\sum_{j=1}^{N-1}\omega_{j}^{2}
(B_{jj})^{-1}(y_{j}-d_{j})^{2} \, ,
\label{eq:35}
\end{eqnarray} 
where $\omega_{j}(>0)$ are frequencies and $d_{j}(>0)$ are the bond lengths. The statistical average of a function 
$F=F({\bf y}_{1},\ldots,{\bf y}_{N-1})$ is: 
 \begin{eqnarray}
 \langle F\rangle= \frac{\int [\rm d{\bf y} ]
  F({\bf y}_{1},\ldots,{\bf y}_{N-1}) \exp\left[-\frac{V_{con}}{k_{B}T}\right]}
{\int  [\rm d{\bf y} ]
\exp\left[-\frac{V_{con}}{k_{B}T}\right]} \;  .
 \label{eq:36}
\end{eqnarray} 
 We shall assume that all $\omega_{j}\rightarrow+\infty$ (say, the limit in which the harmonic springs become 
very stiff), which forces $\exp[-U/(k_{B}T)]$ to equal a constant factor $\lambda^{-1}$ times $\prod_{j=1}^{N-1}\delta(y_{j}-
d_{j})$, where $\delta$ denotes Dirac 's delta function. Thus, the physical bond-length constraints $y_{j}=d_{j}$ are 
recovered in the stiff  harmonic spring limit.   $\lambda$ should diverge as the frequencies do, although  that  will not be  
relevant here. Then, one gets in the stiff harmonic spring limit:  
 \begin{eqnarray}&&
 Z_{hs} = \lambda_{1}\int [{\bf d\Omega}]
  \exp\left[-\frac{V_{con}}{k_{B}T}\right],\ 
\langle F\rangle= \frac{\int [{\bf d\Omega}]
  F \exp\left[-\frac{V_{con}}{k_{B}T}\right]}
{\int  [{\bf d\Omega}]
\exp\left[-\frac{V_{con}}{k_{B}T}\right]} \;  .
\label{eq:38}\\&&[{\bf d\Omega}]   =
 \prod_{s=1}^{N-1}\sin \theta_{s}\rm d\theta_{s} \rm d\varphi_{s} \;  .
\label{eq:37}
\end{eqnarray}
 All constant factors in $Z_{hs}$ have been embodied into a single one, $\lambda_{1}$, which, in turn, cancels out in 
$\langle F\rangle$. Those coordinates (like the $y_{s}$) which, as a result of  the above stiff harmonic spring limit, 
take on constant values, are named hard variables \cite{Go1}. The remaining coordinates, which remain unconstrained after 
having taken such a limit, are called soft variables \cite{Go1}.  The larger the frequencies are, the more 
rapidly and wildly  oscillate the hard variables ($y_{s}$)  about certain values ($d_{s}$), to be regarded as constant 
parameters.  In any (small) period corresponding to those rapid oscillations, the soft variables do not 
change appreciably. Let $V_{con}=0$ and let all $d_{s}=d$, $s=1,\ldots,N-1$, in Eqs. (\ref{eq:38}).  In the stiff 
harmonic spring limit, Eqs. (\ref{eq:38})   yield: $\langle {\bf y}_{s}^{2}\rangle=d^{2}$, 
$\langle {\bf u}_{s} {\bf u}_{j}\rangle=0$ and $\langle({\bf R}_{N}-{\bf R}_{1})^{2}\rangle =(N-1)d^{2}$. These results 
  agree with well known results for the standard Gaussian model \cite{Gros,Doi}.  A more detailed comparison shows that 
the three-dimensional distribution function for the end-to-end vector ${\bf R}_{N}-{\bf R}_{1} $ in 
Fraenkel's classical model with stiff harmonic springs coincides with the   end-to-end distribution given in 
Eqs. (14-83), (14-84) and (14-86) in McQuarrie \cite{McQ}.     However, it does not seem that Eqs. (\ref{eq:38})  
 imply the Gaussian distribution for the individual bond lengths. See, in this connection, the discussion  
in pages $22-27$ in \cite{Freed}. Then,  Fraenkel's classical model with stiff harmonic springs is partially  consistent with the standard Gaussian model. 
\par
We now turn to display paradoxes,  upon comparing Kramers' model and Fraenkel's one. In fact, $Z_{con}$ ( (\ref{eq:28})) 
and $Z_{hs}$ ((\ref{eq:38})) differ clearly from each  
other. The discrepancy has been displayed  neatly for suitable angular probability distributions in the case $N=3$ 
 and $V_{con}=0$ (the trimer) \cite{Gottl,Ral,vanKampen}. In particular, the analysis in \cite{Gottl}  was based on 
molecular dynamics  simulations 
  for $N=3$. That  discrepancy gave rise, in the framework of Classical Mechanics 
(without introducing, as yet, quantum effects and the uncertainty principle!), to a controversy. 
The latter could be stated as follows. If, in the framework of Classical Mechanics, certain coordinates in the macromolecular 
chain ( the bond lengths) have to be regarded as constrained, should one employ Kramers' approach (rigid constraints) or 
Fraenkel's one (infinitely stiff flexible constraints)? 
  Certain simulations   favoured Kramers'  approach, while others  were more consistent with Fraenkel's 
one and, finally, others gave  almost undistinguishable results.  See
  \cite{Gottl,Pear,ErKirk}. 
 
\section{On  the role of quantum effects }
\label{subsec:RFAE.GFC.2.2}

Several  researchers  carried out  further analysis  aimed at  clarifying  the above controversy, by bringing some 
quantum-mechanical ideas into the scene. Helfand \cite{Helfand} recognized clearly that the statistical properties of rigidly 
constrained systems (Kramers') and flexibly constrained ones with large stiffness (Fraenkel's) are different, in general. 
 He also treated a simple two-dimensional quantum-mechanical model. See also Van 
Kampen \cite{vanKampen}. We shall focus below on \cite{Go1,Go2} and, mostly, on \cite{Ral}.

 Go and Scheraga \cite{Go1,Go2} explored the origin of the differences between the  classical  partition functions resulting 
from Kramers'  and Fraenkel's models [say, between Eq. (\ref{eq:28}) and Eq. (\ref{eq:38})], by including from the outset 
quantum effects for the variables which should be constrained (the hard ones): in their analysis, the latter  were 
not only  the $y_{s}$ but also the bond angles.  
  Their first work \cite{Go1} supported  Kramers' approach.
 However, the  second  work \cite{Go2} concluded, that  Fraenkel's approach was, within the approximations involved, more accurate than Kramers' one.
 
   Rallison \cite{Ral} argued that the expectation that rigidity or rigid rods (Kramers ) and flexible stiffness (Fraenkel) be equivalent need be not valid in Statistical Mechanics, and that Quantum Mechanics is required to formulate  the problem. According to him, the paradox had arisen from a failure to recognize the relevance of quantum effects,   
 in the limit  where certain coordinates are to  be constrained to constant values. He also argued that a unique answer 
could be obtained in principle (although 
it could be very difficult to achieve, in practice), by regarding the system as the classical limit of a 
quantum-mechanical 
one. He carried out a detailed analysis for the trimer. According to him, the corresponding result for the trimer 
through  Kramers' model would never appear as a natural limit, when a full quantum-mechanical analysis is performed 
for such a system with quantized vibrations. Accordingly, let $\omega_{s}$ be  frequencies, 
playing a role similar to those in Fraenkel's model. Rallison also tried to characterize the extent of quantum effects by 
the dimensionless parameter $Q\equiv \hbar \omega_{s}/k_{B}T$ ($\hbar$ being Planck' s constant), which is reasonable, and 
stated that, when $Q\rightarrow 0$, the classical stiff spring results are recovered. The last statement  leads to 
 a conflict  with the stiff harmonic spring limit  in Fraenkel's purely classical framework, namely, when going from Eq. 
 (\ref{eq:34})    to Eqs. (\ref{eq:38})  . Thus,  coming back to the formulation of Fraenkel's model after the above comments, one now realizes  that 
 $\omega_{s}$ should not exceed $k_{B}T/\hbar$, if the formulation is to remain in the classical framework, so that to allow 
for  $\omega_{s}\gg k_{B}T/\hbar$ 
would bring in quantum effects. In \cite{Ral},  the degrees of freedom to be constrained (due to the very stiff springs) were treated quantum mechanically, while   
  the remaining unconstrained coordinates were regarded as classical.

  No strictly  definite conclusion about which model was more accurate  seemed to have been reached.   Kramers' model is 
formulated in a mathematically consistent way in the frameworks of  (constrained) Classical Mechanics and 
Classical Statistical Mechanics.   
    A number of researchers  have made the following  choice: they have 
disregarded  Fraenkel's model,  and  they have concentrated on developing  quite vigorously computational approaches based upon constrained Classical  Hamiltonian Dynamics (cCHDa): see \cite{Shake,Ciccotti} and other articles in this volume. On the other hand,  other researchers tended to adhere, with 
various qualifications and reservations (mostly arising from quantum considerations),  to the conclusion in \cite{Go2}, namely, Fraenkel's approach: see Section 1 of    \cite{Pear} (where   Brownian dynamics simulations for classical chains are used). For a  comprehensive  discussion, see the  Introduction in \cite{Eche}, which  includes a lot of additional and updated information. The different effects of the constraints in both Kramers' approach  and Fraenkel's one, on the 
conformational equilibrium distribution for a relevant biomolecule (a dipeptide),  were studied comparatively in  \cite{Eche}, although the question of which model was a better approximation was not specifically addressed.  This work analyzed for the first time the conformational dependences  of 
correction terms (related to the determinants) to that equilibrium distribution, using {\em ab initio} quantum mechanical calculations 
(including electron correlations) and without simplifying assumptions. Their numerical computations (Monte Carlo simulations), which  employed a realistic potential energy function, concluded that those correction terms could be neglected in certain situations, up to peptides of considerable length,  but not in general. They also 
analyzed whether the equilibrium values of the hard coordinates were dependent on the soft ones: 
their computations indicated that such a possibility should be critically considered in  each case. 
See   \cite{SSTtiEche}. 
 
 What about quantizing Kramers' model?  
 The classical Kramers' model for a constrained freely-jointed molecular chain, based upon  Eq.  (\ref{eq:25}), can be quantized through 
certain well established procedure initiated by Podolsky \cite{Podolsk} and developed further by De Witt \cite{deWitt}, with a probabilistic 
interpretation in the 
standard quantum-mechanical framework \cite{MesI,GalPas}. For details, see 
  \cite{AEPR,AE} and references 
therein. Since the quantization through \cite{Podolsk} is formulated,  from the outset,  in terms of  the 
unconstrained coordinates (say, $\theta_{s}$ and $\varphi_{s}$), the former will shed no light on the issue 
of the quantum uncertainties for the constrained coordinates ($y_{s}$) and of those for  their canonically conjugate momenta! 

One could also  follow  the alternative path   in which Fraenkel's 
model is not  disregarded and  to proceed further, by entering definitely into the quantum regime. In such an option, it is worthwhile  to extract the  following summary  from  the comments and studies above.  Thus, in  agreement with \cite{Go1,Go2,Ral},  we remind that: i) the 
best description of the actual molecular chain   is provided, in principle,  through a quantum-mechanical treatment,  ii) 
specifically, one should start with a quantized chain with flexible but stiff springs, iii) at a later stage, one could proceed to the classical limit.  We also recall that   the  main troubles regarding this philosophy  were  practical \cite{Go1,Go2,Ral}: it was very  difficult to carry through a treatment based upon  i) and ii), 
unless a number of approximations were made, the validity of which, in turn,  was rather difficult to control. Thus, as commented in \cite{Go1,Go2,Ral}: a) the 
large quantum zero-point energies  of the constrained hard degrees of freedom (due to the very stiff springs) 
could depend, in  principle, on the remaining unconstrained (soft) coordinates, regarded as classical, b) if such 
dependences were taken into account, the subsequent analysis appeared to be quite hard, while if they were not, 
then the validity of the resulting approximations would be difficult to assess. c) if those quantum zero-point energies were 
sufficiently large (as they would eventually become, for suitably large frequencies of the harmonic springs), it would be 
necessary to treat, at least in principle, the remaining unconstrained coordinates not just as classical variables, but 
through Quantum Mechanics as well: the latter possibility could add even greater difficulties.

\chapter{A  variational quantum-mechanical approach:  General aspects}
\label{sec:RFAE.GFC.3}  
 
 What about quantizing Fraenkel's model? If done adequately,  do  large quantum zero-point energies  of   hard degrees of 
freedom 
 depend  
on  soft ones?

 The models for macromolecular chains based on Quantum Mechanics to be reported  in this work    will  be derived 
through    a different (and less ambitious) strategy:  
   the application of a   variational quantum-mechanical   inequality 
procedure. The latter, not relying on the absence of the 
crossed terms containing second partial derivatives,    will not be  
subject to  the  limitation in \cite{DaCos2}.  Their variational foundations  will imply that the quantum  models  
reported   here should be not be regarded, in principle,  as definitive or final formulations, but as approximate ones.

 The following   
comments should suffice    to support physically the variational approach to be pursued along this paper.
The physical fact that the  relative distances (bond lengths) from any atom to its nearest neighbours  are, within narrow 
limits,  approximately constant, in  generic macromolecular chains (below some maximum temperature ) 
\cite{Leh,Volk,Flo,Gros,McQ,Doi,Elias}, provides justification to the following   picture, which incorporates the 
Born-Oppenheimer approximation \cite{MesII} and the 
  previous remarks \cite{Go1,Ral,Go2}. All atomic constituents in the macromolecule  are subject to an effective potential 
(due, reciprocally and self-consistently, to  all of them), which includes: 1)  electronic contributions 
(the largest ones, denoted as $E_{el}$), 2) potentials (having  magnitudes smaller than those in 1)) which, 
for intermediate distances, resemble qualitatively harmonic-oscillator-like vibrational   ones  and  force atoms  to oscillate  with frequencies $\omega_{i}$ 
about some equilibrium position in  the chain, so that the relative separations  to its nearest neighbors  do not vary appreciably, on the average, and  3)
other  potentials (having magnitude smaller than the vibrational ones  and, hence, not   upsetting the oscillations) and  are, in turn, 
 responsible for other weaker effects: angular variables, etc. As another justification, we recall that for certain simple (small) molecules about room temperature,  individual internal rotations about bonds   are slowly-varying, as their typical energies  are, at least, two orders of magnitude 
smaller than the  vibrational energies  $\hbar\omega_{i}$ \cite{Volk,MesII}: such a qualitative pattern appears to be valid  for macromolecules as 
well, on the average (except, possibly, for some moderate quantitative variations). For generic  vibrational frequencies 
$\omega_{i}$  of typical atomic  constituents  in  physically interesting macromolecular chains, it is a  fact  
\cite{Volk} that the  vibrational energy  $\hbar\omega_{i}$ is larger  (about  one order of magnitude more or less) than
 $k_{B}T$ ($T$  in an interval about  $ 300$ K). Thus, for a chain at thermal 
equilibrium at those $k_{B}T$, generic   vibrational  degrees of freedom are  rapidly-varying,  have to 
be treated quantum-mechanically (in a way radically different from the classical description), and  they appear to be,  
quite approximately, in their  ground states, with all excited vibrational states unoccupied 
\cite{Go1,Ral}.  Electronic  degrees of freedom remain   unexcited in phenomena involving energies in the above range of 
$k_{B}T$ and, hence, 
   $E_{el}$  can  be regarded as approximately constant. 
\par
The above  facts  agree with  certain qualitative statements made by Schr\"{o}dinger in his well known book \cite{Schrod}: 
in fact, his statements (on quantum theory, molecules and how their stability depends on temperature) appear to have 
discussed very  lucidly a good part of these physical issues.
  It  follows that  quantum effects for those vibrational degrees of freedom will be dominant (and even more  
  for   electronic ones). Hence, one should employ Quantum Mechanics as the starting point to deal with them, 
at least in principle. Those dominances  and the fact that electronic and vibrational energies are larger than $k_{B}T$ ($T$  in an interval about  $ 300$ 
K)   are   
ubiquitous   manifestations of quantum 
effects in macromolecules (in particular in biological ones). In turn,  the quantized vibrations about their ground states in macromolecular chains at 
thermal equilibrium  appear to decouple, at least as a zeroth  approximation, from internal rotations.   Our study, with $\omega_{s}\gg 
k_{B}T/\hbar$ could be viewed as a fully quantized version of Fraenkel's one. It will also improve that in \cite{Ral}, since all variables (those to be constrained and the ones which will remain unconstrained) with be 
treated quantum-mechanically from the outset. 
\par
The  variational quantum-mechanical models to be treated  from section  \ref{sec:RFAE.GFC.4} onwards 
 would  be subject, in principle, to  the objections a), b) and c) at the end of subsection \ref{subsec:RFAE.GFC.2.2}. We anticipate that, fortunately,  by virtue  of 
 various crucial exact cancellations, at the end of the computations those    models will turn out to be free of the  
difficulties a), b) and c) at the end of subsection \ref{subsec:RFAE.GFC.2.2}: the expected quantum zero-point energies  of hard degrees of 
freedom will be obtained and, hence, shown to be constant.  
\par
  The  successive analysis of  quantum  macromolecules with      constrained bond lengths and increasingly complicated 
 constraints     are neither  easy nor direct. 
A number of those  generalizations  have  indeed   been  carried out, but at the expense of getting involved into  and 
overcoming considerable difficulties at the quantum level. One finds  new and interesting quantum-mechanical structures. 
 Although it is not easy  
 to handle those structures, they imply  an important consequence: the rotational invariance (
conservation of total  angular momentum) in 
  the 
quantum models. Moreover, those structures 
  provide   a consistent basis to get  classical approximations, at about room temperature, which are more manageable.  
    
 \par
We shall employ both  harmonic-oscillator-like potentials  and Morse potentials \cite{Morse,GalPas}. 
The latter   can be approximated by ( and are less idealized than)   the former, for certain range of intermediate distances. 
   We shall  outline the methods and  avoid the  presentation of  calculation details throughout the main body of the present tutorial review.  Some specific  calculations are given in the Appendixes. 
\par
We will devote a great deal of attention to constructing partition functions for the different chains, which will be well defined and meaningful first at the quantum level and, later, at the classical one. We shall carry out several approximations   in   the more relevant parts of the resulting classical 
partition functions. Specifically, we shall use  those classical partition functions (with or without  those  approximations) in order  to find various physically interesting quantities: internal energies, correlations, probability   distributions, equilibrium  constants. For the latter purposes, the  less relevant contributions to those 
partition functions will not contribute and, hence, will not be given. Thus, the  complete results of the approximate evaluations of the classical 
partition functions will not be strictly needed and, hence, will not be reported: see \cite{AE,Calvo,RaCa,CAEDNA,CAEDNA2,CAEDNA1}.

\section{The  variational  inequality}
\label{subsec:RFAE.GFC.3.1}
 
We continue to use   Eqs. (\ref{eq:Rcm})-(\ref{eq:phii}).  We start from the total quantum Hamiltonian operator: 
\begin{eqnarray} &&  
H_{Q,1}  =  -\frac{\hbar^{2}}{2}\sum_{i=1}^{N} \frac{1}{M_{i}}\nabla_{{\bf R}_{i}}^{2}    +  V({\bf R})\;  , \label{eq:H1}
\end{eqnarray}
where $\nabla$ denotes the gradient operator and  $V({\bf R})$ represents the real potential energy among atoms.  The kinetic energy operator 
 $-(\hbar^{2}/2)\sum_{i=1}^{N} M_{i}^{-1}\nabla_{{\bf R}_{i}}^{2} $ in 
 Eq. (\ref{eq:H1})  is the quantum-mechanical version of the classical kinetic energy in Eq. (\ref{eq:21}). 
Then, using expressions (\ref{eq:Rcm}) and (\ref{eq:y}), (\ref{eq:H1}) gives the quantum Hamiltonian in terms of the new variables as
$ H_{Q,1}  =  - (\hbar^{2}/2M_{tot})\nabla_{{\bf R}_{CM}}^{2} +\tilde{H}_{Q}$.   CM degrees of freedom will always 
be factored out and disregarded (and, hence, so will be $- (\hbar^{2}/2M_{tot})\nabla_{{\bf R}_{CM}}^{2}$), 
for all types of single-stranded macromolecular chains to be studied in Sections \ref{sec:RFAE.GFC.3}-\ref{sec:RFAE.GFC.6}. 
But CM degrees of freedom will be 
taken into account for  double-stranded chains (Sections \ref{sec:RFAE.GFC.7} and \ref{sec:RFAE.GFC.8}). For single-stranded chains, 
we shall always concentrate on  the internal quantum Hamiltonian operator: 
\begin{eqnarray}
\tilde{H}_{Q}  =  
 - \frac{\hbar^{2}}{2}\sum_{i=1}^{N-1} B_{i}\nabla_{{\bf y}_{i}}^{2}    +  
 \hbar^{2}\sum_{i=2}^{N-1}\frac{\nabla_{{\bf y}_{i-1} }\cdot \nabla_{{\bf y}_{i}}}{M_{i}}   +  U({\bf y}) = 
H_{Q,in}  + U({\bf y})\;  , \label{eq:Ha} 
\end{eqnarray}
$H_{Q,in}$ is the quantum-mechanical operator associated to $H_{in} $ in Eq. (\ref{eq:30}). The coefficients $B_{ij}$ in  
Eq. (\ref{eq:30}) are those in  $H_{Q,in}$: in particular,  $B_{i}=M_{i}^{-1} + M_{i+1}^{-1}(=B_{ii})$. See also Eq. (\ref{eq:matrixB}).  For single-stranded chains, we denote the set of atomic coordinates 
$({\bf y}_{1}, {\bf y}_{2}, \ldots, {\bf y}_{N-1})$ by ${\bf y}$ (for double-stranded chains, in 
Sections \ref{sec:RFAE.GFC.7}-\ref{sec:RFAE.GFC.8},  ${\bf y}$ will have 
another meaning). We remark that  all molecular chains are treated in the framework of the Born-Oppenheimer approximation   
\cite{MesII,Calvo}.  Then, by recalling the  previous discussion in this 
 section \ref{sec:RFAE.GFC.3},  the effective potential in the macromolecule is described by  $E_{el}+ U({\bf y}) $. 
$E_{el}(<0)$ is  the electronic energy, corresponding to  the  electronic degrees of freedom (the most rapidly-varying and the most energetic ones), which is regarded, essentially, as  a constant.  We shall suppose that  $E_{el}$  has  already been subtracted out and, then,  omitted from the outset. There 
remains a  total (real) effective potential energy  $U({\bf y})=  V({\bf R})$,  independent on ${\bf R}_{CM}$, which  
accounts for the effective interactions denoted as 2) + 3), previously   in this 
 section. $U({\bf y})$ accounts for  large covalent-bond interactions, 
which are responsible for the very existence of the macromolecule, as an extended and connected object, with 
energies smaller than the electronic ones ($E_{el}$), and other weaker interactions. Specifically,  $U({\bf y})$ includes, first of all,  
the  interactions   which constraint any $\mid{\bf y}_{i} \mid=y_{i}$ to equal  the  fixed    bond lengths,  as well as the (somewhat weaker) interactions hindering 
rotations (constraining bond angles), those associated to the unconstrained angular variables and further weaker residual interactions. 
\par
The three-dimensional momentum operator reads in spherical coordinates:
\begin{eqnarray}
-i\hbar\nabla_{{\bf y}_{i}} &=& - \frac{{\bf a}_{i}}{ y_{i}} -i\hbar {\bf u}_{i} \frac{\partial}{\partial y_{i}}\;  ,\label{eq:Mom}\\ 
 {\bf a}_{i}  &=& i\hbar {\bf u}_{\theta_{i}} \frac{\partial}{\partial \theta_{i}}  + 
i\hbar {\bf u}_{\varphi_{i}} \frac{1}{\sin \theta_{i}} \frac{\partial}{\partial \varphi_{i}} \;  , \label{eq:ai} 
 \end{eqnarray}
 and we recall  Eqs. (\ref{eq:yi})(\ref{eq:ui}), (\ref{eq:thetai}) and (\ref{eq:phii}).
We shall refer to both ${\bf y}_{i}$ and ${\bf u}_{r_{i}}$ as bond vectors. For later use, we denote the set of all 
$\theta_{1}, \ldots , \theta_{N-1}$, $\varphi_{1}, \ldots , \varphi_{N-1}$ by $\theta$ and $\varphi$, respectively. Notice that the ${\bf a}_{i}$'s 
are not Hermitian operators. Using Eq. (\ref{eq:Mom}), the  atomic kinetic energy operator $H_{Q,in} $ becomes:
 \begin{eqnarray}
H_{Q,in} & = & \frac{1}{2}\sum_{i=1}^{N-1} B_{i} \left\{ -\hbar^{2} \frac{\partial^{2}}{\partial y_{i}^{2}} - 2\hbar^{2} \frac{1}{y_{i}} \frac{\partial}{\partial y_{i}} + \frac{{\bf a}_{i} \cdot {\bf a}_{i}}{ y_{i}^{2}} \right\} \nonumber \\
& + & \sum_{i=2}^{N-1}\frac{1}{M_{i}}\left\{ \hbar^{2} {\bf u}_{i-1} \cdot {\bf u}_{i} \frac{\partial^{2}}{\partial 
y_{i-1}\partial y_{i}} - i\hbar {\bf u}_{i-1} \cdot{\bf a}_{i} \frac{1}{y_{i}} \frac{\partial}{\partial y_{i-1}} \right\} \nonumber \\
& - & \sum_{i=2}^{N-1}\frac{1}{M_{i}}\left\{ i\hbar {\bf a}_{i-1} \cdot{\bf u}_{i} \frac{1}{y_{i-1}} 
\frac{\partial}{\partial y_{i}} - \frac{{\bf a}_{i-1} \cdot {\bf a}_{i}}{y_{i-1} y_{i}} \right\} \; .   \label{eq:T1} 
\end{eqnarray}
  Since, by assumption, the large quantum ($Q$) macromolecular chain is in thermodynamical equilibrium, at absolute temperature $T$, we describe its state 
 through  
 the (exact)  quantum partition function $\tilde{Z}_{Q}$, which   is given by \cite{Hua}:
\begin{eqnarray}&&
\tilde{Z}_{Q} = {\rm Tr}[\exp{ [-(k_{B}T)^{-1}\tilde{H}_{Q}]}] = \sum_{\sigma} \exp[-\frac{E_{\sigma}}{k_{B}T}] \;  ,
\label{eq:parfuncquan}
\end{eqnarray}
where  $\sigma$ denotes the set of  all quantum numbers and $E_{\sigma}$ represents the 
eigenvalues of the entire spectrum of the quantum Hamiltonian $\tilde{H}_{Q}$. 
\par
  Let $\Phi_{\lambda}$ be an 
arbitrary orthonormal set of wave functions for the system ($\lambda$ representing a set of quantum numbers). 
It is not required that 
$\Phi_{\lambda}$  coincide with the  exact eigenfunctions of $\tilde{H}_{Q}$. Then, the exact quantum 
partition function for the system $\tilde{Z}_{Q}$ (as given in 
Eq.\ (\ref{eq:parfuncquan})) satisfies  Peierls` variational  inequality \cite{Peie,Hua}:
\begin{eqnarray}&&
\tilde{Z}_{Q} \geq \sum_{\lambda} \exp{[-(k_{B}T)^{-1}(\Phi_{\lambda}, \tilde{H}_{Q} \Phi_{\lambda} ) ]} \;  .
\label{eq:peierls}
\end{eqnarray}
The equality holds if $\Phi_{\lambda}$ is the complete set of exact eigenfunctions of $\tilde{H}_{Q}$. The 
$\Phi_{\lambda}$'s will be the variational trial wave functions. 
The partition function $\tilde{Z}_{Q}$ and the inequality Eq. (\ref{eq:peierls}) will provide the basic framework for the successive  variational computations to be summarized here. We anticipate that,  before performing the variational computation in each case, a very crucial and delicate question   will be  the choice of   suitable trial wave functions (say, the $\Phi_{\lambda}$'s). For that purpose, the following   key fact, genuine of the Born-Oppenheimer approximation \cite{MesII}
 will be very helpful. On physical grounds, the (fast) vibrational and the (slow) rotational degrees of freedom may be decoupled, at least as a zeroth order of 
 approximation, because the vibrational energies are larger than the rotational ones  \cite{MesII}. 
 
 At about room temperature, 
rotational configurations of groups of atoms  in  
 ss macromolecules   can be regarded to be, typically,  in excited rotational states. 
 This fact suggests  that, at about $300$ K,  the slow rotational degrees of freedom  
 could  be  treated  through Classical Statistical Mechanics. Based upon all that,   
the successive approximations   presented in  sections 4, 5 and 6 for ss chains   will follow 
 the following pattern: we shall    start out from a quantum-mechanical formulation, deal 
firstly with and decouple the quantized vibrational degrees of freedom, turn to the rotational ones 
(at the quantum level first) and take the classical limit thereof.  At the end, the classical statistical 
 description will  appear as a justified approximation. Our final purpose will be 
to arrive, following such a route, at a simpler and reliable effective classical hamiltonian and partition function 
for  ss 
  macromolecules, 
depending only  on  the relevant slowly-varying degrees of freedom (all angular variables, in this case). 
That pattern   will be extended to double-stranded chains in   \ref{sec:RFAE.GFC.7}.

\chapter{Single-stranded open   macromolecules }
\label{sec:RFAE.GFC.4}

\section{Freely-jointed chains: radial variational computation }
\label{subsec:RFAE.GFC.4.1}

We shall treat in  \ref{subsec:RFAE.GFC.4.1} and \ref{subsec:RFAE.GFC.4.2}     freely-jointed chains
  (only  the  bond lengths  being
 constrained).  
Freely-jointed chains  constitute  an idealization   because, in addition to bond length,  angles are also 
constrained in real  macromolecules. In spite of that, the study of freely-jointed chains is absolutely 
essential, because the 
interactions which produce  the 
 bond length constraint are the most important contributions to $U({\bf y})$ and, hence, are responsible 
for the existence of the chain as an extended connected object.

We  assume that $U({\bf y})$  includes interactions  among the  atoms   only if they lie in nearest neighbor positions. 
We shall consider two possibilities. 
In the first,  $U({\bf y})$ describes 
  harmonic-oscillator-like  potentials (and equals  $U$, for Fraenkel's model):
\begin{eqnarray}&&
U({\bf y}) =  \sum_{i=1}^{N-1}\frac{\omega_{i}^{2}}{2B_{i}}(y_{i}-d_{i})^{2}\; . \label{eq:Uy}
\end{eqnarray}
  $\omega_{i} (>0)$ are frequencies, while $d_{i} (>0)$  are  bond lengths (the equilibrium distances between two successive atoms). 
In the second possibility, we use the Morse potential \cite{Morse}:
\begin{eqnarray}
V_{M}(y) = D \left\{ \exp{[-2\alpha(y-d)]} - 2\exp{[-\alpha(y-d)]} \right\} \; .
\label{eq:morse}
\end{eqnarray}
 $D$ is the dissociation energy corresponding to the equilibrium distance $d$ and 
$\alpha^{-1}$  represents  the range of the potential. Then, instead of (\ref{eq:Uy}), we consider: 
\begin{eqnarray}
U({\bf y}) =  \sum_{i=1}^{N-1}V_{M}(y_{i}) \; , \label{eq:morsei}
\end{eqnarray}
each $V_{M}(y_{i})$ with its corresponding $\alpha_{l}$ and $D_{l}$. 
  We shall introduce: $\omega_{l}\equiv \alpha_{l}\sqrt{2D_{l} B_{l}}$.  
 
  Our purpose is to arrive, through a variational computation [via Eq. (\ref{eq:peierls})], at a model 
for a microscopic chain, in which all $y_{i}$
  be  constrained at the quantum-mechanical level.
     We shall 
follow the presentation in \cite{RaCa}, to which we refer for further details.
By recalling the difference between fast vibrational (hard) and  slow rotational (soft) degrees of freedom, 
 we  choose  the trial variational  wave function $ \Phi ({\bf y})$  as a product of two wave functions 
$\phi (y)$ and $\psi_{\sigma} (\theta , \varphi ) $ depending, respectively, on the radial and 
angular variables ($\sigma$ denotes a set of quantum numbers). 
  Our choice of the 
 trial radial wave 
function   is: 
\begin{eqnarray}
\phi (y) = \prod_{l=1}^{N-1} \phi_{l} (y_{l})\;  .\label{eq:rad}
\end{eqnarray}
 Then, the (normalized) trial variational  (radial plus angular) wave function of the chain reads:
\begin{eqnarray}
\Phi=\Phi ({\bf y}) = \left[ \prod_{l=1}^{N-1} \phi_{l} (y_{l}) \right] \psi_{\sigma} (\theta , \varphi ) \; .
\label{eq:tot}
\end{eqnarray}
 Throughout this work, we shall suppose that all vibrational frequencies  $\omega_{l}$   are 
very large (and, more or less, of similar orders of magnitude) and fulfill: 
\begin{eqnarray}
\hbar\omega_{l}\gg k_{B}T, \hbar\omega_{l}\gg d_{l}^{-2}\hbar^{2}B_{l} \; ,
\label{eq:frec}
\end{eqnarray}
with the corresponding $\omega_{l}$ for either (\ref{eq:Uy}) or (\ref{eq:morsei}). 
For Eq. (\ref{eq:Uy}), each $\phi_{l} (y_{l})$ is chosen as  a Gaussian \cite{MesI}, approximating the ground state of the 
corresponding 
 harmonic-oscillator-like  potential:
\begin{eqnarray}
\phi_{l} (y_{l}) = d_{l}^{-1} \left( \frac{\omega_{l}}{\hbar \pi B_{l}} \right)^{1/4} \exp{ \left[ -\frac{\omega_{l}}{2 \hbar B_{l}} 
(y_{l} - d_{l})^2 \right] }, \hspace*{1cm} l = 1, \ldots, N-1 \; . 
\label{eq:osc}
\end{eqnarray}
   In order to perform calculations to leading order in frequencies, we shall make use of the following property of 
Dirac's delta function ($\delta(y-d)$):
\begin{eqnarray}
\lim_{\lambda \to \infty} \left( \frac{\lambda }{\pi} \right)^{1/2} \frac{\exp{ \left[ -\lambda (y - d)^2 \right] }}{d^2} = 
\frac{\delta (y - d)}{d^2}\;  .
\label{eq:Dirac}
\end{eqnarray} 
For Eq. (\ref{eq:morse}), we consider the   normalized radial wavefunction $\phi_{M,n=0}(y)$ \cite{Morse,GalPas},  
  associated  to its  ground state. It corresponds to 
the discrete ground state energy $E_{M,n=0}$: 
\begin{eqnarray}
E_{M,0}= -D +   \frac{\hbar\omega}{2} -
\frac{\hbar^{2}\omega^{2}}{16D},\, \omega=\alpha\sqrt{2D B_{j}}\;  ,
\label{eq:levels}
\end{eqnarray}
and so on for ~(\ref{eq:morsei}), with  ground state energy $\sum_{l=1}^{N-1}E_{M,l,n=0}$.  $E_{M,l,n=0}$ is given by
 ~(\ref{eq:levels}), with $D_{l}$, $\alpha_{l}$, $\omega_{l}$. We shall notice the following useful property
\begin{eqnarray}
\left|\phi_{M,n=0}(y)\right|^{2} \rightarrow   \frac{\delta(y-d)}{d^{2}}\; ,
\label{eq:dirac}
\end{eqnarray} 
when  $\alpha \rightarrow +\infty$ and $D \rightarrow +\infty$, while the dimensionless ratio 
$\sqrt{2D}/(\hbar\alpha[B _{j}]^{1/2})$ 
equals a finite  constant, so that $\omega_{l}\equiv \alpha_{l}\sqrt{2D_{l} B_{l}}$ grows  (say, when the Morse potential is very deep).  
Eq. ~(\ref{eq:dirac}) plays for the 
Morse potential the same role as Eq. ~(\ref{eq:Dirac}) for the harmonic-oscillator-like one. We have obtained a  mathematical 
 justification of  ~(\ref{eq:dirac}), which is a bit lengthy and, so, will be omitted.  
An important feature of the Morse potential is that it is a bounded potential, so that the Schroedinger equation containing it has a finite 
  number of bound states and  
 it can account for both 
 bound as well as unbound states. This clearly distinguishes it from   the harmonic oscillator-like 
potential, which rise to an infinite number of bound states, but no  unbound ones.
 
Hence,  as each $\omega_{l} \rightarrow \infty$ ($l = 1,\ldots, N-1$), both  Eqs.\ (\ref{eq:Dirac}) and \ (\ref{eq:dirac}) yield:
\begin{eqnarray} 
\left| \phi (y) \right|^{2} \rightarrow  \prod_{l=1}^{N-1} \frac{\delta (y_{l} - d_{l})}{d_{l}^{2}} \; .
\label{eq:Del2}
\end{eqnarray}
The normalization condition $\int [d{\bf y}]\mid \Phi ({\bf y})\mid ^{2}=1$ ( with   $[d{\bf y}]$  given in ~(\ref{eq:32})) becomes  in the limit $\omega_{l} \rightarrow \infty$:
\begin{eqnarray}
\int [{\bf d\Omega}] |\psi_{\sigma} (\theta , \varphi )|^{2} = 1 \;  , 
\label{eq:normaliz}
\end{eqnarray}
with  $[{\bf d\Omega}]$ given in Eq. ~(\ref{eq:37}). 
 We emphasize that, except for the above normalization condition, $ \psi_{\sigma} (\theta , \varphi )$ is fully arbitrary. 
 We evaluate, as all frequencies $\omega_{i}$, $i=1,\ldots,N-1$, grow very large, 
the quantum expectation value 
$(\Phi, \tilde{H}_{Q} \Phi )\equiv \int  [d{\bf y}] \Phi ({\bf y})^{*}\tilde{H}_{Q} \Phi ({\bf y})$, by 
using Eqs.\ (\ref{eq:Del2}), (\ref{eq:tot}) and (\ref{eq:normaliz}). The details of the computation are summarized in appendix 
A. The result can be cast as:
\begin{eqnarray}
(\Phi, \tilde{H}_{Q} \Phi ) &=  &  E_{0}+ (\psi_{\sigma}, H_{Q}^{(o)}   \psi_{\sigma}) +  {\cal O} ^{( o)}(\hbar)\;  ,
\label{eq:hfrec}\\
( \psi_{\sigma}, H_{Q}^{(o)}   \psi_{\sigma})  &=&  \int[{\bf d\Omega}] \psi_{\sigma}^{*} (\theta , \varphi) H_{Q }^{(o)} 
\psi_{\sigma}(\theta , \varphi ) \;  ,\label{eq:definham}
\end{eqnarray}
for any  $\psi_{\sigma}(\theta , \varphi )$.  Here, $E_{0}$ stands for 
$\sum_{l=1}^{N-1}E_{M,l,n=0}$ and $\sum_{i=1}^{N-1}\frac{\hbar \omega_{i}}{2}$ (the sum of all zero-point energies) for all Morse and harmonic-oscillator-like potentials, respectively. The new quantum (angular) Hamiltonian $H_{Q}^{(o)}$ reads:
\begin{eqnarray}
H_{Q}^{(o)}  =  \sum_{i=1}^{N-1} \frac{B_{i}}{2} \left[ \frac{({\bf e}_{i} \cdot {\bf e}_{i})}{d_{i}^{2}} \right]  -
 \sum_{i=2}^{N-1}\frac{1}{M_{i}} \left[ \frac{({\bf e}_{i-1} \cdot {\bf e}_{i})}{d_{i}d_{i-1}} \right]  ,
\label{eq:HQANG}
\end{eqnarray}
 with ${\bf e}_{l}$ being the following   operator: ${\bf e}_{l}\equiv i\hbar {\bf u}_{l} -{\bf a}_{l}$, $l=1,\ldots,N-1$. Through direct partial integrations, 
 ${\bf e}_{l}$ can be shown to be a Hermitean operator. 
  Notice that  $H_{Q}^{(o)} $ is expressed in terms of  the 
${\bf e}_{l}$'s, which could be regarded as some sort of quantized transverse momenta.  Finally,  
${\cal O} ^{( o)}(\hbar)$   denotes the  set of all remaining contributions  which do not depend on the frequencies $ \omega_{i}$: 
they are   proportional to some positive  power of $\hbar$. Contributions of this sort will also arise 
in the analysis of other quantum chains, with more complicated constraints. In this tutorial review, we shall not give 
explicitly the 
various ${\cal O} (\hbar)$'s which arise successively, because they are increasingly complicated,  can be obtained from 
 \cite{AE,Calvo,RaCa,CAEDNA,CAEDNA1} and 
become eventually irrelevant  when one takes $\hbar\rightarrow 0$. We shall concentrate on the $H_{Q}^{(o)} $'s, due to their interesting structures 
and because they give a nonvanishing classical  limit.  

It is important to notice that all  dependences of $ (\Phi, \tilde{H}_{Q} \Phi )$  on the frequencies  appear only  in the constant 
$ E_{0}$, in the right hand side of Eq.\ (\ref{eq:hfrec}). This solves the difficulties a), b) and c) at the end of subsection \ref{subsec:RFAE.GFC.2.2}.
\par
We 
shall  state very briefly  the following  
mathematical properties of the ${\bf e}_{l}$'s, which correct and complement adequately a previous discussion in \cite{AE}, and which will
 be physically relevant at the end.  
Let us  write ${\bf e}_{l}=( e_{l,1},e_{l,2},e_{l,3})$, $e_{l,1} $ and so on being  the three Cartesian components of ${\bf e}_{l}$. 
Let ${\bf l}_{l}={\bf y}_{l}\times[-i\hbar\nabla_{{\bf y}_{l}}]=( l_{l,1},l_{l,2},l_{l,3})$ be the quantized Hermitean) orbital 
 angular momentum associated 
to ${\bf y}_{l}$. Let $[A,B]=AB-BA$ be the commutator of the operators $A$ and $B$. The 
commutation relations of  $l_{l,1} $,   $l_{l,2} $ 
  and $l_{l,3} $  are well known     in Quantum Mechanics \cite{MesI,MesII}: 
$[l_{k,\alpha},
l_{j,\beta} ]=i\hbar\delta_{k,j}\epsilon_{\alpha\beta\gamma}l_{j,\gamma}$, $\alpha,\beta=1,2,3$ \cite{MesI,MesII}. Here, 
$\delta_{k,j}$ and $\epsilon_{\alpha\beta\gamma}$ are, respectively,
 the Kronecker delta and the standard totally antisymmetric tensor with three indices ($\epsilon_{123}=+1$, etc.) 
    One  
could ask what are the   
commutation  relations  of the  operators $e_{l,1}$, $e_{l,2}$ and $e_{l,3}$. Such a question is quite natural, at least from a 
purely mathematical  standpoint, as an attempt to characterize the ${\bf e}_{l}$'s somewhat more. 
Through some lengthy algebra, one  gets ($\alpha,\beta=1,2,3$): 
\begin{eqnarray} 
[e_{k,\alpha},
e_{j,\beta}]=-i\hbar\delta_{k,j}\epsilon_{\alpha\beta\gamma}l_{j,\gamma} \; ,
[l_{k,\alpha},
e_{j,\beta}]=i\hbar\delta_{k,j}\epsilon_{\alpha\beta\gamma}e_{j,\gamma} \; ,
 \label{eq:comrel}
\end{eqnarray}
 to which the above 
commutation relations of $l_{l,1} $,
 $l_{l,2} $ and 
$l_{l,3} $   should be added.  A simple, but  important, consequence   is that $H_{Q}^{(o)} $ is a Hermitean 
operator. 
   One also has:  ${\bf e}_{j}^{2}={\bf l}_{j}^{2}+\hbar^{2}$ and   ${\bf e}_{j}.{\bf l}_{j}={\bf l}_{j}.{\bf e}_{j}=0$. 
    Then, Eqs. (\ref{eq:HQANG}),     (\ref{eq:comrel}) and the 
commutation relations of $l_{l,1} $,
 $l_{l,2} $ and 
$l_{l,3} $  imply, consistently, that the quantum total orbital angular momentum $\sum_{j=1}^{N-1} {\bf l}_{j}$  commutes with  
$H_{Q}^{(o)} $,  which is physically important. See \cite{AE} and references therein. Thus,    the
conservation of the total orbital angular momentum  fully justifies, a posteriori, 
the  previous 
(painstaking) excursion into those commutation relations and algebraic matters. We shall add here the following  curious 
feature (not commented before) of the above algebra for the quantum freely-jointed chain. The  
commutation relations in (\ref{eq:comrel}) together with  those of $l_{l,1} $,
 $l_{l,2} $ and $l_{l,3} $  coincide with the ones for 
 the generators of the homogeneous Lorentz group, which, in turn,  play a crucial role 
in  Special Relativity  \cite{Schweb}.  Non-relativistic quantum-mechanical approaches 
to macromolecular chains and Special Relativity appear to be   subjects very  
disjoint and disconnected from each other. 
So, the unexpected appearance of the same algebraic structures (the same closed algebra, formed by the above commutation relations) 
in both subjects is certainly curious and surprising. Then, $H_{Q}^{(o)} $ and  the above algebraic  structure at each bond vector 
describe an extended, connected and flexible 
quantum system  (some sort of non-relativistic quantum ``string'', loosely speaking).  
For the quantized version of Kramers' model, one can introduce a   set of variables   which, even if different from $e_{k,\alpha}$'s, 
 play  a role similar to the  latter.  Such variables in Kramers' model, together with  an adequate formulation of the 
corresponding quantum angular momentum  variables, can be 
shown to generate a closed algebra identical to the above one for $e_{k,\alpha}$'s  and $ l_{l,\beta}$'s.  We omit details 
and refer to  \cite{AEPR}.  
\par

 We proceed to implement Eq. (\ref{eq:peierls}) in this  case, with 
  $\Phi_{\lambda} = \Phi ({\bf y})$ and  $(\Phi_{\lambda}, \tilde{H}_{Q} \Phi_{\lambda})$ being 
 given in Eqs.\ (\ref{eq:tot}) and Eq. (\ref{eq:hfrec}).
 Then, Eq. (\ref{eq:peierls}) becomes:
\begin{eqnarray}
\tilde{Z}_{Q} & \geq & \exp\left[-(k_{B}T)^{-1}E_{0}\right]\cdot Z_{Q}^{(o)}  \;  ,\label{eq:peierls1} \\  
Z_{Q}^{(o)}  & \equiv &
\sum_{\sigma} \exp{\left[-(k_{B}T)^{-1}\int[{\bf d\Omega}] \psi_{\sigma}^{*} (\theta , \varphi )  H_{Q} ^{(o)} \psi_{\sigma}
(\theta , \varphi )+{\cal O}^{( o)} (\hbar)\right]}\; .
\label{eq:peierls2}
\end{eqnarray}
 $Z_{Q}^{(o)}$ can be regarded as the approximate three-dimensional  quantum partition function for the 
slowly-varying angular degrees of freedom of the open ($o$) 
 chain (in the   framework of   
Peierls' variational inequality). The bond lengths are   constants and, so,  constrained in the  
quantum-mechanical $Z_{Q}^{(o)}$. Notice that if ${\cal O}^{( o)} (\hbar)$ is disregarded and the angular wave functions  
$\psi_{\sigma}(\theta, \varphi )$ are taken 
as  the complete set of all orthonormal eigenfunctions of $ H_{Q}^{(o)}$, then  Eq.\ (\ref{eq:peierls2}) becomes 
$Z_{Q}^{(o)}= {\rm Tr}[\exp{ [-(k_{B}T)^{-1}H_{Q}^{(o)}]}$. This $Z_{Q}^{(o)}$ includes the contributions of all possible 
quantum states and, hence, for 
all possible values of the squared total angular momentum ($(\sum_{j=1}^{N-1} {\bf l}_{j})^{2}$) and of its third component 
($\sum_{j=1}^{N-1}  l_{j,3}$). Those values being  the standard discretized values \cite{MesI,GalPas}. In recent  years, there are  interesting researches on   classical partition functions of single-stranded chains
 with configurations 
corresponding to a given value of the  total angular momentum. 
\cite{Deut1,Deut2}.   
The  construction, through the 
above variational procedure, of  quantum 
partition functions including only states  with prescribed  values of $(\sum_{j=1}^{N-1} {\bf l}_{j})^{2}$ and 
$\sum_{j=1}^{N-1}  l_{j,3}$ stands as an open problem.

Let all atoms   be identical to one another ($M_{i}=M$ in the  chain,  $i=1,\ldots,N$) and, for simplicity,  
 let all bond lengths be equal
 ( $d_{j}=d$, $j=1,\ldots,N-1$). Then,  the very existence of the bond length constraints, forcing identical atoms 
to lie successively along the chain  destroys  their  indistinguishability to a very large extent, but  not completely. A 
discussion for  the quantum Kramers' model (for both bosons and fermions)  appears in
 \cite{AEPR}, which appears to be also  valid for the actual (variational) quantization of Fraenkel's model. The trial 
variational wave function \ (\ref{eq:tot}) and its radial part ($\left[ \prod_{l=1}^{N-1} \phi_{l} (y_{l}) \right]$) 
 also hold,  while
$\psi_{\sigma} (\theta , \varphi )$   has to fulfill the surviving indistinguishability restrictions.

\par   
 We shall apply briefly the  argument leading  to  Eq.\ (\ref{eq:peierls2}) to  a 
one-dimensional quantum macromolecular chain, at thermal equilibrium. By assumption, the quantum 
Hamiltonian is given by the right-hand-side (rhs) of  Eq.\ (\ref{eq:Ha}) with 
${\bf y}_{i}$ and  $\nabla_{{\bf y}_{i}}$ replaced by  $ y_{i}$ and $\partial/\partial  y_{i}$, respectively, 
the one-dimensional relative coordinate $ y_{i}$ being positive or negative. Also, the potential energy 
among the atoms in the chain is provided by the rhs of  Eq.\ (\ref{eq:Uy}), with $ y_{i}$ substituted by 
$\mid y_{i}\mid$. The variational trial function is given by the rhs of  Eq.\ (\ref{eq:osc}), with 
$d_{l}^{-1}$ replaced by $2^{-1}$ and $ y_{l}$ substituted by $\mid y_{l}\mid$. Then, the actual 
counterpart of  Eq.\ (\ref{eq:hfrec}) becomes, simply: $(\Phi, \tilde{H}_{Q} \Phi ) =  E_{0} + 
{\cal O}(\hbar)$, with $E_{0}= \sum_{i=1}^{N-1}\frac{\hbar \omega_{i}}{2}$ and ${\cal O}(\hbar)$ being frequency-independent and 
of order $\hbar$, at least.  Eq.\ (\ref{eq:peierls1}) also holds, now with 
$Z_{Q}^{(o)}=\exp[-(k_{B}T)^{-1}{\cal O}(\hbar)]$ (which, in turn, approaches unity as 
$\hbar\rightarrow 0$). Notice that $\exp[-(k_{B}T)^{-1}\sum_{i=1}^{N-1}\frac{\hbar \omega_{i}}{2}]$ is the 
partition function for a set of harmonic oscillators in the high-frequency limit, and, so, it differs 
drastically from that for classical oscillators \cite{MesI}.  

So, for large vibrational frequencies,  we have obtained $H_{Q}^{(o)}$ from $\tilde{H}_{Q} $ in (\ref{eq:Ha}) and  $Z_{Q}^{(o)}$ 
from  (\ref{eq:parfuncquan}), through    (\ref{eq:peierls}).  
 We remind here  certain mathematically  rigorous results (as equalities in asymptotic limits, not  
 as variational inequalities) 
for other different models by several authors: see  \cite{Simon,Poly1,Poly2,Fink,Enc} and references therein. Their  results 
bear qualitative  similarities to ours, although there are also considerable differences.  In fact, those authors start from certain 
hamiltonians (depending on both hard and soft variables) and, without using   the variational inequality  (\ref{eq:peierls}), 
   take  the large coupling constant limit (akin to our large vibrational frequencies), so as to obtain: i)  asymptotic behaviors 
of (low-lying) eigenvalues of Schr\"{o}dinger operators \cite{Simon}, ii)  simpler hamiltonians and partition functions for spin 
chains (Haldane-Shastry chains), spins being here the soft variables \cite{Poly1,Poly2,Fink,Enc}.  

  Let two external stretching forces ${\bf f}$ and  
$-{\bf f}$ act upon the two atoms at  ${\bf R}_{1}$ and ${\bf R}_{N}$, respectively. Then, $\tilde{H}_{Q}$ in  
\ (\ref{eq:Ha})  is replaced by    $\tilde{H}_{Q,{\bf f}}=\tilde{H}_{Q}+{\bf f}\sum_{i=1}^{N-1}{\bf y}_{i}$. 
The quantum-mechanical  variational computation goes through as above and 
yields a new quantum partition function: $Z_{Q,{\bf f}}^{(o)}= {\rm Tr}[\exp [-(k_{B}T)^{-1}(H_{Q}^{(o)}+
 {\bf f}\sum_{i=1}^{N-1}d_{i}{\bf u}_{i})]]$.

\section{Freely-jointed chains: classical partition function }
\label{subsec:RFAE.GFC.4.2}
Eq. (\ref{eq:HQANG})  and    
Eq.\ (\ref{eq:peierls2}) simplify  enormously, if one proceeds to the classical ($\hbar\rightarrow 0$) limit  for the 
angular degrees of freedom, under the  additional assumptions:
\begin{eqnarray}
 k_{B}T\gg d_{l}^{-2}\hbar^{2}B_{l}\; ,
\label{eq:Claslim}
\end{eqnarray}
so that excited rotational states are occupied and quantum operators and statistics can be approximated by 
classical ones.  Then, $ {\cal O}^{( o)}(\hbar) $ and all quantities of order $\hbar $ or higher ( and  not containing any $\omega_{l}$) can be neglected, 
so that  Eq.\ (\ref{eq:frec}) be respected. 
 Eqs.\ (\ref{eq:HQANG}) and \ (\ref{eq:ai}) indicate  that the Hamiltonian $H_{Q} ^{(o)}$  becomes, in the classical limit: 
\begin{eqnarray}
H_{c}^{(o)} & = & \sum_{i=1}^{N-1} B_{i} \frac{({\bf a}_{i,c } \cdot {\bf a}_{i,c })}{2d_{i}^{2}} - 
\sum_{i=2}^{N-1}\frac{1}{M_{i}}  \frac{({\bf a}_{i,c } \cdot {\bf a}_{i-1,c })}{d_{i}d_{i-1}} \;. \label{eq:GClassa}
\end{eqnarray}
 The ${\bf a}_{i,c}$'s are no longer operators but classical variables (  the 
classical limit of  $ {\bf e}_{i}$). They are:  
 \begin{eqnarray}
{\bf a}_{i,c } = - {\bf u}_{\theta_{i}}P_{\theta_{i}} - \frac{{\bf u}_{\varphi_{i}}P_{\varphi_{i}}}{\sin{ \theta_{i}}}\;  . \label{eq:aClass}
\end{eqnarray}
 $P_{\theta_{i}}$, $P_{\varphi_{i}}$ are the classical momenta canonically conjugate to $\theta_{i}$ and $\varphi_{i}$, $i = 1, \ldots , N-1$. 
  
\par

The  three-dimensional classical partition function (the classical limit of Eq.\ (\ref{eq:peierls2}))  is 
($[dP_{\theta}dP_{\varphi}] = \prod_{l=1}^{N-1} (\sin{\theta_{l}})^{-1}dP_{\theta_{l}}
dP_{\varphi_{l}}$) \cite{AE,RaCa}: 
\begin{eqnarray}
Z_{c }^{(o)}  & \simeq & \frac{1}{(2\pi \hbar)^{2(N-1)}} \int [{\bf d\Omega}] \int [dP_{\theta}dP_{\varphi}] 
   \left[ \exp{-\frac{H_{c }^{(o)}}{k_{B}T}} \right] \; .\label{eq:oppar2}
\end{eqnarray}
 Are there    further classical constraints ( besides $y_{l}=d_{l}$), contained somehow in 
(\ref{eq:GClassa}))  and (\ref{eq:oppar2})), which   would be the counterparts of those expressing   
$\dot{  y }_{i}=0$ in   Kramers' classical model?. Let us introduce, for a short while, the classical momenta 
$\mbox{\boldmath{$\pi$}}_{j,c,F}=\pi_{j,c,F}
{\bf u}_{j}-
d_{j}^{-1}{\bf a}_{j,c } $ ( not to be confused with $\mbox{\boldmath{$\pi$}}_{i,c}$, employed and integrated over  in 
Fraenkel's classical  model in  \ref{subsec:RFAE.GFC.2.1},  $\pi_{j,c,F}$  being a radial momentum. 
   One can  develop a mathematical argument implying that $\pi_{j,c,F}=0$ \cite{AEM}. 

 The integrations over all the classical momenta in $Z_{c }^{(\text{O})}$ are 
Gaussian. Upon   performing  them \cite{AEM}, one finds: 
\begin{eqnarray}
&&Z_{c }^{(o)}  = \left[ \frac{k_{B}T}{2\pi \hbar^{2}} \right]^{N-1} 
\frac{\left[ \prod_{l=1}^{N-1} d_{l}^{2} \right]}{({\rm det} B)^{3/2}} Z_{R}\;  ,\label{eq:FinalPF0}\\&&
Z_{R}=\int [{\bf d\Omega}][\Delta_{N-1}]^{-1/2}  \;  , 
 \label{eq:FinalPF}\\&&
[\Delta_{N-1}]^{-1/2}= \left[ {\rm det} ({\bf u}_{i}(B^{-1})_{ij}{\bf u}_{j})\right]^{-1/2}\; ,\label{eq:determ}
\end{eqnarray}
The elements $B_{ij}$ of   tridiagonal matrix $B$, of order $(N-1) \times (N-1)$, are:
\begin{eqnarray}
B_{ij} = \left\{ \begin{array}{lll}  
\frac{1}{M_{i}} + \frac{1}{M_{i+1}} & \mbox{if $i = j$}\\
-\frac{1}{M_{i}} & \mbox{if $j = i-1$ or $j = i+1$} \\
0 & \mbox{otherwise}
\end{array} \right. \label{eq:matrixB}
\end{eqnarray}
On the other hand,  the matrix $({\bf u}_{i}(B^{-1})_{ij}{\bf u}_{j})$ is given by:  
\begin{eqnarray}
({\bf u}_{i}(B^{-1})_{ij}{\bf u}_{j}) = \left\{ \begin{array}{lll}  
\frac{1}{M_{tot}} \left[ \sum_{k=1}^{i} M_{k} \right] \left[ \sum_{l=i+1}^{N} M_{l} \right] & \mbox{if $i = j$}\\
\frac{1}{M_{tot}} \left[ \sum_{k=1}^{i} M_{k} \right] \left[ \sum_{l=j+1}^{N} M_{l} \right] ({\bf u_{r_{i}}} \cdot {\bf u_{r_{j}}}) &  \mbox{if $i<j$} \\
\frac{1}{M_{tot}} \left[ \sum_{k=1}^{j} M_{k} \right] \left[ \sum_{l=i+1}^{N} M_{l} \right] ({\bf u_{r_{i}}} \cdot {\bf u_{r_{j}}}) &  \mbox{if $i>j$}
\end{array} \right. \label{eq:matrixinvB}
\end{eqnarray}
 $M_{tot} = \sum_{k=1}^{N} M_{k}$  and  
$({\bf u}_{i}(B^{-1})_{ij}{\bf u}_{j})$ is a symmetric matrix of order $(N-1) \times (N-1)$.
 The determinant  $[\Delta_{N-1}]^{-1/2}$  is manifestly 
rotationally invariant (as it depends on the scalar products $({\bf u}_{ i} \cdot  {\bf u}_{ j})$ but not 
directly on angles). However, the matrix $({\bf u}_{i}(B^{-1})_{ij}{\bf u}_{j}) $ is not tridiagonal, 
  unlike $(B_{ij})$.  The contribution of  $[\Delta_{N-1}]^{-1/2}$ differs from those of other 
 determinants  appearing in \cite{Kram,Fix,Go2,Mazars1,Eche,SSTtiEche}.  The equilibrium statistical average of a function $F=F({\bf u}_{1},...,
{\bf u}_{N-1})$ is: 
\begin{eqnarray}
\langle F\rangle=\frac{1}{Z_{c }^{(o)}}\left[ \frac{k_{B}T}{2\pi \hbar^{2}} \right]^{N-1} 
\frac{\left[ \prod_{l=1}^{N-1} d_{l}^{2} \right]}{({\rm det} B)^{3/2}}  \int [{\bf d\Omega}] \frac{F({\bf u})}{[\Delta_{N-1}]^{1/2}}\;  . 
 \label{eq:avePF}
\end{eqnarray}
 The internal energy $U$ of the linear polymer, which is an interesting thermodynamical property,  does not require the evaluation of $[\Delta_{N-1}]^{-1/2}$. 
Thus,  $U$  is obtained from $U=A-T(\partial A/\partial T)$, where $A$ is the free energy ($Z_{c }^{(o)} =\
exp[-(k_{B}T)^{-1}A]$). One finds readily energy equipartition, namely: $U=(N-1)k_{B}T$ \cite{AE,RaCa}.   
\par
We shall suppose   that the region  in which the ss open freely-jointed chain  moves (inside which  
 its center-of-mass  ${\bf R}_{CM}$ varies) is  a sphere of very  large radius $R_{0}$. Let  $M_{i}=M$ (with $i=1,\ldots,N$) and $d_{j}=d$ (with $j=1,\ldots,N-1$), namely, 
equal masses for all atoms and equal bond lengths in the  chain. Let us introduce the probability distribution function for the position vectors ${\bf  R}_{1}$,\ldots,${\bf  R}_{N}$:
\begin{eqnarray}
W({\bf  R}_{1},\ldots,{\bf  R}_{N})=\left\langle \prod_{l=1} ^{N-1}\delta^{(3)}\left({\bf R}_{l+1}-{\bf R}_{l}-d{\bf u}_{l}\right)\right\rangle\; ,
\label{eq:Gfj1} 
\end{eqnarray} 
$\delta^{(3)}$ being the three-dimensional delta function. 
Similarly, the probability distribution function for the end-to-end position vector is:
\begin{eqnarray}
W({\bf  R}_{N}-{\bf  R}_{1})=\left\langle \delta^{(3)}\left({\bf R}_{N}-{\bf R}_{1}-d\sum_{l=1}^{N-1}{\bf u}_{l}\right)\right\rangle \; .
\label{eq:Gfj12} 
\end{eqnarray} 
We remind the standard Gaussian model for ss open  freely-jointed chains: 
\begin{eqnarray}
W_{eq}({\bf  R}_{1},\ldots,{\bf  R}_{N}) &=&W_{eq}=\prod_{l=1}^{N-1}W_{G}({\bf R}_{l+1}-{\bf R}_{l};2d^{2})\;  ,
 \label{eq:Gfj2}\\
  W_{G}({\bf R}_{l+1}-{\bf R}_{l};2d^{2})&=&
[3/(2\pi d^{2})]^{3/2}
\exp[-3({\bf R}^{(r)}_{l+1}-{\bf R}^{(r)}_{l})^{2}/(2d^{2})]\;  .
\label{eq:Gfj3} 
\end{eqnarray} 
$W_{G}({\bf R}_{l+1}-{\bf R}_{l};2d^{2})$ is  the Gaussian distribution for  the $l$-th bond vector. See \cite{Gros,McQ,Doi} 
for   accounts  of the standard Gaussian 
model. Can  \ (\ref{eq:Gfj1}) be approximated by  \ (\ref{eq:Gfj2})? 
\par

For $N=3,4,5$, $[\Delta_{N-1}]^{-1/2}$  is given in   
  \cite{AEM}. Based upon the latter and through further 
  approximations on $[\Delta_{N-1}]^{-1/2}$ in \ (\ref{eq:determ}) for large $N$, one gets, for the   single  
open freely-jointed chain,   results for various physical quantities: correlations among 
bond vectors, squared end-to-end distance, probability distribution for the end-to-end vector, behaviour of the chain under 
weak  external stretching forces and ''rubber elasticity``, 
structure factor for small wave vector. Such results do agree with those implied by the standard Gaussian model (say, by \ (\ref{eq:Gfj2})), 
which provides a check of consistency: see \cite{AE} for details. Further   analysis on $[\Delta_{N-1}]^{-1/2}$,  
more detailed and considerably improved,  have been carried out in 
\cite{RaCa,CAEDNA,CAEDNA2} also for  the case $M_{i}=M$ ($i=1,\ldots,N$) and $d_{j}=d$ ($j=1,\ldots,N-1$), thereby providing 
further 
confirmation of the  consistency with the standard Gaussian 
model. Thus, one finds the following approximations \cite{AE,CAEDNA,CAEDNA2}:
\begin{eqnarray}
W({\bf  R}_{1},...,{\bf  R}_{N})&\simeq& 
W_{eq},\,  W({\bf  R}_{N}-{\bf  R}_{1})\simeq W_{G}({\bf R}_{N}-{\bf R}_{1};2(N-1)d^{2}) \; ,  
\label{eq:Gfj2bis}\\ 
 Z_{c }^{(o)}&\simeq&  \left[ \frac{k_{B}T}{2\pi \hbar^{2}} \right]^{N-1}\left[\frac{d^{2(N-1)}}{(N/M^{N-1})^{3/2}}\right] Z_{R,app} Z\; ,\label{eq:Gfj4}\\
Z&=& [\frac{4\pi R_{0}^{3}}{3}]^{-1}  \int  \left[ d^{3}{\bf R}_{N} d^{3}{\bf R}_{1}\right] G(N-1)\, ,
\label{eq:Gfj5} 
\\
G(N-1)&=&G({\bf  R}_{N};{\bf  R}_{1};N-1)=\int \left[\prod_{l'=2}^{N-1}d^{3}{\bf R}_{l'}\right] W_{eq}\;  .
\label{eq:Gfj6} 
\end{eqnarray} 
 Notice that $Z$ and $G(N-1)$ are manifestly rotationally invariant. Also, the factor $[(4\pi R_{0}^{3}/3)]^{-1}$ cancels out a 
similar contribution arising from the  integrations over both ${\bf R}_{N}$ and ${\bf R}_{1}$ in $Z$.  $Z_{R,app}$ ($T$-
independent) is an approximation estimate for $Z_{R}$ (see \cite{AE,RaCa} and Eqs. (C.4) and (C.2) in  ~\cite{CAEDNA}), 
 and it  will not be relevant here.  The  representation in \ (\ref{eq:Gfj4})-\ (\ref{eq:Gfj6}) displays quite directly 
the connection with the standard Gaussian model.   
One  finds \cite{AE,CAEDNA,CAEDNA2}: a) $\langle{\bf u}_{i}\cdot{\bf u}_{j}\rangle\simeq 0$ for $i\neq j$, b) 
$\langle({\bf R}_{N} - {\bf R}_{1} )^{2}\rangle\simeq (N-1)d^{2}$. Eq \ (\ref{eq:Gfj4}) includes  $Z$ explicitly,  
 even if $Z=1$:  such a redundancy is justified because (\ref{eq:Gfj4})-(\ref{eq:Gfj6}) will help to understand  the 
  generalizations \ (\ref{eq:dsDNA})-\ (\ref{eq:green0})   for ds chains (\ref{sec:RFAE.GFC.8}), in which the counterpart of 
$Z$ will be $\neq 1$. 
 
Let us revisit briefly the case with   external stretching forces  included at the end of \ref{subsec:RFAE.GFC.4.1}. Then, 
    $Z_{Q,{\bf f}}^{(o)}$ becomes, in the classical limit, $Z_{c, {\bf f}}^{(o)} $, given by 
(\ref{eq:FinalPF0}) and (\ref{eq:FinalPF}), provided that the latter includes, in its integrand,   a 
factor $\exp [-(k_{B}T)^{-1} {\bf f} 
\sum_{i=1}^{N-1}d_{i} {\bf u}_{i} ]$.  For applications, see \cite{AE}.

\section{Chain with weak next-to-nearest-neighbours interaction}
\label{subsec:RFAE.GFC.4.3}

We continue to assume   $d_{i}=d$ and $M_{i}=M$, $i=1,\ldots,N-1$. Here. we shall suppose that, besides the strong  
 nearest-neighbour interactions     through   $V_{j}=(2B_{j})^{-1}\omega_{j}^{2}(y_{j}-d)^{2}$ with large $\omega_{j}$ in 
(\ref{eq:Uy}), other  weak   potentials ($ V_{j,j+1} =  v(\mid {\bf y_{i}}+{\bf y_{i+1}}\mid )$) exist between  
atoms which are next-to-nearest 
neighbours ($nnn$), $v$ being a real function.  Then, the total interaction potential is now: $U({\bf y}) = \sum_{j=1}^{N-1}V_{j}+
\sum_{j=1}^{N-2} v(\mid {\bf y_{i}}+{\bf y_{i+1}}\mid)$.  
The case in which all $V_{j,j+1} $  be adequately strong (and, then,  constraint the bond angles)  will be analyzed in 
section \ref{sec:RFAE.GFC.6}.  The analysis in subsections \ref{subsec:RFAE.GFC.4.1} and \ref{subsec:RFAE.GFC.4.2} can be  extended straightforwardly to this case. We shall proceed to treat this chain in the classical limit. Then, \ (\ref{eq:FinalPF0})-\ (\ref{eq:FinalPF}) hold, with $Z_{R}$  replaced by: 
\begin{eqnarray}
&&Z_{R}=\int [{\bf d\Omega}][\Delta_{N-1}]^{-1/2}  
\exp\left[-(k_{B}T)^{-1}
U_{nnn}\right] \; ,
\label{eq:FinalPFw} 
 \end{eqnarray} 
with $U_{nnn}=
\sum_{i=1}^{N-2}v(d\mid {\bf u}_{i}+{\bf u}_{i+1}\mid )$. 
Let  $W({\bf r})$ be the end-to-end distribution   (${\bf r}={\bf  R}_{N}-{\bf  R}_{1}$),  given in (\ref{eq:Gfj12}), and let 
$G({\bf q})$ be its Fourier transform, 
${\bf q}(=(0,0,q))$ being the wavevector (compare with \cite{Flo}).  One has: 
\begin{eqnarray}
G({\bf q}) & = & \int  d^{3}{\bf r} W({\bf r})\exp(i{\bf q}{\bf r}) \nonumber \\
& = & \frac{\int [{\bf d\Omega}] [\Delta_{N-1}]^{-1/2}  \exp\left[id{\bf q}\cdot\sum_{j=1}^
{N-1}{\bf u}_{j}\right]\exp\left[-(k_{B}T)^{-1}U_{nnn}\right]}{\int
[{\bf d\Omega}] [\Delta_{N-1}]^{-1/2}  \exp\left[-(k_{B}T)^{-1}U_{nnn}\right]} \; .
 \label{eq:Gq}
\end{eqnarray}
We suppose that $v$ is not deep but very smooth, so that its minimum times $(k_{B}T)^{-1}$ is  negative and, in absolute 
magnitude, smaller than unity. 
 According to  \cite{Hua}, an important necessary
condition for the replacement of quantum partition functions by classical ones
be justified is that interparticle distances be appreciably larger than
thermal wavelengths (`` the classical
limit restriction''). Let  $\lambda_{th}=(2\pi\hbar^{2}/Mk_{B}T)^{1/2}$  be the thermal wavelength. The above 
classical limit  restriction reads $\lambda_{th}\ll d$, which holds for adequate  $T$. Let us consider  the distances 
$\left[({\bf R}_{l} - {\bf R}_{j})^{2}\right]^{1/2}$ (with  $l=1,\ldots,N-2$ and $N\geq j>l+1$) between the $l-th$ and the $j-th$ atoms. 
For a better characterization of the model, the integrations in Eq.\ (\ref{eq:FinalPFw}) should be performed, by assumption, 
 over all angles consistent with the classical limit restrictions: 
$\left[({\bf R}_{l} - {\bf R}_{j})^{2}\right]^{1/2}>\lambda_{th}^{2}>0$, in general. These would seem to amount to some sort of 
excluded-volume effects. Then, $({\bf R}_{i+2} - {\bf R}_{i})^{2} =2 d^{2}[1+{\bf u}_{i}\cdot{\bf  u}_{i+1}]>\lambda_{th}^{2}$. Then,  one 
can reasonably  conjecture that, for large $N$, $Z_{R}$ gets important contributions  from small integration domains where:  
a) ${\bf u}_{i}\cdot{\bf  u}_{i+1}$ is  close to $+1$, and b) ${\bf u}_{i}\cdot{\bf  u}_{i+1}$    is larger than  about $-1+\lambda_{th}^{2}/2 d^{2}$, with, in turn,  $1\gg\lambda_{th}^{2}/2 d^{2}>0$. 
Let  $\rho_{Q} (\leq 1)$ account
for a small excluded-volume effect arising from the above classical limit restriction. A reasonable estimate is $\rho_{Q}\simeq
1-\lambda_{th}^{2}/(2d^{2}\epsilon_{max})$, with $\epsilon_{max}$ ($0<\epsilon_{max}< 1$) being
a  cutoff value for  the integrations near ${\bf u}_{i}\cdot
 {\bf u}_{i+1} = - 1$. Clearly,  $\rho_{Q}=1$ amounts to
the absence of  classical limit
restrictions.  We shall estimate $ \lambda_{th}$, at room
temperature ($k_{B}T\simeq 0.02$ electronvolts), for $d\simeq 5 \AA$,
$M\simeq 12$ proton masses ( about  the mass of a Carbon atom). One finds:  $\lambda_{th}\simeq 0.6$ \AA. One has  $\rho_{Q}\simeq 0.84 $ for   
 $\epsilon_{max}\simeq 0.05$, while 
    $\rho_{Q}\simeq 0.986 $ if  $\epsilon_{max}\simeq 0.5$. 
 We introduce $\rho\equiv\rho_{Q}\exp\{(k_{B}T)^{-1}[v(2d)-v(\lambda_{th}/(2^{1/2} d))]\}(>0)$. In principle, $\rho$ could be  larger 
or smaller than unity, but not much in both cases, due to the assumed smoothness of  the actual $v$. 

\begin{figure}[t]
\begin{center}
{
\includegraphics[scale=0.4]{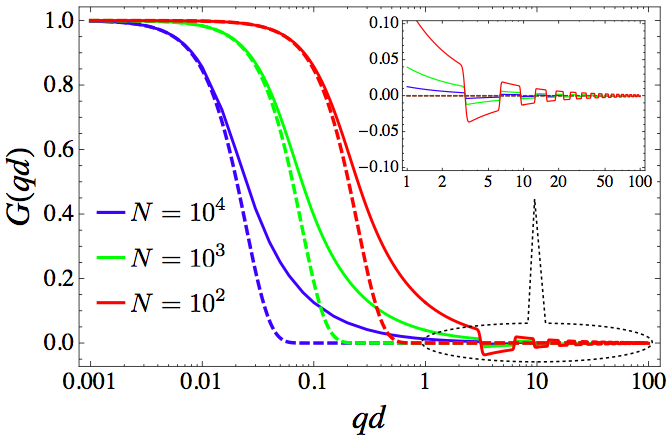}}
\end{center}
\vspace*{-3mm}
\caption{Comparison of $G(qd)$,  obtained from Eq. (\ref{eq:Gqapprox})  (dashed line), to  that given by the exact formula in Eq. 
(\ref{eq:Gq3}) (solid line), when one sets $\rho=1$ and increasing values of $N$. The inset shows the details of the curves within the dotted region.}
\label{fig:RFAE.GFC.2}       
\end{figure}

The squared end-to-end distance  $\langle ({\bf x}_{N} - {\bf
  x}_{1})^{2}\rangle$ for the chain with $N$ atoms can be obtained from $G({\bf q})$ upon expanding for small $q$ through:
$G({\bf q}) \simeq 1 - 6^{-1}q^{2}\langle ({\bf x}_{N} - {\bf
  x}_{1})^{2}\rangle + \cdots $. A somewhat lengthy calculation, gives:
\begin{eqnarray}
\langle({\bf x}_{N} - {\bf x}_{1})^{2}\rangle=d^{2}\left[ 1 +
  \frac{N}{\rho} + \frac{(1-\rho)^{N} -
  (1+\rho)^{N}}{2\rho^{2}(1+\rho)^{N-2}}\right]  . \label{eq:endtoendF} 
\end{eqnarray}
See appendices B  and C for   the derivation of Eq. (\ref{eq:endtoendF}) in outline. Then, for small-$q$: 
     \begin{eqnarray}
G({\bf q})\simeq  \exp\left[ - \frac{q^{2}\langle ({\bf x}_{N} - {\bf
  x}_{1})^{2}\rangle}{6} \right] \; .
\label{eq:Gqapprox} 
\end{eqnarray}
Figure \ref{fig:RFAE.GFC.2} shows a comparison of $G(qd)(=G({\bf q}))$ as obtained
from Eq. (\ref{eq:Gqapprox}) and from Eq. (\ref{eq:Gq3}), when we set in both $\rho = 1$. In general, as $N$
increases, there is a significant improvement of the agreement between formulas (\ref{eq:Gq3}) and (\ref{eq:Gqapprox}).

Inverting Eq.(\ref{eq:Gq}), with (\ref{eq:Gqapprox}), one 
finds  the  end-to-end   distribution: 
\begin{eqnarray}
W({\bf r})  \simeq  \left[ \frac{3}{2\pi\langle ({\bf x}_{N} - {\bf
  x}_{1})^{2}\rangle}\right]^{3/2}\exp\left[ - \frac{3r^{2}}{2\langle ({\bf x}_{N} - {\bf
  x}_{1})^{2}\rangle} \right] \; .
\label{eq:Wapprox} 
\end{eqnarray}
Some interesting consistency checks of Eqs. (\ref{eq:endtoendF})-(\ref{eq:Wapprox})
are: 1) for $\rho=1$ ( by omitting the classical limit restrictions
and assuming $U_{nnn}\simeq 0$), one gets $\langle({\bf x}_{N} - {\bf
  x}_{1})^{2}\rangle=d^{2}(N-1)$ (random coil); 2) for  $\rho$ close to $ 0$
(that is, going to the limit of  physical reliability  of our computation and
perhaps beyond, as $v$ has been assumed to be smooth)  and fixed $N$, some
algebra gives $\langle({\bf x}_{N} - {\bf x}_{1})^{2}\rangle\simeq
d^{2}(N-1)^{2}$ (fully stretched chain); 3) as $N\rightarrow +\infty$, with
fixed $\rho$ in $0<\rho\leq 1$, one gets  $\langle({\bf x}_{N} - {\bf
  x}_{1})^{2}\rangle\simeq d^{2}(N/\rho)$. We emphasize that when
$\rho=1$, Eq. (\ref{eq:Wapprox}) coincides with the well known result for the
Gaussian end-to-end probability distribution \cite{Volk,Flo,Gros,McQ} and also  provides a check of consistency,
 a posteriori,  of the approximations 
made in appendix B.   On the other hand,  2) and 3) indicate that the limits   $N\rightarrow +\infty$ and 
 $\rho\rightarrow 0$ cannot be interchanged. Clearly,  a  behavior for 
$\langle({\bf x}_{N} - {\bf x}_{1})^{2}\rangle$ proportional to $d^{2}N^{6/5}$ for arbitrarily large $N$ 
(say, Flory`s behavior) cannot be obtained analytically in the  actual model with next-to-nearest neighbors interactions 
only. However, for small $\rho$ and  a restricted but interesting range of (large) values for $N$, Eq.  (\ref{eq:endtoendF}) 
gives values which could resemble Flory`s behavior numerically. In fact, take, for instance,  
$\rho\simeq 0.2$ and $10^{6}\leq N\leq 10^{8}$: then, $\ln[\langle({\bf x}_{N} - {\bf x}_{1})^{2}\rangle/d^{2}]/(\ln N)$ 
varies between $1.116$ and $1.087$. It is also interesting to compare  Eq. (\ref{eq:endtoendF}) with the squared end-to-end distance 
$\langle({\bf x}_{N} - {\bf x}_{1})^{2}\rangle_{V}$, obtained in another
model for an open linear chain, treated as a cooperative system \cite{Volk}:
the latter model in \cite{Volk} is not based on Quantum Mechanics and is
formulated directly on an analogy with the one-dimensional Ising model. In
general, $\langle({\bf x}_{N} - {\bf x}_{1})^{2}\rangle_{V}$ differs from
Eq.  (\ref{eq:endtoendF}), but in the limit $N\rightarrow +\infty$,
$\langle({\bf x}_{N} - {\bf x}_{1})^{2}\rangle_{V}\simeq d^{2}(N/\rho)$,
that is, it agrees with the above result 3) (when, in the latter, we set
$\rho_{Q}=1$, $v(2d)-v(\lambda_{th}/(2^{1/2} d))<0$), which provides a partial
check of physical consistency for  Eq.  (\ref{eq:endtoendF}).

 \section{Star freely-jointed polymer}
\label{subsec:RFAE.GFC.4.4}

Another interesting class of macromolecules is formed by the so-called open  star polymers  
\cite{Elias}. The procedures for dealing with quantum constraints discussed in    \ref{subsec:RFAE.GFC.4.1}
can be generalized,  without  
essential difficulties, and have led to a  model for 
  a  three-dimensional open  star polymer  \cite{RaCa}, as we now summarize. 
We treat   the case of $n$ arms ($n\geq 3$), all of which start from a certain atom 
(``the vertex or origin of the star''), with position vector $ {\bf R}_{0}$ and mass $M_{0}$. The   $r$-th arm has $N_{r}$ atoms (without counting the ``vertex''), forming a linear subchain ( $1\leq r\leq n$). The mass and the  position vector of the $i$-th atom  along the $r$-th arm are   $M^{(r)}_{i}$ and  ${\bf  R}^{(r)}_{i}$,  $1\leq i\leq N_{r}$).   Then, let  the center-of-mass (CM) position vector and the relative (``bond'') ones  along the  $r$-th arm  be denoted by ${\bf R}_{CM}$,
  and ${\bf y}^{(r)}_{i}$, respectively. One defines:
\begin{eqnarray}&&
{\bf R}_{CM} = \frac{ M_{0} {\bf R}_{0} +\sum_{r=1}^{n}\sum_{i=1}^{N_{r}} M^{(r)}_{i} {\bf R}^{(r)}_{i} }{M_{tot} }\; , 
M_{tot} = M_{0}+\sum_{r=1}^{n}\sum_{i=1}^{N_{r}} M^{(r)}_{i}\; ,\label{eq:Rcms}
\end{eqnarray}
 and ${\bf y}^{(r)}_{i} = {\bf R}^{(r)}_{i} - {\bf R}^{(r)}_{i-1}      $ for  $(  1 \leq i \leq N_{r})$, with the understanding that, 
for any $r$:  ${\bf R}^{(r)}_{0}\equiv  {\bf R}_{0}$ (the position vector of ``the vertex''). Below, we denote the set of atomic coordinates $({\bf y}^{(r)}_{i})$ by ${\bf y}$. 
The starting quantum Hamiltonian  is also  Eq.\ (\ref{eq:H1}), with obvious replacements of indices. 
 Then, using expressions (\ref{eq:Rcms}) and ${\bf y}^{(r)}_{i}$, (\ref{eq:H1}) gives 
(after   various   cancellations) the quantum Hamiltonian in terms of the new variables as: 
$ H_{Q,1}  =  - (\hbar^{2}/2M_{tot} )\nabla_{\bf R_{CM}}^{2} +\tilde{H}_{Q}$, with :
\begin{eqnarray}&&
\tilde{H}_{Q}  =   - \frac{\hbar^{2}}{2M_{0}}(\sum_{r=1}^{n}\nabla_{{\bf y}^{(r)}_{1}})^{2} 
 - \sum_{r=1}^{n}\sum_{i=1}^{N_{r}} \frac{\hbar^{2}}{2M^{(r)}_{i}}( \nabla_{{\bf y}^{(r)}_{i}}-\nabla_{{\bf y}^{(r)}_{i+1}})^{2}  
 +  U({\bf y}) \; ,\label{eq:Has} 
\end{eqnarray}
with the understanding that $\nabla_{{\bf y}^{r}_{N_{r}+1}}\equiv 0$. 
The total potential energy interaction  $U({\bf y})=    
 V({\bf R})$ 
 is independent on ${\bf R}_{CM}$. We shall assume it to be: 
\begin{eqnarray}&&
U({\bf y}) =   \sum_{r=1}^{n}\sum_{i=1}^{N_{r}}\frac{(\omega^{(r)}_{i})^{2}}{2B^{(r)}_{i}}(y^{(r)}_{i}-d^{(r)}_{i})^{2}\; . \label{eq:Uys}
\end{eqnarray} 
$\omega^{(r)}_{i}$ are large frequencies and $d^{(r)}_{i}$ are the bond lengths. One employs: $B^{(r)}_{i}\equiv  (M^{(r)}_{i-1})^{-1}+(M^{(r)}_{i})^{-1}$, also with $M^{(r)}_{0}\equiv M_{0} $. Also, $y^{(r)}_{i}\equiv\mid {\bf y}^{(r)}_{i}\mid$. We shall assume that the star polymer is in thermodynamical equilibrium at absolute temperature $T$.  
 The analysis for large $\omega^{(r)}_{i}$,  
 leading to an effective description in terms of slowly-varying angular degrees of freedom,   follows the 
same pattern as that  in  \ref{subsec:RFAE.GFC.4.1}.
 We assume a variational wavefunction having a similar structure, now 
 with $\prod_{r=1}^{n}\prod_{i=1}^{N_{r}} \left[ \phi^{(r)}_{i} (y^{(r)}_{i}) \right] \psi_{\sigma} (\theta , \varphi ) $, where 
$\phi^{(r)}_{i}$ is the corresponding ground-state wave function for the harmonic-oscillator-like potential, similar to  
Eq.\ (\ref{eq:osc}). $\theta , \varphi $ denote the set of angular variables for all ${\bf y}^{(r)}_{i}$. We perform the 
variational computation and employ Peierls'  inequality and so on. Finally,  the classical Hamiltonian reads:
\begin{eqnarray}&&
H_{c }^{(os)}   =  \frac{1}{2M_{0}}(\sum_{r=1}^{n}\frac{{\bf a}^{(r)}_{1, c} \cdot{\bf a}^{(r)}_{1,c }}
{(d^{(r)}_{1})^{2}} )^{2} 
 + \sum_{r=1}^{n}\sum_{i=1}^{N_{r}} \frac{1}{2M^{(r)}_{i}}(
  \frac{{\bf a}^{(r)}_{i,c }}{d^{(r)}_{i}}-   \frac{{\bf a}^{(r)}_{i+1,c }}{d^{(r)}_{i+1}} )^{2}\; , \label{eq:GClassas}
\end{eqnarray} 
with the understanding that ${\bf a}^{(r)}_{N^{(r)}+1,c }=0$. The classical partition function for the open star polymer  is 
($N=\sum_{r=1}^{n}N^{(r)}$): 
\begin{eqnarray}&&
Z_{c }^{(os)} = \left[ \frac{k_{B}T}{2\pi \hbar^{2}} \right]^{N} \frac{\left[\prod_{r=1}^{n} \prod_{l=1}^{N^{(r)}}( d^{r}_{l})
^{2} \right]}{({\rm det} B)^{3/2}} \int [{\bf d\Omega}] \left[ {\rm det} ({\bf u}^{(r)}_{i}(B^{-1})^{(rs)}_{ij}{\bf u}^{(s)}_
{j})\right]^{-1/2} \; . \label{eq:FinalPFSP}
\end{eqnarray}
The elements $B^{(rs)}_{ij}$ of the actual $(N) \times (N)$ matrix $B$ are readily obtained from 
 Eq.\ (\ref{eq:GClassas}), and we shall omit them.
   The internal energy $U$ of the star polymer  is obtained 
through  procedures similar to those  in the previous section. 
One again finds  energy equipartition: $U=Nk_{B}T$.

\chapter{Single-stranded closed-ring freely-jointed chain: getting  constant  quantum zero-point energies  of hard degrees of freedom}
\label{sec:RFAE.GFC.5}
 
The relative simplicity of  open freely-jointed chains, summarized in \ref{subsec:RFAE.GFC.4.1}, may cause the 
impression that the  generalizations of those methods   to  interesting quantum  
macromolecules,   with other kinds of 
constraints,  be always easy or direct. 
 Unfortunately, such an impression is  not correct.  
We shall now proceed to another  kind of constraint in  quantum  macromolecules, which poses  new and considerable 
difficulties and  which  will have to be tackled, in order that   those  generalizations could  be 
accomplished, at least up to certain degree.
\par
\begin{figure}[b]
\begin{center}
{
\includegraphics[scale=0.34]{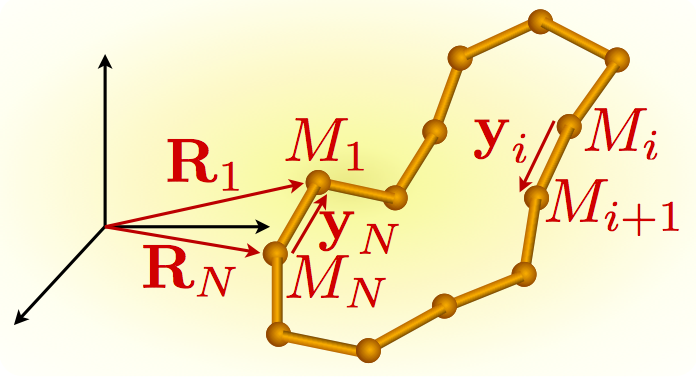}}
\end{center}
\vspace*{-3mm}
\caption{\small Single-stranded closed-ring macromolecular chain.}
\vspace*{-4mm}
\label{fig:RFAE.GFC.3}
\end{figure}
We shall construct a quantum-mechanical model for a three-dimensional  closed-ring 
(freely-jointed) molecular 
chain. To fix the ideas, we shall suppose that the atoms in the chain and/or the bond lenghts are not all identical to one another. 
Let $y_{N}= \left[\sum_{i,j=1}^{N-1} {\bf y}_{i} {\bf y}_{j}\right]^{1/2}$ denote the distance 
 between the atoms $1$ and $N$ (see figure \ref{fig:RFAE.GFC.3}). Notice that $y_{N}$ is not an independent variable.  A first warning of the difficulties to be met 
is that 
all variables $y_{l}$ ($l = 1, \ldots, N-1$), which are independent,  and $y_{N}$, which is not,  
should be treated on the same footing. That is, even for non-identical atoms and/or bond lengths, the 
choice of where the chain was closed to become a ring should be completely arbitrary 
and physically irrelevant.  We shall  choose the Hamiltonian $ \tilde{H}_{Q} $ to be  also given in  Eq.\ (\ref{eq:Ha}).
 If our purpose is to extend the quantum-mechanical 
variational computation of section 4 so that we end up with  a quantum model for a closed-ring chain: what should $U({\bf y})$ be? 
We shall restrict to harmonic-oscillator-like potentials. It seems reasonable that 
$U({\bf y})$ now 
includes  the right-hand-side of  Eq.\ (\ref{eq:Uy}) plus  an interaction potential  between the atoms $1$ 
and $N$, namely, $(2B_{N})^{-1}\omega_{N}^{2}(y_{N}-d_{N})^{2}$ ( $B_{N}\equiv M_{N}^{-1} + M_{1}^{-1}$).  $d_{N}$ 
and $\omega_{N}$   are another bond length and a large frequency, respectively.   Another    question is: 
What should  the trial variational   function $\phi (y)$ (the counterpart of \ (\ref{eq:rad})) be ?   
A reasonable variational wave  function $\phi ({\bf y})$ for a chain, which should become a 
closed ring in the $\omega_{i} \rightarrow \infty$ limit ($i=1,.., N-1, N$), could be chosen as:  
\begin{eqnarray}&&
\phi (y_{1}, y_{2},\ldots, y_{N}) = \prod_{l=1}^{N} \phi_{l} (y_{l}) \; .
\label{eq:rad2}
\end{eqnarray}
Thus, both $U({\bf y})$ and $\phi (y_{1}, y_{2},\ldots, y_{N})$ 
include all variables $y_{l}$ ($l = 1, \ldots, N-1, N$)   on the same footing.
Each $\phi_{l} (y_{l})$ is chosen as  the  ground state wave function for the corresponding 
harmonic-oscillator-like potential see  Eq.\ (\ref{eq:osc}).
Then, the (normalized) total trial  radial-angular wave function of the chain reads:
\begin{eqnarray}&&
\Phi=\Phi ({\bf y}) = \left[ \prod_{l=1}^{N} \phi_{l} (y_{l}) \right] \psi_{\sigma} (\theta , \varphi ) \; .
\label{eq:totcl}
\end{eqnarray}
 The normalization condition $\int [d{\bf y}]\mid \Phi ({\bf y})\mid ^{2}=1$
( with  $[d{\bf y}] \equiv \prod_{i=1}^{N-1} d^{3}{\bf y}_{i}$) becomes  
 in the limit $\omega_{l} \rightarrow \infty$ (by virtue of  Eq.\ (\ref{eq:Dirac})):
\begin{eqnarray}&&
\int [{\bf d\Omega}] |\psi_{\sigma} (\theta , \varphi )|^{2} \frac{\delta (y_{N} - d_{N})}{d_{N}^{2}} = 1\; ,
\label{eq:normalizationcl}
\end{eqnarray}
 $[{\bf d\Omega}]$ being given in  \ (\ref{eq:37})). 
Except for  \ (\ref{eq:normalizationcl})),   
$ \psi_{\sigma} (\theta , \varphi )$ is fully arbitrary.  The restriction  given by $\delta (y_{N} - d_{N})$
  (after having imposed those related to $l = 1, \ldots, N-1$) is  interpreted as: 
\begin{eqnarray}
y_{N} = y_{N}(\theta , \varphi) = \left[\sum_{i,j=1}^{N-1}d_{i}d_{j} {\bf u}_{i} {\bf u}_{j}\right]^{1/2}=d_{N}\;  ,
\label{eq:constraint3}
\end{eqnarray}
which establishes the closed-ring constraint   among all $2(N-1)$ angles contained in $\theta , \varphi$. This 
displays, at this level,    the 
difficulties of the quantum closed-ring chain.

Our first aim is to evaluate, as all frequencies $\omega_{i}$, $i=1,..,N-1,N$ grow very large, 
$  (\Phi, \tilde{H}_{Q} \Phi)\equiv \int  [d{\bf y}] \Phi ({\bf y})^{*}\tilde{H}_{Q} \Phi ({\bf y})$, by 
using Eqs. \ (\ref{eq:Ha}), \ (\ref{eq:Dirac}),  (\ref{eq:totcl}) and 
(\ref{eq:normalizationcl}).   After lenthy computations 
and cancellations, one finds (see appendix D in this  review and \cite{Calvo}):
\begin{eqnarray}\nonumber
(\Phi, \tilde{H}_{Q} \Phi ) &=&  E_{0}  
 +  \sum_{i=1}^{N-1} \frac{B_{i}}{2} \left\{ \int [{\bf d\Omega}] \psi_{\sigma}^{*} (\theta , \varphi ) 
\left[ \frac{({\bf a}_{i} \cdot {\bf a}_{i})}{d_{i}^{2}}\psi_{\sigma}(\theta, \varphi )\right] \frac{\delta (y_{N} - d_{N})}
{d_{N}^{2}} \right\} \\
&-&  \sum_{i=2}^{N-1}\frac{1}{M_{i}} \left\{ \int [{\bf d\Omega}] \psi_{\sigma}^{*} (\theta , \varphi ) \left[ \frac{({\bf a}_{i-1} 
\cdot {\bf a}_{i})}{d_{i}d_{i-1}}\psi_{\sigma} (\theta , \varphi ) \right] \frac{\delta (y_{N} - d_{N})}{d_{N}^{2}} \right\} \nonumber\\
  &+&   \bigg \langle  {\cal O} ^{( c)}(\hbar) \bigg \rangle \; ,
\label{eq:Hamiltoniancl}
\end{eqnarray}
 for any  $\psi_{\sigma}(\theta , \varphi )$.   
  $ {\bf a}_{l}$ are the same as in  Eq.\ (\ref{eq:ai}). Here, 
 $\bigg \langle {\cal O} ^{( c)}(\hbar) \bigg \rangle$ denotes the total contribution of the 
 set of all   terms which do not depend on the frequencies and which do include, at least, one operator acting 
on the delta function: they and some additional calculational details are given in appendix A in \cite{Calvo}. Therefore, all 
contributions contained in  $\bigg \langle {\cal O}^{( c)} (\hbar) \bigg \rangle$ are 
proportional to, at least, the first power of $\hbar$. We use here the specific notation $\bigg \langle... \bigg \rangle$ 
(not employed for similar contributions ${\cal O} (\hbar)$ for other chains) in order to facilitate  comparisons with   \cite{Calvo}. The meaning of, say, 
$\left[({\bf a}_{i} \cdot{\bf a}_{i})\psi_{\sigma}(\theta , \varphi )\right] \delta (y_{N} - d_{N})$ in  
Eq.\ (\ref{eq:Hamiltoniancl}) is the following. First,  the 
differential operator $({\bf a}_{i} \cdot{\bf a}_{i})$ acts upon  $\psi_{\sigma}$  by  regarding  all angular 
variables $\theta_{l}$, $\varphi_{l}$, $l=1,..,N-1$, as if they were independent on one another (that is, 
as if  the constraint $\delta (y_{N} - d_{N})$ were not operative). After the operator has acted upon 
$\psi_{\sigma}$ with that understanding, then  $\delta (y_{N} - d_{N})$ acts and implies that there is one 
relationship among those $2(N-1)$ angles. A similar interpretation applies for  
$\left[({\bf a}_{i-1} \cdot{\bf a}_{i})\psi_{\sigma}(\theta , \varphi )\right] \delta (y_{N} - d_{N})$.  
These operations are genuine  consequences of  the closed-ring constraint at the quantum level.

It is crucial  to notice that all  dependences of $ (\Phi, \tilde{H}_{Q} \Phi )$  on the frequencies  
appear only  in  the constant $ E_{0}=\sum_{i=1}^{N}\frac{\hbar \omega_{i}}{2}$, in the right hand side of 
Eq.\ (\ref{eq:Hamiltoniancl}).  It is   rewarding that, after  the computation and  the cancellations (outlined in appendix D),   
$ E_{0}$    is  independent on angles, equals  
the sum of all zero-point 
energies associated to all harmonic-oscillator-like potentials (including the $N$-th ),  and that  $ E_{0}$ 
 depends on all 
frequencies $\omega_{i}$, $i=1,....N-1,N$
on the same footing. The relevance of all that is explained in the following paragraph.

     A priori, one could have expected that, due to
$ \delta (y_{N} - d_{N})$, the calculation of 
$(\Phi, \tilde{H}_{Q} \Phi )$ would give rise to terms linear in the frequencies multiplied by integrals 
containing functions $D_{i}$ depending on, at least, some of the angles 
$\theta_{1}, \ldots , \theta_{N-1}$, $\varphi_{1}, \ldots , \varphi_{N-1}$ (say, to pieces like 
 
$\omega_{i}\int [{\bf d\Omega}] \psi_{\sigma}^{*} (\theta , \varphi )D_{i}
(\theta , \varphi )\psi_{\sigma}(\theta , \varphi)$, with $i=1,..,N-1,N$). Then, these integral terms would 
have implied  wild variations of the integrand in $\int [{\bf d\Omega}]$, as they would  become amplified by the 
diverging vibrational frequencies. This would upset the reliability of the result for 
$(\Phi, \tilde{H}_{Q} \Phi )$, as commented in \cite{Go1,Ral}: see also  
\cite{AE} and references therein. However, if, by virtue of the algebra and cancellations 
in the variational computation, 
it  would turn out that all functions $D_{i}$ be constant (independent on the angles), then, the 
result of our calculation for $(\Phi, \tilde{H}_{Q} \Phi )$ would be  reliable, and we could separate, 
in a physically meaningful way, the very large vibrational frequency contributions (which would become 
constant, by virtue of the normalization condition in  Eq.\ (\ref{eq:normalizationcl}) ) from the smaller 
rotational contributions (which do imply angular variations). The important result  displayed 
by Eq.\ (\ref{eq:Hamiltoniancl}) is that, in fact,  all terms linear in the frequencies go multiplied with 
factors which are, indeed,  constant, as a consequence of our choice of the symmetric interaction $U({\bf y})$, the 
variational trial wave function in (\ref{eq:totcl}), 
the algebra,  the cancellations   and Eq.\ (\ref{eq:normalizationcl}) \cite{Calvo}. This solves the difficulties a), b) and c) at the end of subsection 2.2. 
\par
   
Let us  make use of the   variational  inequality  (\ref{eq:peierls}). Then, the  quantum partition function $\tilde{Z}_{Q}$ for the 
actual closed-ring chain fulfills:
\begin{eqnarray}\tilde{Z}_{Q}
& \geq & \exp\left[-(k_{B}T)^{-1}E_{0}\right]\cdot Z_{Q}^{(c)} \; ,\label{eq:peierls3} \\  
Z_{Q}^{(c )} & = &
\sum_{\sigma} \exp{\left[-\frac{1}{k_{B}T}\int[{\bf d\Omega}] \frac{\delta (y_{N} - d_{N})}{d_{N}^{2}} \psi_{\sigma}^{*} (\theta , \varphi ) 
 H_{Q}^{(c)} \psi_{\sigma}(\theta , \varphi)+\bigg \langle {\cal O}^{( c)} (\hbar) \bigg \rangle\right]}\; .
\label{eq:peierls4}
\end{eqnarray}
The quantum angular Hamiltonian  $H_{Q}^{(c)}$ for the closed-ring chain   reads:
\begin{eqnarray}&&
H_{Q}^{(c)} =  \sum_{i=1}^{N-1} \frac{B_{i}}{2} \left[ \frac{({\bf a}_{i} \cdot{\bf a}
_{i})}{d_{i}^{2}} 
\right]  - \sum_{i=2}^{N-1}\frac{1}{M_{i}} \left[ \frac{({\bf a}_{i-1} \cdot{\bf a}_{i})}{d_{i}d_{i-1}} \right] 
 \; .
\label{eq:HQANGCL}
\end{eqnarray}
with the operational meaning for $({\bf a}_{i} \cdot{\bf a}_{i})$ and $({\bf a}_{i-1} \cdot{\bf a}_{i})$ indicated above.  We emphasize that 
$Z_{Q}^{(c)}$ can be regarded as the three-dimensional angular quantum partition function for the closed-ring 
 chain.  A posteriori and by grouping  
 the resulting terms  in Eq.  \ (\ref{eq:Hamiltoniancl}),   the latter   can be 
reformulated.  
After some lengthy algebra, Eq.  \ (\ref{eq:Hamiltoniancl}) becomes \cite{CAEDNA1}:
\begin{eqnarray}
(\Phi, \tilde{H}_{Q} \Phi )  &=&  E_{0} +  {\cal O} ^{( c)}(\hbar)_{1}  
 \nonumber \\
 &+&  \int[{\bf d\Omega}]  \psi_{\sigma}^{*} (\theta , \varphi)
 \frac{1}{2} \sum_{i,j=1}^{N-1}\frac{ B_{ij}}{d_{i}d_{j}} 
\left\{ {\bf e}_{i} \cdot \left[ \frac{\delta (y_{N} - d_{N}) }{d_{N}^{2}}   {\bf e}_{j}\psi_{\sigma} (\theta , \varphi ) \right]\right\} \; .
   \label{eq:Hamiltoniancl11}
\end{eqnarray}
(with ${\cal O} ^{( c)}(\hbar)_{1}\rightarrow 0 $ as $\hbar\rightarrow 0$). From \ (\ref{eq:Hamiltoniancl11}),   
  invariance  under overall rotations in three-dimensional space can be shown to hold for 
 $Z_{Q}^{(c)}$. See \cite{CAEDNA1}. 
\par
    We have also carried out 
(as an unpublished work) the extension for a quantum closed-ring freely-jointed chain, by using Morse potentials, instead of 
harmonic-oscillator-like ones. Then, $U({\bf y})$ now 
 equals   the right-hand-side of  Eq.\ (\ref{eq:morsei}) plus  another  Morse  potential  between the atoms $1$ 
and $N$, namely, $V_{M}(y_{N})$. The resulting computations and cancellations follow the same pattern as those for 
harmonic-oscillator-like potentials (although they are more complicated). Eqs. \ (\ref{eq:Hamiltoniancl})-\ (\ref{eq:HQANGCL}) 
continue to hold, now with the constant $E_{0}$ standing for $\sum_{i=1}^{N}E_{M,i,n=0}$. 

Let all atoms   be identical to one another ($M_{i}=M$ in the  chain,  $i=1,\ldots,N$) and, for simplicity,  
 let all bond lengths be equal ($d_{j}=d$, $j=1,\ldots,N$). The comments on indistinguishability in  \ref{subsec:RFAE.GFC.4.1} keep their validity here. A new deep issue  
related to indistinguishability   appears here, arising from the fact that any atom can be considered as the first one. In order 
to deal with it, an overall  prefactor 
(scaling as $1/N$,  eventually) could   be introduced in the quantum partition function, which would  not alter the variational 
computation.

Eqs.\ (\ref{eq:peierls4}) and (\ref{eq:HQANGCL})  solve the difficulties a), b) and c) at the end of \ref{subsec:RFAE.GFC.2.2} 
 for the actual 
chain (which justifies their derivation), but their complicated structures make their practical use   
 very difficult.  Eq.\ (\ref{eq:peierls4}) will become  considerably reduced, if one proceeds to the classical limit 
approximation for the angular degrees of freedom (by assuming  Eq.\ (\ref{eq:Claslim}) for $l=1,..,N$). Notice that 
 $\langle{\cal O} ^{( c)} (\hbar)\rangle$ vanishes in that limit.
The quantum  Hamiltonian $H_{Q }^{(c )}$ for the closed-ring chain  can be  
approximated, in the classical limit, by the same  $H_{c }^{(o)}$  in  Eq.\ (\ref{eq:GClassa}) 
[with  the same classical   ${\bf a}_{i,c }$   in  Eq.\ (\ref{eq:aClass})] as 
for the open chain: of course, $H_{c }^{(o)}$  goes now multiplied by $\delta (y_{N} - 
d_{N})/d_{N}^{2}$.  An   analysis of    this classical  limit has been carried out in \cite{Calvo}, and we shall quote  
the  result.  $Z_{Q}^{( c)}$   can be approximated by the following 
 three-dimensional classical partition function for the closed-ring chain: 
\begin{eqnarray}&&
Z_{C}^{(c)} = \left[ \frac{k_{B}T}{2\pi \hbar^{2}} \right]^{N-1} \frac{\left[ 
\prod_{l=1}^{N-1} d_{l}^{2} \right]}{({\rm det} B)^{3/2}} \int [{\bf d\Omega}] \frac{\delta (y_{N} - 
d_{N})}{d_{N}^{2}} [\Delta_{N-1}]^{-1/2}    \; ,
\label{eq:FinalPFCl}
\end{eqnarray}
with the same  
$({\rm det} B)^{3/2}$ and $ [\Delta_{N-1}]^{-1/2}$  as 
in  Eqs.\ (\ref{eq:matrixB}) and \ (\ref{eq:matrixinvB}).   $\delta (y_{N} - 
d_{N})/d_{N}^{2}$ appeared in the exponent in the quantum Eq.\ (\ref{eq:peierls4}), but it is located 
downstairs in the classical approximation.

 We shall comment briefly  about an  important necessary condition for approximating  quantum partition functions by classical ones
  \cite{Hua}:  
interparticle distances should be appreciably larger than thermal wavelengths ('' the classical limit  restriction'',  
 like in  \ref{subsec:RFAE.GFC.4.3}).  
Let us employ  the distances  $\left[({\bf R}_{l} - {\bf R}_{j})^{2}\right]^{1/2}$  between the $l-th$ and the $j-th$ atoms, which  are 
to be  expressed in terms of ${\bf y}_{k}$ ( Eq.\ (\ref{eq:y}) ). In general,  the integrations in  
Eq.\ (\ref{eq:FinalPFCl}) should be performed over all angles consistent with the  classical limit 
restriction:  all  the atomic distances  $\left[({\bf R}_{l} - {\bf R}_{j})^{2}\right]^{1/2}$ should  be 
appreciably larger than all thermal wavelengths  $\lambda_{th,j}=(2\pi\hbar^{2}/M_{j}k_{B}T)^{1/2}$, 
$j=1,..,N$.  The approximation   of   $Z_{Q}^{( c)}$ by   
$Z_{C}^{(c)}$ is justified  only when they hold. An estimate of thermal wavelengths at room temperature is given 
\ref{subsec:RFAE.GFC.4.3}. Then, those restrictions would imply the exclusions of some (eventually rather  small) angular domains in the 
integrations in  \ (\ref{eq:FinalPFCl}). See  \cite{Calvo} and \cite{RaCa}.

    Eq.\ (\ref{eq:FinalPFCl}) yields  energy equipartition: $U=(N-1)k_{B}T$, as for the open chain.  

Topological constraints and knots play an important role in the properties of closed-ring chains \cite{Topol1,Topol2}. We  would 
expect 
that the variational quantum-mechanical computation keeps its validity regardless  topological constraints, and that 
$Z_{Q}^{(c )}$ would count and include all kinds of topological conformations (knots)  that a closed-ring chain may adopt 
in space, and so forth for  $Z_{C}^{(c)}$ (under the conditions of validity of the classical approximation). 
 Such studies lie outside 
our scope, as commented in  \cite{Calvo}.

 \chapter{Single-stranded open freely-rotating chain: getting  constant  quantum zero-point energies   of   hard degrees of freedom  }
\label{sec:RFAE.GFC.6}

The Hamiltonian $ \tilde{H}_{Q} $ continues to be  given in  Eq.\ (\ref{eq:Ha}). 
We shall suppose that, not only  nearest-neighbour atoms interact  through stiff harmonic-oscillator-like 
potentials $V_{j}=(2B_{j})^{-1}\omega_{j}^{2}(y_{j}-d_{j})^{2}$ ($ y_{j} = \mid {\bf y}_{j}\mid$) with large vibrational frequencies 
$\omega_{j}$
 and  lengths $d_{j}(>0)$ \cite{CAEDNA1}, but also that   similar   potentials exist between  atoms which are next-to-nearest neighbours:  
$V_{j,j+1}=2^{-1}B_{j,j+1}\omega_{j,j+1}^{2}(\mid {\bf y}_{j}+{\bf y}_{j+1}\mid-d_{j,j+1})^{2}$. Here, $\omega_{j,j+1}$ are other frequencies, $d_{j,j+1}$ are lengths such that $\mid d_{j}-d_{j+1}\mid\leq d_{j,j+1}\leq  d_{j}+d_{j+1}$ and  
$B_{j,j+1}(M_{j}^{-1} + M_{j+1}^{-1})=1$.  Then, we suppose: 
\begin{eqnarray}
U({\bf y}) = \sum_{j=1}^{N-1}V_{j} + \sum_{j=1}^{N-2}V_{j,j+1} \; , \label{eq:UyM}
\end{eqnarray} 
For suitably  large $\omega_{j,j+1}$,  $ \sum_{j=1}^{N-2}V_{j,j+1}$ hinders part of the allowed 
 internal rotations in the macromolecular chain,   which becomes a freely-rotating one. 
 On physical grounds,   $ \sum_{j=1}^{N-2}V_{j,j+1}$  could  approximate   for the effect of the covalent 
bonding due to successive single   pairs of shared electrons, which produce precisely those hindrances 
\cite{Volk,Flo,Gros}. Weak  next-to-nearest neighbour interactions  were treated in \ref{subsec:RFAE.GFC.4.3} , while strong ones are considered 
in this section. 
\par
\begin{figure}[t]
\begin{center}
{
\includegraphics[scale=0.4]{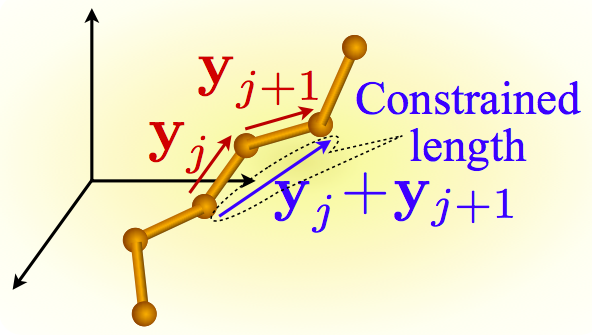}}
\end{center}
\vspace*{-3mm}
\caption{\small Freely-rotating single-stranded macromolecular chain.}
\vspace*{-4mm}
\label{fig:RFAE.GFC.4}
\end{figure}
We shall  suppose: 
\begin{eqnarray}
\hbar\omega_{l} \gg k_{B}T \ ,\hbar\omega_{l} \gg \frac{\hbar^{2}B_{l}}{d_{l}^{2}}  \; ,  \label{eq:freca} \\
\hbar\omega_{l,l+1} \gg k_{B}T \ ,\hbar\omega_{l,l+1} \gg \frac{\hbar^{2}B_{l}}{d_{l}^{2}}\; ,\label{eq:frecb} 
\end{eqnarray}
for any $l$. Physically, one would expect that $\hbar\omega_{l} $ be larger than $\hbar\omega_{l,l+1} $, because the interaction responsible for the 
existence of bond 
lengths should be certainly stronger than that giving rise to constrained bond angles. Based upon \ (\ref{eq:parfuncquan}), \ (\ref{eq:UyM})
\ (\ref{eq:freca}) and  \ (\ref{eq:frecb}),  the following   variational computation will lead to   a model for an open  
freely-rotating   chain, in which all $d_{j}$ and $d_{j,j+1}$ be given constants.   We  choose the variational  (radial-angular) wave  function  $\Phi ({\bf  y})$ as: 
\begin{eqnarray}
\Phi ({\bf  y})=\phi_{nn} (y) \phi_{nnn} ({\bf y})\psi_{\sigma} (\theta ,
\varphi  ) \ , \phi_{nn}(y) =  \prod_{l=1}^{N-1}  \phi_{l}(y_{l}) \; ,
 \label{eq:Phi} 
\end{eqnarray}
$\phi_{l}(y_{l}) $ being given in \ (\ref{eq:osc}). In turn, we choose:
$\phi_{nnn} ({\bf y})  =  \prod_{l=1}^{N-2}  \phi_{l,l+1} $.  Each  $\phi_{l,l+1}$  is  also  chosen  as the real  (Gaussian)   wave
function associated to   $V_{l,l+1}$: 
 \begin{eqnarray}
\hspace*{-1.0cm}
\phi_{l,l+1}= \left[\frac{\omega_{l,l+1} B_{l,l+1}}{\pi\hbar}\right]^{1/4}\exp\left[-\frac{\omega_{l,l+1} B_{l,l+1}}{2\hbar}\left(\mid 
{\bf y}_{l}+{\bf y}
_{l+1}\mid-d_{l,l+1}\right)^{2}\right] \; .
\label{eq:phiAn}
\end{eqnarray}
As all frequencies $\omega_{l}$ and $\omega_{l,l+1}$ become suitably large,  Gaussians approach Dirac delta functions ($w_{rad}=[\prod_{l=1}^{N-1}d_{l}^{-2}\delta(y_{l}-d_{l})]$): 
\begin{eqnarray} 
\left\vert \phi_{nn} (y) \phi_{nnn} ({\bf y})\right\vert^{2}\rightarrow  w_{rad} w_{ang} 
 \; , \label{eq:wrad,ang}\\
 w_{ang}=\prod_{l=1}^{N-2}\delta\left(\left\vert d_{l}{\bf u}_{l}+d_{l+1}{\bf u}_{l+1}\right\vert-d_{l,l+1}\right)\; . 
\label{eq:wang}
\end{eqnarray}
     $\delta(\mid d_{l}{\bf u}_{l}+d_{l+1}{\bf u}_{l+1}\mid-d_{l,l+1})$ is equivalent to constraining the bond angle between 
${\bf u}_{l}$ 
and ${\bf u}_{l+1}$: see   Eq.\ (\ref{eq:angcons}). 
The complex functions $\psi_{\sigma} (\theta , \varphi  ) $ ($\sigma$ now being a set of quantum numbers) 
are
arbitrary, except that: i) they are  periodic in each $\varphi_{i}$ with period $2\pi$ and  independent on   any $y_{l}$, and 
ii) they are normalized  through  the scalar product:
\begin{eqnarray}
(\psi_{1},  \psi_{2})\equiv \int [{\bf d\Omega}]   w_{ang} \psi_{1} (\theta , \varphi )^{*}\psi_{2} (\theta , \varphi )  \; . 
\label{eq:normalization1}
\end{eqnarray}
One evaluates, as all frequencies $\omega_{i}$ $i=1,\ldots,N-1$,  and $\omega_{l,l+1}$ $l=1,..,N-2$, 
grow very large, the quantum expectation value 
$\langle\Phi, \tilde{H}_{Q} \Phi \rangle\equiv \int  [d{\bf y}] \Phi ({\bf y})^{*}\tilde{H}_{Q} 
\Phi ({\bf y})$, by using Eqs.\  (\ref{eq:Phi})-(\ref{eq:normalization1}).  
The computations and cancellations \cite{CAEDNA1}, which are quite lengthy,  follow the same pattern as those in \cite{Calvo} and embody 
similar cancellations, 
now applying the angular constraints  in  \  (\ref{eq:wang}). One novel  aspect is that, by the end 
of the computations, one 
manages to group the resulting terms into some, quite compact,   structures. 
  The result is \cite{CAEDNA1}:
\begin{eqnarray}&&
\langle \Phi, \tilde{H}_{Q}  \Phi \rangle =    
   E_{0} +
 ( \psi_{\sigma}, H_{Q}^{(fr)} \psi_{\sigma}) +{\cal O}^{(fr)}(\hbar)   \; ,\label{eq:hfrec1} \\&&
(\psi_{\sigma}, H_{Q}^{(fr)} \psi_{\sigma} ) = 
\int[{\bf d\Omega}]  \psi_{\sigma}^{*} (\theta , \varphi )
   w_{ang}H_{Q}^{(fr)}  \psi_{\sigma} (\theta , \varphi ) \; ,\label{eq:definham1}
\end{eqnarray}
for any  normalized $\psi_{\sigma}(\theta , \varphi)$ fulfilling the above requirements. The constant $E_{0}$  equals 
$ \sum_{i=1}^{N-1}2^{-1}\hbar\omega_{i}+\sum_{i=1}^{N-2}2^{-1}\hbar\omega_{i,i+1}$ (namely, the sum of the zero-point  energies  
associated to  all $V_{i}$ and  $V_{j,j+1}$). In turn, the angular Hamiltonian $H_{Q}^{(fr)}  $ is given through:
\begin{eqnarray}
 w_{ang}H_{Q}^{(fr)}  = \frac{1}{2} \sum_{i,j=1}^{N-1}\frac{ B_{ij}}{d_{i}d_{j}} 
\left[ {\bf e}_{i} \cdot ( w_{ang}{\bf e}_{j}) \right]  \; ,
\label{eq:HQANG00} 
  \end{eqnarray}
 with ${\bf e}_{l}\equiv i\hbar {\bf u}_{l} -{\bf
  a}_{l}$, $l=1,\ldots,N-1$.  
   $ {\cal O}^{(fr)}(\hbar)  $  denotes the total contribution of the set of all remaining
 terms which  do not
depend on any of the frequencies $\omega_{0,i}$ and $\omega_{0,i,i+1}$.  $ {\cal O}^{(fr)}(\hbar)  $  is  
 proportional to  $\hbar^{2}$ and it   does not  contain differential operators acting upon  
$\psi_{\sigma}(\theta , \varphi)$ (that is,  $ w_{ang}{\cal O}_{ang}(\hbar)$ acts multiplicatively on 
 $\psi_{\sigma}(\theta , \varphi)$).  The explicit form of  $ {\cal O}^{(fr)}(\hbar)  $ , which is not relevant here, is given in appendix A 
in 
\cite{CAEDNA1}. The structures in \  (\ref{eq:hfrec1})-\  (\ref{eq:HQANG00}) solve the difficulties a), b) and c) at the end 
of subsection 2.2. and ressemble formally those which 
 appear formally for the freely-jointed chain (recall \  (\ref{eq:hfrec})-\  (\ref{eq:HQANG}) ), which appears 
 to be rewarding. Notice also the formal analogy with \  (\ref{eq:Hamiltoniancl11}). We emphasize     the crucial importance of having chosen  
the variational trial wave functions   adequately (through \  (\ref{eq:Phi}), \  (\ref{eq:rad}), $\phi_{nnn} ({\bf y})$ 
 and \  (\ref{eq:phiAn})). Otherwise, one may obtain either contributions depending  on the unconstrained soft 
variables or results  which, 
even if independent on the  latter variables, do not yield the  physically expected  quantum zero-point 
energies. An example of the latter situation was given  in \cite{AE} (subsection 6.2)), where   a   variational trial 
wave function  containing  factors
differing from  \  (\ref{eq:phiAn})) was used. 

\par

 We continue to apply Eq. (\ref{eq:peierls}). In our case, $\Phi_{\sigma} = \Phi ({\bf
 y})$ and $\langle\Phi_{\sigma}, \tilde{H}_{Q} \Phi_{\sigma}\rangle$ are given in Eqs.\
 (\ref{eq:Phi}) and  (\ref{eq:hfrec1}).

Then:
\begin{eqnarray} &&
\tilde{Z}_{Q}  \geq  \exp[-(k_{B}T)^{-1}E_{0}]\cdot Z_{Q}^{(fr)}  \; ,
\label{eq:peierlsang1} \\ && 
Z_{Q}^{(fr)}   \equiv  
\sum_{\sigma} \exp{\left[-(k_{B}T)^{-1}[\int[{\bf d\Omega}] \psi_{\sigma}^{*}
  (\theta , \varphi )  w_{ang} H_{Q}^{(fr)} \psi_{\sigma}(\theta , \varphi
  )+{\cal O}^{(fr)}(\hbar) ]\right]} \; ,
\label{eq:peierlsang2} 
\end{eqnarray} 
where $Z_{Q}^{(fr)} $ can be regarded as the effective three-dimensional quantum partition function for the 
slow motions of the
unconstrained  angular degrees of freedom of the  freely-rotating  chain.   $H_{Q}^{(fr)} $  is a  
    Hermitian operator (with respect to the scalar product in  
Eq.\ (\ref{eq:normalization1})): $(\psi_{1}, H_{Q}^{(fr)} \psi_{2})=(H_{Q}^{(fr)} \psi_{1}, \psi_{2})$, 
for any $\psi_{1}, \psi_{2}$.   One has:
\begin{eqnarray} 
\hspace*{-2.0cm}
( \psi_{1},H_{Q}^{(fr)}  \psi_{2})=\int[{\bf d\Omega}] 
\frac{1}{2} \sum_{i,j=1}^{N-1}\frac{ B_{ij}}{d_{i}d_{j}}  ({\bf e}_{i}\psi_{1}
  (\theta , \varphi ))^{*}\cdot w_{ang}({\bf e}_{j}\psi_{2}(\theta , \varphi))\; . \label{eq:mehang0} 
\end{eqnarray}
It can be justified that $H_{Q}^{(fr)}$ and, hence, $Z_{Q}^{(fr)} $ (with ${\cal O}^{(fr)}(\hbar)$ discarded) are  invariant under 
overall rotations in three-dimensional space.
 The above 
properties  have  indeed been established, at the price of performing various partial integrations and  handling carefully  
$w_{ang}$ (as the latter  involves Dirac delta functions). See  appendix B 
in 
\cite{CAEDNA1}.

We shall suppose that all angular wave functions $\psi_{\sigma}(\theta, \varphi )$ are  the complete set of all orthonormalized  
eigenfunctions of $H_{Q}^{(fr)}$ so that, upon discarding ${\cal O}^{(fr)}(\hbar)$,   Eq.\ (\ref{eq:peierls2}) becomes 
$Z_{Q}^{(fr)}= {\rm Tr}[\exp [-(k_{B}T)^{-1}H_{Q}^{(fr)}]$. It is convenient 
to express the $2(N-1)$ angular variables  in $\theta$ and $\varphi$ in terms of another 
set of  $2(N-1)$ more  suitable ones. The latter will be chosen to be: 
 $\theta_{1}, \ldots , \theta_{N-1}$, $\varphi_{0}(\equiv (N-1)^{-1}\sum_{j=1}^{N-1}\varphi_{j})$ and 
$\beta_{j,j+1}\equiv {\bf u}_{j}{\bf u}_{j+1}
(=\cos \theta_{j}\cos \theta_{j+1}+\sin \theta_{j}\sin \theta_{j}\cos (\varphi_{j+1}-\varphi_{j}))$, $j=1,..,N-2$. One has: $[{\bf d\Omega}]=[\prod_{j=1}^{N-2}d\beta_{j,j+1}]d\varphi_{0}[\prod_{l=1}^{N-1}d\theta_{l}] J$, the Jacobian $J$  depending on all $\theta_{j}$ and $\beta_{j,j+1}$. The angular constraint reads ($({\bf u}_{j}{\bf u}_{j+1})^{(0)}=(2d_{j}d_{j+1})^{-1}(d_{j,j+1}^{2}-d_{j}^{2}-d_{j+1}^{2})$):
\begin{eqnarray}
w_{ang}=\left[\prod_{j=1}^{N-2}\frac{d_{j,j+1}}{d_{j}d_{j+1}}\right]
\left[\prod_{j=1}^{N-2}\delta(\beta_{j,j+1}-({\bf u}_{j}\cdot{\bf u}_{j+1})^{(0)})\right]\; .
\label{eq:angcons}
\end{eqnarray} 
 Specifically, the constrained bond angles are all those given by  $\cos^{-1}[({\bf u}_{j}\cdot{\bf u}_{j+1})^{(0)}]$, $j=1,\ldots,N-2$. 
Under  \ (\ref{eq:freca}) and \ (\ref{eq:frecb}), we shall
study the transition to the classical limit,  which is simpler  if one has got rid of all $\beta_{j,j+1}$, previously. 
Then, we start from 
$Z_{Q}^{(fr)}= {\rm Tr}[\exp [-(k_{B}T)^{-1}H_{Q}^{(fr)}]$,   apply  to it  Peierls` variational  inequality      \ (\ref{eq:peierls}), 
employing now the complete set of all 
$\psi_{\sigma}(\theta, \varphi )$'s depending on $\theta_{1}, \ldots , \theta_{N-1}$ and  $\varphi_{0}  $ (but not on  
$\beta_{j,j+1}$) and 
use  Eq.  \ (\ref{eq:angcons}), for integrating over and  getting  rid of all $\beta_{j,j+1}$. The resulting quantum partition function has 
an adequate  form for taking $\hbar\rightarrow 0$. 
 About room temperature, the individual slowly-varying internal rotations about bonds, which remain unconstrained,  
have  typical energies which can be estimated to be  smaller than the  vibrational energies
 $\hbar\omega_{i}$ and  $\hbar\omega_{l,l+1} $\cite{Volk,MesII,McQ}. Then,  we shall suppose that $k_{B}T\gg (\hbar^{2}B_{l})/(d_{l}^{2})$, 
so that an appreciable number of excited  states for those unconstrained rotations may be occupied and quantum operators and
statistics can be approximated by classical ones. Accordingly, 
all quantities of order $\hbar $ or higher (like $ {\cal O}^{(fr)}(\hbar)  $ ) can be neglected 
directly. We shall omit the analysis,   which is given in \cite{CAEDNA1}.  One finds   the
following classical ($c$) Hamiltonian:
  \begin{eqnarray} &&
H_{c  }^{(fr)}=\frac{1}{2}\sum_{i,j=1}^{N-1}\frac{B_{ij}}{d_{i}d_{j}}\,{\bf a}_{i,c}^{(fr)}\cdot{\bf a}_{j,c}^{(fr)}\; .
\label{eq:Hangrescl} 
\end{eqnarray} 
${\bf a}_{i,c}^{(fr)}$'s are  classical variables ( arising from the classical limit of ${\bf e}_{i}$):
 ${\bf a}_{i,c}^{(fr)}=- [{\bf u}_{\theta_{i}} P_{\theta_{i}}+((N-1)\sin \theta_{i})^{-1}{\bf u}_{\varphi_{i}} P_{\varphi_{0}}]$.  
  $P_{\theta_{i}}$ and $P_{\varphi_{0}}$ are the classical momenta canonically conjugate to $\theta_{i}$, $i = 1, \ldots , N-1$, and $\varphi_{0}$. 
 In the classical limit, one gets the 
   classical partition function $Z_{c  }^{(fr)} $. 
\begin{eqnarray}&& 
Z_{c  }^{(fr)} =\frac{1}{(2\pi \hbar)^{N}}\int \left[\prod_{l=1}^{N-1}d\theta_{l}dP_{\theta_{i}}\right]d\varphi_{0}dP_{\varphi_{0}}
 \exp{\left[-\frac{H_{c  }^{(fr)}}{k_{B}T} \right]}\; .
\label{eq:} 
\end{eqnarray} 
One performs all Gaussian integrations in  $Z_{c  }^{(fr)}$   over  the classical momenta  $P_{\theta_{i}}$ 
and $P_{\varphi_{0}}$ and over $\varphi_{0}$.  Then:
\begin{eqnarray}&&
Z_{c  }^{(fr)}=\frac{1}{(2\pi\hbar)^{N}}2\pi[2\pi k_{B}T]^{N/2} \int \left[\prod_{l=1}^{N-1}d\theta_{l}\right]
\left[ {\rm D}_{N} \right]^{-1/2} \; . \label{eq:parfunclres2}
\end{eqnarray}
$ {\rm D}_{N}$ is the $N\times N$ symmetric matrix formed by the coefficients of $P_{\varphi_{0}}$ and 
$P_{\theta_{i}}$, $i=1,\ldots,N-1$ in $H_{c  }^{(fr)}$. The internal energy $U$ of the classical freely-rotating chain
 can be obtained directly from \ (\ref{eq:parfunclres2}) ( like for the classical freely-jointed chain). One finds readily energy equipartition, namely: $U=(N/2)K_{B}T$ \cite{AE,RaCa}.

A simple and interesting  example of an  open freely-rotating chain is polyethylene (the synthetic polymer
  $...-CH_{2}-CH_{2}-CH_{2}-CH_{2}-....$,  each  $CH_{2}$ being regarded, for simplicity, as one of the 
``atoms'').  We accept that  bond lengths  do not vary appreciably along the chain ($d_{j}\simeq d$) and that the same holds for   bond angles. 
 For polyethylene, the bond length is $d=1.54\times 10^{-1}$ nanometers and the bond angle   
$\cos^{-1}\beta_{j,j+1}$ is about $70$ degrees and $32$ minutes (for any $j$) \cite{McQ}. 
 \par
Let $d_{j}=d$, $j=1,..,N-1$, and  $M_{i}=M_{0}$, $i=1,..,N$. It now  follows (using   some  
long-distance approximations ) that  Eq.  \ (\ref{eq:parfunclres2}) implies the existence of a 
persistence length $d_{pl}$ ($>d$): see \cite{AE} (subsection 6.4) for a qualitative   estimate. So,  on a  suitably large length scale,  the classical freely-rotating 
chain described by  Eq.  \ (\ref{eq:Hangrescl}) can be  approximated by a freely-jointed  one  having 
$N_{pl}$ bonds ($N_{pl}<N-1$) and  a new bond length $d_{pl}$.

The   procedures in this section  leading to Eqs. \ (\ref{eq:hfrec1}),  \ (\ref{eq:definham1})  and  
\ (\ref{eq:HQANG00})       have been generalized, in outline, to single-stranded 
closed-ring freely-rotating chains  and to single-stranded  open freely-rotating chains with further 
constraints (helical-like, star-like...), all at the quantum level \cite{CAEDNA1}. 
   We remark that one always finds structures 
similar to 
those in Eqs. \ (\ref{eq:hfrec1}),  \ (\ref{eq:definham1})  and  
\ (\ref{eq:HQANG00}), with different $ w_{ang}$ associated to the corresponding constraints other than those for 
constrained bond lengths. That is, the fact that bond vectors are constrained is not represented by any contribution to $ w_{ang}$, 
but by 
the very structure $2^{-1} \sum_{i,j=1}^{N-1}( B_{ij}/d_{i}d_{j}) 
 {\bf e}_{i} \cdot       {\bf e}_{j}$ in \ (\ref{eq:HQANG00}). Thus, \ (\ref{eq:HQANG00})  also includes the freely-jointed 
case, if  $ w_{ang}\equiv 1$:  compare with. \ (\ref{eq:HQANG}).  Eq. \ (\ref{eq:Hamiltoniancl11}) was one 
 particular case of \ (\ref{eq:definham1}). As an example, we 
remind that the DNA of the bacteriophage $\phi$ $X$ $174$ is single-stranded and forms a ring \cite{Elias} (and, of course, 
it has angle constraints, so that to  regard it as freely-jointed chain would be too crude).

\chapter{Double-stranded open chain}
\label{sec:RFAE.GFC.7}

Double-stranded (ds) macromolecular chains  (specifically, dsDNA) have an absolutely crucial importance  in 
Molecular Biology \cite{Leh,Volk,Gros,Frank,Gotoh}. This fact has  given rise to an enormous and permanent research activity on 
Chemical Physics and Physics of dsDNA: see, for instance \cite{Pol2,Ritort,Pey,Pey04,Pey1,Ares,Yaku} and references therein. 
Motivated by dsDNA,  
we shall undertake in this section, in outline, a  study of double-stranded  open macromolecular chains, by extending the methods of the previous sections.  
We shall apply its consequences  to open  dsDNA, in    \ref{sec:RFAE.GFC.8}. 

\section{Double-stranded open quantum chain: Some general aspects}
 \label{subsec:RFAE.GFC.7.1}
 We shall consider  a model for a single  ds open  macromolecular chain, also based upon 
Quantum Mechanics. Each of the two strands or individual chains has $N$ atoms.  The mass and the  
position vector of the $i$-th atom ($1\leq i\leq N$) in the $r$-th strand  ($r=1, 2$) are   $M^{(r)}_{i}$ and 
 ${\bf  R}^{(r)}_{i}$.   We start from the quantum Hamiltonian operator: 
\begin{eqnarray}   
H_{Q,1}  =  -\frac{\hbar^{2}}{2}\sum_{r=1}^{2}\sum_{i=1}^{N} \frac{1}{M^{(r)}_{i}}\nabla_{{\bf R}^{(r)}_{i}}^{2} +
 \sum_{r=1}^{2} U^{(r)}_{b}  + \sum_{r=1}^{2} U^{(r)}_{a} + \sum_{r=1}^{2} V^{(r)}_{1} + V_{ds}\;  .\label{eq:H1DS}
\end{eqnarray}
$ U^{(r)}_{b} $ is the total potential energy among neighbouring  atoms in the $r-th$ strand. It 
can  be   chosen as   either Eq.\ (\ref{eq:Uy})  ( harmonic-oscillator-like potentials)  or Eq.\ (\ref{eq:morsei}) 
(  Morse potentials).     We recall that the Morse potential gives a 
 qualitatively adequate effective interaction between two nucleotides. 
$U_{a}^{(r)}$ is the total potential ( due to 
next-to-nearest neighbours interactions) accounting for  the most important    angular constraints in the 
$r-th$ strand:
\begin{eqnarray}
 U_{a}^{(r)}=\sum_{j=1}^{N-2}V^{(r)}_{j,j+1}(\vert {\bf y}^{(r)}_{j}+{\bf y}^{(r)}_{j+1}\vert) \, .
\label{eq:V1}
\end{eqnarray}
This   $ V_{j,j+1}$ can be either weak, like in   \ref{subsec:RFAE.GFC.4.3}, or strong 
(like that in  Eq.\ (\ref{eq:UyM})).  
We shall assume that: i) 
  $ U^{(r)}_{b} $, which is the strongest potential (with the largest absolute magnitude) in 
\ (\ref{eq:H1DS})),  will give rise 
to constant bond 
lengths ($y^{(r)}_{j}=d^{(r)}_{j}$), ii)  the nearest-neighbour potentials $V^{(r)}_{j,j+1}(\vert d^{(r)}_{j}{\bf u}^{(r)}_{j}
+d^{(r)}_{j+1}{\bf u}^{(r)}_{j+1}\vert)$ can be either weak or moderately   strong, with 
    a unique deep minimum , so as to  constrain ${\bf u}^{(r)}_{j}\cdot{\bf u}^{(r)}_{j+1}$ to some 
fixed value $\beta^{(r,0)}_{j}$.
   $ V^{(r)}_{1}$ is a  residual   interaction in the $r$-th strand, 
not included in either of the stronger 
interactions 
 $ U^{(r)}_{b}$ or $ U^{(r)}_{a}$ (and weaker than $U_{a}^{(r)}$). 
\par
\begin{figure}[b]
\begin{center}
{
\includegraphics[scale=0.38]{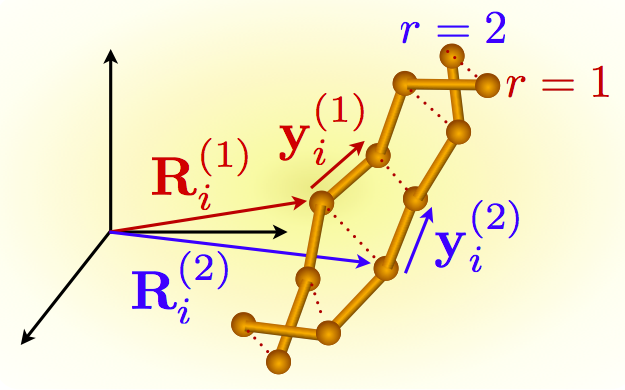}}
\end{center}
\vspace*{-3mm}
\caption{\small Double-stranded open macromolecular chain.}
\vspace*{-4mm}
\label{fig:RFAE.GFC.5}
\end{figure}

$V_{ds}= V_{ds}({\bf R}^{(1)}_{1},..,{\bf R}^{(1)}_{N},{\bf R}^{(2)}_{1},..,{\bf R}^{(2)}_{N}) $ is the 
total potential energy among atoms belonging to different strands: it is supposed to depend on the relative 
distances thereof, so that it displays   overall rotational invariance. On physical grounds,   $ V_{ds} $ 
should  approach zero as all  relative distances between any atom of one strand and any other atom in the 
other strand become very large. This condition is  fulfilled if one takes  $ V_{ds} $ to be  a sum of 
Morse  potentials. Physically, the physically relevant values of $ V_{ds} $  are certainly smaller than those in  
 $\sum_{r=1}^{2} U^{(r)}_{b}  + \sum_{r=1}^{2} U^{(r)}_{a}$.  

  Let  the overall  center-of-mass (CM) position vector of the ds chain and    the  
center-of-mass (CM)  of the $r$-th  strand be denoted by ${\bf R}_{CM}$ and  ${\bf R}^{(r)}_{CM}$, respectively. 
The relative position vectors between both  ${\bf R}^{(r)}_{CM}$ is ${\bf y}={\bf R}^{(2)}_{CM}-{\bf R}^{(1)}_{CM}$.  The relative 
(``bond'') position vectors    along the  $r$-th strand  are  denoted by ${\bf y}^{(r)}_{j}$, $j=1,.., N-1$. One has:
\begin{eqnarray}&&
{\bf R}_{CM} = \frac{ \sum_{r=1}^{2}\sum_{i=1}^{N} M^{(r)}_{i} {\bf R}^{(r)}_{i} }{M_{tot} }\; , M_{tot}=\sum_{r=1}^{2}\sum_{i=1}^{N} 
M^{(r)}_{i}\; ,\label{eq:RcmDS}\\
 &&
{\bf R}^{(r)}_{CM} = \frac{ \sum_{i=1}^{N} M^{(r)}_{i} {\bf R}^{(r)}_{i} }{\sum_{i=1}^{N} M^{(r)}_{i} }\; , 
{\bf y}^{(r)}_{j} = {\bf R}^{(r)}_{j+1} - {\bf R}^{(r)}_{j}    \;  ,    \label{eq:yDS}
\end{eqnarray}
 to be compared to Eqs. (\ref{eq:Rcm}) and (\ref{eq:y}) for a ss chain. Having started with the set formed by all  ${\bf  R}^{(r)}_{i}$, 
 $1\leq i\leq N$, $r=1, 2$, we shall find  also interesting  the new  set of independent position vectors formed by  
 ${\bf R}_{CM}$, all  ${\bf y}^{(r)}_{j}$, $j=1,.., N-1$, $r=1, 2$ and ${\bf y}$.
 Then, by using  (\ref{eq:RcmDS}) and (\ref{eq:yDS}), Eq. (\ref{eq:H1DS})  can be recast,   
in terms of the  variables of the new   set, as: $
   H_{Q,1}  =   - (\hbar^{2}/2M_{tot})\nabla_{{\bf R}_{CM}}^{2} +\tilde{H}_{Q}$, with: 
\begin{eqnarray}
\tilde{H}_{Q} & = &  - \frac{\hbar^{2}}{2M_{red}}(\nabla_{{\bf y}})^{2}   - \sum_{r=1}^{2} \frac{\hbar^{2}B^{(r)}_{1}}{2 }
( \nabla_{{\bf y}^{(r)}_{1}})
^{2}- \sum_{r=1}^{2}\sum_{i=2}^{N-2} \frac{\hbar^{2}B^{(r)}_{i}}{2 }( \nabla_{{\bf y}^{(r)}_{i}}-\nabla_{{\bf y}^{(r)}_{i+1}})^{2} 
\nonumber \\ & - & 
\sum_{r=1}^{2} \frac{\hbar^{2}B^{(r)}_{N}}{2 }( \nabla_{{\bf y}^{(r)}_{N-1}})^{2}  + \sum_{r=1}^{2} U^{(r)}_{b}  + 
\sum_{r=1}^{2} U^{(r)}_{a} + \sum_{r=1}^{2} V^{(r)}_{1}  + V_{ds} \; ,
  \label{eq:HaDS} 
\end{eqnarray}
($M_{red}^{-1}=\sum_{r=1}^{2}(\sum_{i=1}^{N} M^{(r)}_{i})^{-1}$).  Here,  the 
contribution 
$- \frac{\hbar^{2}}{2M_{red}}(\nabla_{{\bf y}})^{2}$,  represents the kinetic energy associated 
to the relative motion of the centers of mass of both chains.

We shall suppose that the ds chain is in thermal equilibrium, at absolute temperature $T$ in an interval about  and not far from 
$  300$ K), and   
  that $ U^{(r)}_{b} =\sum_{i=1}^{N-1}V_{M}(y^{(r)}_{i})$,    $V_{M}$ being the Morse potential 
 with  parameters $D^{(r)}_{j}$, $\alpha^{(r)}_{j}$ and $\omega^{r}_{j}$. 
We shall consider  typical orders of magnitude,  which may apply for dsDNA, at least qualitatively. We suppose that  all 
$D^{(r)}_{j}$  are of similar orders of magnitude and that the  same holds  for all  $\omega^{r}_{j}$ and for all 
$v^{(r)}_{j,j+1;max}(= \textrm{Max}\vert V^{(r)}_{j,j+1}\vert $). We assume  that  $D^{(r)}_{j}$ are appreciably larger 
than all  $v^{(r)}_{j,j+1;max}$ and 
that, in turn,  all  $v^{(r)}_{j,j+1;max}$  are  larger than  all $\hbar\omega^{(r)}_{j}$. Physically,  
the interaction responsible for the 
existence of constant bond 
lengths should always be adequately stronger than that giving rise to constrained bond angles (recall the comments in  
\ref{sec:RFAE.GFC.6} ). 
Moreover, we focus on the case where $k_{B}T$ is less or, at most, of the order of $\hbar\omega^{(r)}_{j}$ and 
\begin{eqnarray} 
\hbar\omega^{(r)}_{j}\gg \frac{\hbar^{2}B^{(r)}_{j}}{(d^{(r)}_{j})^{2}}\, .
\label{eq:frecc}
\end{eqnarray}
Typical values employed in various  analysis of macromolecules and of DNA are consistent with the above assumptions 
 for  $T\simeq 300$ K and somewhat above \cite{Volk,McQ,Proh,Yaku}. For instance (with $0.6$ kcal/mol $\simeq 0.025$ eV, corresponding 
to  $T\simeq 300$ K):  $D^{(r)}_{j}$ about $100$ kcal/mol (or larger), $d^{(r)}_{j}\simeq 1$ to a few \AA,  
$\alpha^{(r)}_{j}\simeq$ a few  \AA$^{-1}$,  $\hbar\omega^{(r)}_{j}$ about $3$-$10$ kcal/mol (1 nanometer= $10 \AA$)  
 and    
atomic masses typical in Organic Chemistry \cite{Yaku}.  On the other hand, all $v^{(r)}_{j,j+1;max}$
 are supposed to be somewhere between $12$ and some value smaller than about $100$ kcal/mol~\cite{McQ}.   
\par

\section{Radial variational computation}
\label{subsec:RFAE.GFC.7.2}

 We have  performed two radial  variational computations.   The (normalized) trial 
variational  total (radial-angular) wave function of the open ds chain reads:
\begin{eqnarray}
\Phi = \left[\prod_{r=1}^{2} \prod_{l=1}^{N-1} \phi^{(r)}_{l} (y^{(r)}_{l}) \right] \psi_{ \sigma} ({\bf y};\theta , \varphi ) \;  .
\label{eq:totDS}
\end{eqnarray}
   $\theta^{(r)}_{i}$, $\varphi^{(r)}_{i} $ are   the angles of  ${\bf y}^{(r)}_{i} $. $\theta , 
\varphi$ denote the set of all   $\theta^{(r)}_{i}$, $\varphi^{(r)}_{i} $. In the first radial variational computation \cite {RaCa}, 
$\phi^{(r)}_{l}$ are similar to  Eq.\ (\ref{eq:osc}) (harmonic-oscillator-like).  In the second radial  variational computation \cite {CAEDNA}, $\phi^{(r)}_{l}$  correspond to 
    Morse potentials. 
The normalization condition $\int [d{\bf y}]\mid \Phi\mid ^{2}=1$
(with  $[d{\bf y}] \equiv d^{3}{\bf y} [\prod_{r=1}^{2} \prod_{i=1}^{N-1} d^{3}{\bf y}^{(r)}_{i}  $) becomes  
 in the limit in which \ (\ref{eq:dirac}) or \ (\ref{eq:Dirac}) hold:
\begin{eqnarray}
\int [ d{\bf y}]\int[{\bf d\Omega}] |\psi_{\sigma} ({\bf y} ;\theta , \varphi )|^{2} = 1\;  ,
\label{eq:normalizationDS}
\end{eqnarray}
where $[{\bf d\Omega}] \equiv  \prod_{r=1}^{2} \prod_{l=1}^{N-1} d\varphi^{(r)}_{l} d\theta^{(r)}_{l} \sin \theta^{(r)}_{l}$. 
The first radial  computation   of $(\Phi, \tilde{H}_{Q} \Phi )$ (when all vibrational  frequencies $ \omega^{(r)}_{i}$ of 
the harmonic-oscillator-like potentials 
$ U_{b}^{(r)}$  grow)  is    similar to that leading to  Eq.\ (\ref{eq:hfrec}).  
The  second radial computation \cite {CAEDNA} of $(\Phi, \tilde{H}_{Q} \Phi )$, using Morse potentials, 
   has been   more difficult to accomplish.
In both radial variational computations,  one  applies   \ (\ref{eq:peierls}) for the quantum partition function of the 
ds macromolecular chain, $\tilde{Z}_{Q}$, determined by Eq.\ (\ref{eq:HaDS}). All these   lead to \cite {RaCa,CAEDNA}:
\begin{eqnarray}
\tilde{Z}_{Q} & \geq & \exp[-(k_{B}T)^{-1}E_{0}]\cdot Z_{Q}^{(ds)} \; ,\label{eq:peierls1DS} \\  
Z_{Q}^{(ds)} & \equiv &
\sum_{\sigma} \exp{\!\left[-\frac{1}{k_{B}T}\!\int[{\bf d\Omega}]\! \int d^{3} {\bf y}\psi_{\sigma}^{*} ({\bf y}; \theta , \varphi ) 
 H_{Q}^{(ds)} \psi_{\sigma}({\bf y};\theta , \varphi ) +{\cal O} ^{( ds)}(\hbar)\right]} 
 \; .
\label{eq:peierls2DS}
\end{eqnarray}
Here,   $E_{0}$ denotes $  \sum_{r=1}^{2}
 \sum_{i=1}^{N-1}\frac{\hbar \omega^{(r)}_{i}}{2}$ for the first  (harmonic-oscillator-like) radial variational computation. For the 
second  radial  variational computation, we have that $E_{0}=  \sum_{r=1}^{2}
 \sum_{i=1}^{N-1}E_{M,i,n=0}^{(r)}$, with $E_{M,i,n=0}^{(r)} $ standing for the energy ($<0$) of the lowest ($n=0$) bound state  
for the Morse potential, constraining the $i$-th 
bond length in the $r$-th strand. The quantum angular Hamiltonian  $H_{Q }^{(ds)}$, which  is  exactly the same  for both radial 
variational computations, reads:
\begin{eqnarray}
\nonumber 
H_{Q }^{(ds)}  &=&    - \frac{\hbar^{2}}{2M_{red}}(\nabla_{{\bf y}})^{2}   + \sum_{r=1}^{2} \frac{B^{(r)}_{1}}{2 d^{(r)}_{1}}
( {\bf e}^{(r)}_{1})^{2}+ \sum_{r=1}^{2}\sum_{i=2}^{N-2} \frac{B^{(r)}_{i}}{2 }\left(\frac{{\bf e}^{(r)}_{i}}{d^{(r)}_{i}}-\frac{{\bf e}^{(r)}_{i+1}}
{d^{(r)}_{i+1}}\right)^{2} \\ 
&+& \sum_{r=1}^{2} \frac{B^{(r)}_{N}}{2d^{(r)}_{N-1} }({\bf e}^{(r)}_{N-1})^{2}+
\sum_{r=1}^{2} U^{(r)}_{a}+\sum_{r=1}^{2} V^{(r)}_{1} + V_{ds} \; . 
  \label{eq:HQANGDS}
\end{eqnarray}
The operators ${\bf e}^{(r)}_{i}$ are analogous to those in  Eq.\ (\ref{eq:HQANG}). $\sum_{r=1}^{2} U^{(r)}_{a}+\sum_{r=1}^{2} V^{(r)}_{1} + V_{ds}   $ in  
Eq.\ (\ref{eq:HQANGDS})
is the restriction of $\sum_{r=1}^{2} U^{(r)}_{a}+\sum_{r=1}^{2} V^{(r)}_{1} + V_{ds}   $ in  Eq.\ (\ref{eq:HaDS}),  when  $y^{(r)}_{l}=d^{(r)}_{l}$, for any $r=1, 2$ and 
$l=1,\ldots,N-1$, the $d^{(r)}_{i}$'s being  the bond lenghts. 
The remainder  ${\cal O}^{( ds)}((\hbar)$ is  the  set of all remaining contributions,  which do not depend on the frequencies 
 and  are   proportional to some positive  power of $\hbar$:  ${\cal O}^{( ds)}((\hbar)\rightarrow 0$ 
  as $\hbar\rightarrow 0$, and we shall disregard it in what follows.  By taking 
the     $\psi_{ \sigma}({\bf y};\theta, \varphi )$'s  as  the 
complete set of all orthonormal eigenfunctions of $ H_{Q}^{(ds)}$, one finds: 
\begin{eqnarray}
 Z_{Q}^{(ds )}
& = & {\rm Tr}[\exp [-(k_{B}T)^{-1}H_{Q}^{(ds)}]]\; .
\label{eq:peierls2DS1}
\end{eqnarray}
 See \cite{CAEDNA} for the   case in which  more bound states are employed in the second radial variational computation.  
  $Z_{Q}^{(ds )}$ can be regarded as the effective  quantum partition function 
for the  three-dimensional double-stranded chain, in terms of slowly-varying degrees of freedom, provided that 
 those associated to bond angles could be treated as soft variables (say, like in  \ref{subsec:RFAE.GFC.4.3} ).
 We shall be concerned  with  bond angles as either soft or hard variables in \ref{subsec:RFAE.GFC.7.2}.  

 Let four external stretching forces ${\bf f}_{1}^{(r)}$ and  ${\bf f}_{N}^{(r)}$, $r=1,2$, 
  act upon the  atoms at  ${\bf R}_{1}^{(r)}$ and ${\bf R}_{N}^{(r)}$, respectively. We suppose that $\sum_{r=1}^{2}
({\bf f}_{1}^{(r)}+{\bf f}_{N}^{(r)})=0
 $, so that there is  no net force upon the overall CM. Then, $\tilde{H}_{Q}$ in  
\ (\ref{eq:HaDS})  is replaced by:   
\begin{eqnarray}&&
\tilde{H}_{Q,{\bf f}}=\tilde{H}_{Q}+\sum_{r=1}^{2}\sum_{i=1}^{N-1} ({\bf f}_{1}^{(r)}\alpha_{1,i}^{(r)}+
{\bf f}_{N}^{(r)}\alpha_{N,i}^{(r)}){\bf y}_{i}^{(r)}+({\bf f}_{1}^{(2)}+{\bf f}_{N}^{(2)}){\bf y}\; .
\label{eq: HaDSf}
\end{eqnarray}
Here, we have employed: ${\bf R}_{i}^{(r)}={\bf R}{(r)}_{CM}+\sum_{i=j}^{N-1} \alpha_{i,j}^{(r)}{\bf y}_{j}^{(r)}$, where
$(\sum_{s=1}^{N}M^{(r)}_{s}) \alpha_{i,j}^{(r)}=\sum_{h=1}^{j}M^{(r)}_{h}$ and $-\sum_{h=j+1}^{N}M^{(r)}_{h} $, for 
$j=1,..,i-1$ and $j=i,..,N-1$, respectively \cite{CAEDNA}. 
The quantum-mechanical  variational computation goes through as above and 
yields a new quantum partition function: $Z_{Q,{\bf f}}^{(ds)}= {\rm Tr}[\exp [-(k_{B}T)^{-1}H_{Q,{\bf f}}^{(ds)} 
  ]]$, with: 
\begin{eqnarray}&&
H_{Q,{\bf f}}^{(ds)}= H_{Q}^{(ds)}+\sum_{r=1}^{2}\sum_{i=1}^{N-1} ({\bf f}_{1}^{(r)}\alpha_{1,i}^{(r)}+
{\bf f}_{N}^{(r)}\alpha_{N,i}^{(r)})d_{i}^{(r)}{\bf u}_{i}^{(r)}+({\bf f}_{1}^{(2)}+{\bf f}_{N}^{(2)}){\bf y}\; .
\label{eq:HaDSf}
\end{eqnarray}

\section{Constraining bond angles and  classical limit}
\label{subsec:RFAE.GFC.7.3}

First,    under the assumption that $\sum_{r=1}^{2} U^{(r)}_{a}$ be weak like in   \ref{subsec:RFAE.GFC.4.3} ( the bond angles 
 being regarded as soft variables   ), let us  turn to the classical  limit. Then, we treat  the variables 
${\bf e}^{(r)}_{i}$, $\theta , \varphi$ in each strand like in the freely-jointed chain (subsection \ref{subsec:RFAE.GFC.4.2}). 
  One can also proceed to the classical limit  for  ${\bf y}$  and $-i\hbar\nabla_{\bf y}$,  as each individual chain is a 
very  massive object, for large $N$ \cite{RaCa}.  Then,  $H_{Q
}^{(ds)} $ in \ (\ref{eq:HQANGDS}) becomes, in the classical limit:  
\begin{eqnarray}
H_{c }^{(ds)}  &=&  \frac{\mbox{\boldmath{$\pi$}}_{c}^{2} }{2M_{red}}   +\sum_{r=1}^{2} \frac{B^{(r)}_{1}}{2 d^{(r)}_{1}}
({\bf a}^{(r)}_{1,c} )^{2} + \sum_{r=1}^{2}\sum_{i=2}^{N-2} \frac{B^{(r)}_{i}}{2d^{(r)}_{i} }\left({\bf a}^{(r)}_{i,c }-{\bf a}^{(r)}_{i+1,c }\right)^{2}\nonumber \\ 
 &+&    \sum_{r=1}^{2} \frac{B^{(r)}_{N}}{2d^{(r)}_{N} }({\bf a}^{(r)}_{N-1,c })^{2} + \sum_{r=1}^{2} U^{(r)}_{a}+\sum_{r=1}^{2} V^{(r)}_{1} + V_{ds}  \; .  
 \label{eq:GClassaDS}
\end{eqnarray}
The ${\bf a}^{(r)}_{i,c}$'s  like  in 
 Eq.\ (\ref{eq:aClass}), with components $P_{\theta^{(r)}_{i}},  P_{\varphi^{(r)}_{i}}$. 
The classical vector $\mbox{\boldmath{$\pi$}}_{c} $ corresponds to  the classical limit of $-i\hbar\nabla_{\bf y}$. 
We shall omit the associated 
classical partition function, which can be written by extending straightforwardly ( it can be recovered from  
 Eqs.~(\ref{eq:FIclPFDS}) and (\ref{eq:zred}) below, provided that  the function  $F^{(r)}_{j}$ in (\ref{eq:zred})  
be  proportional to 
$\exp [-(k_{B}T)^{-1}V^{(r)}_{j,j+1}(\vert d^{(r)}_{j}{\bf u}^{(r)}_{j}+d^{(r)}_{j+1}{\bf u}^{(r)}_{j+1}\vert) ]$).  

Second, we now come  to the physically very important case in which $\sum_{r=1}^{2} U^{(r)}_{a}$ (even if weaker 
than $\sum_{r=1}^{2} U^{(r)}_{b}$,  constraining  bond vectors) 
are still adequately  strong so as to constrain the bond angles (hard variables, as well). Thus, let  
$\sum_{r=1}^{2} U^{(r)}_{a}$ be a sum of harmonic-oscillator-like 
potentials.    The  
radial-angular quantum-mechanical variational computation  for  double-stranded 
open freely-rotating chains, in which  the degrees of freedom for $\sum_{r=1}^{2} V_{b}^{(r)}+  \sum_{r=1}^{2}V_{a}^{(r)}$ are treated  
quantum-mechanically both 
 on the same footing, has been carried out in \cite{CAEDNA1},  in outline. Alternatively, 
  $\sum_{r=1}^{2} U^{(r)}_{a}$  could be a sum of other Morse potentials having  sharp minimum for 
$\vert {\bf y}^{(r)}_{j}\vert= d^{(r)}_{j}$ and ${\bf u}^{(r)}_{j}{\bf\cdot u}^{(r)}_{j+1}=\beta^{(r,0)}_{j}$: the corresponding radial-angular variational computation has been outlined in 
 \cite{CAEDNA}, with   results similar to those in \cite{CAEDNA1}. 

The last   step in the second  radial-angular variational computations  with bond angles  as hard variables 
 is to obtain, upon performing the 
transition to the classical 
limit,  a classical partition function which could be handled without unsurmountable difficulties. This is 
 more difficult than in the transition  yielding Eq.\ (\ref{eq:GClassaDS}).  We shall limit ourselves to  the case 
$\beta^{(r,0)}_{j}$ close to $+1$ for any $r$ and $j$, to be assumed below: $\beta^{(r,0)}_{j}\simeq 0.8$  for any $r$ and $j$  corresponds
  approximately to B-DNA  \cite{Volk}). Then,  we  
 arrive, as a result of  the   radial-angular variational computation,  at  a reasonable  classical partition function, 
given below in   Eqs.
 \ (\ref{eq:FIclPFDS})-\ (\ref{eq:zred}). See \cite{CAEDNA} (appendix B), for  an outline.  
In order to get \ (\ref{eq:FIclPFDS})-\ (\ref{eq:zred}), one performs the integrations over all the classical momenta 
 $P_{\theta^{(r)}_{l}},  P_{\varphi^{(r)}_{l}}$ 
( like  in \ref{subsec:RFAE.GFC.4.2} ) and the one over  $\mbox{\boldmath{$\pi$}}_{c} $, which   is also Gaussian.
Then,  the effective  classical partition function for the three-dimensional double-stranded open macromolecular chain, 
with constrained bond lengths and angles reads \cite{CAEDNA}: 
\begin{eqnarray} 
Z_{c }^{(ds)} &=& \left[ \frac{k_{B}T}{2\pi \hbar^{2}} \right]^{2(N-1)} 
\frac{\left[ \prod_{r=1}^{2} \prod_{l=1}^{N-1}( d^{(r)}_{l})^{2} \right]}{[ \prod_{r=1}^{2}{({\rm det}
 B^{(r)})^{3/2}}]} \left[ \frac{M_{red}K_{B}T}{2\pi \hbar^{2}} \right]^{3/2} Z_{red} \;  , \label{eq:FIclPFDS}\\
 Z_{red}&=&
\int  d^{3}{\bf y}\int[{\bf d\Omega}] \left[\prod_{r=1}^{2}\prod_{j=1}^{N-2}F^{(r)}_{j}\right]\left[\prod_{s=1}^{2} [\Delta^{(s)})_{N-1}\right]^{-1/2}\nonumber\\
&\times&  \exp\left[-\frac{ \sum_{r=1}^{2} V^{(r)}_{1} +V_{ds}  }{k_{B}T}\right] \; .\label{eq:zred}
\end{eqnarray}
The $[\Delta^{(s)})_{N-1}]^{-1/2}$'s  are given, for each $s=1,2$, 
by the right-hand-sides of  Eqs. \ (\ref{eq:determ}),  \ (\ref{eq:matrixB}) and  \ (\ref{eq:matrixinvB}).
    The function $F^{(r)}_{j}$, arising from strong $ V_{a}^{(r)}$ (hard bond angles), is sharply peaked at 
 $\vert d^{(r)}_{j}{\bf u}^{(r)}_{j}+d^{(r)}_{j+1}{\bf u}^{(r)}_{j+1}\vert\simeq d^{(r)}_{j,j+1}$, where
 $(d^{(r)}_{j,j+1})^{2}=(d^{(r)}_{j})^{2}+(d^{(r)}_{j+1})^{2}+2d^{(r)}_{j}d^{(r)}_{j+1}\beta^{(r,0)}_{j}$. As an approximation, 
we take   
$F^{(r)}_{j}$ to be  approximately proportional to $ 
\delta({\bf u}^{(r)}_{j}{\bf\cdot u}^{(r)}_{j+1}-\beta^{(r,0)}_{j})$, $\delta$  denoting Dirac's delta function. 
 $\sum_{r=1}^{2} V^{(r)}_{1} +V_{ds} $ in Eq.  (\ref{eq:zred}) is  the restriction of the previous $\sum_{r=1}^{2} V^{(r)}_{1} +V_{ds}  $, 
 when   all bond lengths and angles are constrained.  Through similar arguments, Eq.
 \ (\ref{eq:HaDSf}) would lead to classical models for a ds chain, subject to stretching forces.

 Eqs.
 \ (\ref{eq:FIclPFDS})-\ (\ref{eq:zred}), with   
     further physical approximations,  have  provided a  basis for certain models of dsDNA ~\cite{CAEDNA,CAEDNA2}: 
see section \ref{sec:RFAE.GFC.8} for an outline.
\par

\chapter{Classical effective models for  double-stranded open DNA}
\label{sec:RFAE.GFC.8}

 Let us consider   a   typical   three-dimensional ds open DNA   macromolecule (dsDNA)
(say,  B-DNA \cite{Leh,Volk}),  at thermal equilibrium at temperature  $T$ in an interval from about room temperature  up to 
 about  the melting or
 (thermal ) denaturation 
one, $T_{m}\simeq 360$ $K$. That dsDNA is formed by two open single strands of DNS  (ssDNA).
For $T<T_{m}$, both ssDNA are bound to each other, forming dsDNA. For $ T>T_{m}$, 
dsDNA becomes two separate ssDNAs, each of which retaining still its separate existence as an extended and connected  structure,  
provided that $ T$ be not too high. 

   Each single strand (ssDNA) of real dsDNA is   formed by a very large number $N$ of  nucleotides. 
 $N$ can vary much from one species of dsDNA to another. To fix the ideas, let us take:  $ N\sim 10^{10}$. In turn, 
each nucleotide is formed by a  sugar, a phosphate and a base (either A or C or G or T). The masses of the 
four bases $A$, $C$, $G$ and $T$  differ from their average mass  by less than about $5$, $13$,  $18$ and $11$ per 
cent, respectively. To  simplify the picture,  we shall regard each open  ssDNA  as a single 
discretized chain  formed by  $N(\gg1)$ basic units, also referred to here  as nucleotides, 
all  with equal mass $M$ (such that $NM$ equals the total mass of the DNA single strand). Thus,  $M$ 
includes, in an average or effective sense,  the  contributions  of the  masses of sugars, phosphates and bases. So, $M$  is  
larger than the  average mass of the four bases in  the strict sense (say, of $A$, $C$, $G$ and $T$).  
\par
\begin{figure}[t]
\begin{center}
{
\includegraphics[scale=0.4]{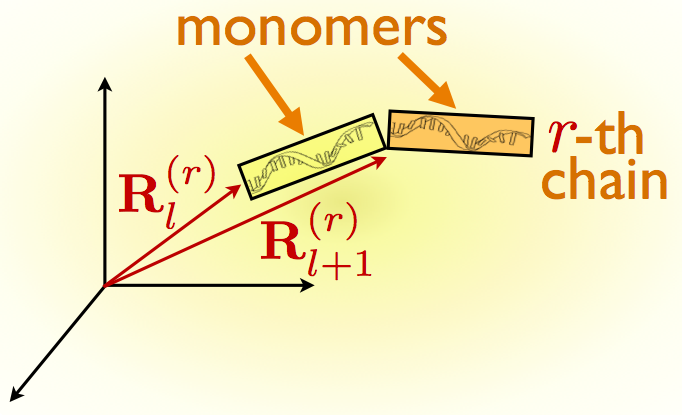}}
\end{center}
\vspace*{-3mm}
\caption{\small Monomers in the $r$th-strand of a double-stranded open macromolecular chain.}
\vspace*{-4mm}
\label{fig:RFAE.GFC.6}
\end{figure}

As stressed previously (see, for instance, 
\cite{Frank}), dsDNA behaves, as far as a variety of phenomena is concerned, as a macromolecule  ressembling,  as a first or rough 
approximation, 
 a Gaussian chain 
in which  certain effective interactions should also be taken into account.  
It is certainly 
most fortunate that those  
simplifying  (Gaussian-like) features hold approximately for dsDNA macromolecules,    which play an absolutely priviledged and unique role in 
Molecular Biology. 
Of course, there are other complicated  phenomena for dsDNA,  which cannot be described, even as a rough 
approximation,  on the basis 
of  an effective Gaussian chain with interactions.     The latter 
 simplifications  are  not valid for  
other macromolecules.

 Based upon Eqs. \ (\ref{eq:FIclPFDS}) and \ (\ref{eq:zred}) (with the  $F^{(r)}_{j}$'s corresponding to hard bond angles) 
for the  classical 
partition function of   three-dimensional  open  dsDNA at thermal equilibrium, we have performed  a detailed analysis and  further approximations, 
 in   $ T< T_{m}$ and as $T$ approaches $T_{m}$ from below    ~\cite{CAEDNA,CAEDNA2}. Below, we shall outline the main outcomes  of 
those studies, 
which seem to be   
  supported, up to certain extent,  by   previous wisdom   in \cite{Frank} (and try to implement it).

 Let $T<T_{m}$. Covalent forces ( the conterpart of  $U({\bf y})$ in  \ref{sec:RFAE.GFC.3}) constrain  approximately constant bond lengths   and bond angles.
 In each   single  ssDNA in dsDNA, all  bond lengths are taken as approximately equal to one another ( $d$),  as are  
  the cosines of all bond angles (  denoted as   
$\beta^{(0)}$). Specifically for 
B-DNA,  the interactions 
$\sum_{r=1}^{2} U^{(r)}_{a}$  and $\sum_{r=1}^{2} U^{(r)}_{a}$ ( section \ref{sec:RFAE.GFC.7}) yield  $d\simeq 0.7$ nanometers and 
$  \beta^{(0)} \simeq 0.8$, respectively.  

   In each   ssDNA of dsDNA at thermal equilibrium,  also for $T<T_{m}$, 
we have   carried  out      medium and  large distance Gaussian-like approximations, 
that lead to  certain  effective monomers (e-monomers),  
as natural molecular blocks  for medium and  large length scales ~\cite{CAEDNA,CAEDNA2}.  
The analysis and   Gaussian-like approximations based upon Eqs. \ (\ref{eq:FIclPFDS}) and \ (\ref{eq:zred}) make  those e-monomers 
to appear explicitly. 
That  enables to regard each 
 ssDNA in dsDNA as formed by $L(=(N-1)/n_{e}\simeq  N/n_{e}\gg 1)$ effective ($e$-)monomers.  
 Each $e$-monomer is   formed by $n_{e}(>1)$  nucleotides and has effective length  $d_{e}$. Lower limits are $ n_{e}=20 $ and  
 $d_{e}\simeq 10$ $nm$, while upper limits of $  d_{e} $ and $ n_{e}  $ could be relatively  close to the  
persistence length $d_{ds}$ ($\simeq 50$ $nm$) of dsDNA \cite{CAEDNA,CAEDNA2}. 
Different  
$e$-monomers behave as statistically independent from one another, 
in some approximate way (except for some weak effective residual interaction $V$ among them, to be given in  Eq. \ (\ref{eq:Vmon}) and  discussed below).  
    In the $r$-th chain,  $r=1,2$,   ${\bf R}^{(r)}_{l}$ and  ${\bf R}^{(r)}_{l+1}$ will denote   the  three-dimensional 
position vectors of the origin and the end of the $l$-th  $e$-monomer,  $l=1,...,L$: no confusion should arise  between the actual 
${\bf R}^{(r)}_{l}$'s  for monomers and the ${\bf R}^{(r)}_{i}$'s ($i=1,...N$) employed in 
\ref{subsec:RFAE.GFC.7.1} 
for the position vectors of atoms.

  All   $e$-monomers  in dsDNA 
    are subject to    effective or residual  intra-chain and  inter-chain interactions (weaker than all covalent ones), all of which 
are    described by an effective potential $V$ ( Eq. \ (\ref{eq:Vmon}) ),  in the domain of validity of     Gaussian and 
long-distance approximations ~\cite{CAEDNA}. $V$ depends on all  ${\bf R}^{(r)}_{l}$. 
 
On the other hand,  in real dsDNA below $T_{m}$,    residual    (  intra-chain and  inter-chain) interactions do operate and lead to longer (super-)monomers. 
The sizes of the super-monomers in the double-stranded system would be characterized by the  persistent length $d_{ds}$ (about $50$ 
nanometers, amounting to $150$ nucleotides, for  B-DNA) or, equivalently, by the  Kuhn length (about $100$ nanometers, for  B-DNA)
~\cite{Volk,Gros,Frank}. The scales of 
$d_{e}$ and $d_{ds}$ are similar, although  $d_{e}$ is certainly  smaller than  $d_{ds}$. A ds-monomer has a length equal to the  
   persistence  length $  d_{ds}$ and  is formed by two parallel substrands of ssDNAs. We emphasize that a ds-monomer should not to be confused with the  
single e-monomers  in each ssDNA  (with  length $  d_{e}  $ ). 
 The fact that $d_{ds}>d_{e}$ can be understood as arising from some effective interactions included in  $V$, which 
would give rise to some repulsion in the double-stranded structure.   In this connection, we remind that the interactions between phosphates  in DNA are 
repulsive. For further comments, see subsection 3.4 in  ~\cite{CAEDNA}. 
  
  Typical   potential energies of  the $A-T$ pair  differ  appreciably from those for the  $C-G$ pair \cite{Yaku}. Specifically, 
 the potential energiy of  a A-T 
pair is   smaller, in absolute value, than that  of  a  C-G pair.
However, the influence of those  inhomogeneities in  the effective residual interaction $ V$ among the 
effective  monomers in dsDNA could possibly be rather weak and be smoothed out, due to the averaging over the  
interactions of the  nucleotides  included in each monomer,  and we shall disregard such inhomogeneities.
  Thus, in our main analysis,     dsDNA is regarded as homogeneous.

 The total  effective residual  intra-chain and inter-chain  potential $  V  $ among all  e-monomers in dsDNA  is taken as:
 \begin{eqnarray} 
V= V_{0}+\sum_{r=1}^{2} V^{(r)}_{1,e}+ V_{2} \; . \label{eq:Vmon}
\end{eqnarray}
  $  V_{0} $ is the potential   between all pairs of  (complementary or mate) e-monomers  at the same positions  in the different  
strands (different ssDNAs).
$    \sum_{r=1}^{2} V^{(r)}_{1,e}  $  is the potential between different  e-monomers  in the same ssDNA. It could be regarded as the 
result of  averaging $\sum_{r=1}^{2} V^{(r)}_{1}$ over monomers (recall \ (\ref{eq:H1DS})).   It also 
takes into account excluded-volume effects. 
   $  V_{2} $  is the potential between pairs of  e-monomers at  unequal or non-complementary positions in the different ssDNAs.

  $  V_{0} $,  $  \sum_{r=1}^{2} V^{(r)}_{1,e} $ and $  V_{2} $ are  repulsive for
    short distances.  
  $  \sum_{r=1}^{2} V^{(r)}_{1,e} $ and $  V_{2} $  take care of stacking interactions.  $    V_{0} $ and, 
eventually, $  V_{2} $ are  attractive at intermediate and large distances.  
 
  As a result of a series of approximations,  Eqs. \ (\ref{eq:FIclPFDS}) and \ (\ref{eq:zred}) yield  the following effective   classical 
partition function for  three-dimensional  open  dsDNA at thermal equilibrium, in terms of configurations of e-monomers, for  $T< T_{m}$ ~\cite{CAEDNA,CAEDNA2}: 
\begin{eqnarray}\nonumber 
Z_{2}&=&\left[ \frac{k_{B}T}{2\pi \hbar^{2}} \right]^{2(N-1)}\left[ \frac{MNK_{B}T}
{4\pi \hbar^{2}}\frac{M_{tot}K_{B}T}{2\pi \hbar^{2}} \right]^{3/2}
\left[\frac{d^{4(N-1)}}{(N/M^{N-1})^{3}}\right]\\
&\times&\left[\prod_{r=1}^{2}Z_{R,app}\right][\frac{4\pi R_{0}^{3}}{3}]Z \; ,\label{eq:dsDNA}\\
Z&=& \frac{3}{4\pi R_{0}^{3}} \int  \left[\prod_{r=1}^{2} d^{3}{\bf R}^{(r)}_{L+1} d^{3}{\bf R}^{(r)}_{1}\right] G(L)\, .
\label{eq:green0}
\end{eqnarray}
  $M_{tot}=2MN$ is the total mass of dsDNA. The region  in which the ds open   chain  moves   is  a sphere of 
very  large radius $R_{0}$ (compare with  \ (\ref{eq:Gfj5})).  An additional factor $(4\pi R_{0}^{3})/3$   has now been included, in order to 
facilitate the comparison with \cite{CAEDNA2}. Such a factor, 
arising from the overall 
CM degrees of freedom of the whole dsDNA, has been  factored out in all previous calculations, up to and including those in 
\ref{subsec:RFAE.GFC.7.1}. 
 $Z_{R,app}$ (one per ssDNA) has appeared previously in Eq. \ (\ref{eq:Gfj4}). 
$Z_{R,app}$'s are $T$-independent, and   they  will either cancel out  or    not be 
relevant here. Anyway, $Z_{R,app}$  is given in Eqs. 
(C.4) and (C.2) in  ~\cite{CAEDNA}. 
For the actual  open ds system:
\begin{eqnarray}
&& G(L)=G({\bf  R}^{(1)}_{L+1},{\bf  R}^{(2)}_{L+1};{\bf  R}^{(1)}_{1},{\bf  R}^{(2)}_{1};L)=\int \left[\prod_{r=1}^{2}\left(\prod_{l'=2}^{L}d^{3}{\bf R}^{(r)}_{l'}\right)\right] W_{eq}
\label{eq:green1}\\&&W_{eq}=\left[\prod_{r=1}^{2}\prod_{l=1}^{L}W_{G}({\bf R}^{(r)}_{l+1}-{\bf R}^{(r)}_{l};2d_{e}^{2})\right]
\exp[-(k_{B}T)^{-1} V]\;  .\label{eq:green2}
\end{eqnarray}
  On the other hand, 
  $W_{G}({\bf R}^{(r)}_{l+1}-{\bf R}^{(r)}_{l};2d_{e}^{2})$ denotes the Gaussian distribution for  
the $l$-th monomer in the $r$-th strand, namely, $[3/(2\pi d_{e}^{2})]^{3/2}\exp[-3({\bf R}^{(r)}_{l+1}-
{\bf R}^{(r)}_{l})^{2}/(2d_{e}^{2})]$. Notice that $Z$ and $G(L)$ are manifestly rotationally invariant.  
Eqs. \ (\ref{eq:green1}) and 
 \ (\ref{eq:green2}) generalize not only \ (\ref{eq:Gfj5}) and 
 \ (\ref{eq:Gfj6}) but also other one-dimensional models of dsDNA \cite{Pey,Pey1}. Through similar arguments, Eqs.
 \ (\ref{eq:HaDSf}),  \ (\ref{eq:green1}) and 
 \ (\ref{eq:green2}) would lead to classical effective models for a ds chain,  with   e-monomers at the ends subject 
to 
 stretching forces. That could possibly provide  some  basis for comparisons with  experimental results observed in DNA stretching experiments 
\cite{Strick,Kumar} and their interpretation by means of the 
worm-like chain model \cite{Marko,Bou}. There appears to be overall consistency between  values of  persistence length obtained 
in  
\cite{Bou} with those in \cite{Gros,Frank}, as it should be. A closer analysis     lies outside our scope here.

We remark that, for $  T<T_{m}$, $Z$ and, hence, $Z_{2}$ include, in principle,  not only contributions from the bound ds  
structure (denoted with the superscript $bo$)  but also from unbound configurations, with the two separate 
single strands unbound from each other (denoted with the superscript $ub$). Accordingly, it is reasonable 
to write $Z=Z^{(bo)}+Z^{(ub)}$ and, hence,  $Z_{2}=Z_{2}^{(bo)}+Z_{2}^{(ub)}$, where $Z^{(bo)}$ ($Z_{2}^{(bo)}$) and $Z^{(ub)}$ ($Z_{2}^{(ub)}$) 
are the contributions due to  $bo$  and  $ub$ configurations. Physically,  $Z^{(bo)}$ should dominate over $Z^{(ub)}$ the more, the 
larger $  T_{m}-T (>0)$. As 
$  T_{m}-T\rightarrow 0$, one may expect that such a dominance of $Z^{(bo)}$ disappears. The decomposition $Z_{2}=Z_{2}^{(bo)}+Z_{2}^{(ub)}$ and 
those features are  not transparent,  a priori, in  \ (\ref{eq:dsDNA}), but they  can  be 
supported  by other  calculations.

 Above thermal  denaturation ($  T>T_{m} $), 
 dsDNA, having melt, becomes  two separate ssDNAs. 
 Then, for $T>T_{m}$, the  effective residual interactions contained in $V$  among monomers  belonging to different ssDNAs become  
 negligible (as both ssDNA's are far from 
each other). Then, for both  $V_{0}$ and $V_{2}$ in Eq. \ (\ref{eq:Vmon}), one has:  
  $ (K_{B}T)^{-1}(V_{0}+V_{2})\simeq 0 $. For further  discussion of ssDNA  for $  T>T_{m} $,  see \cite{CAEDNA2}.
 We have also studied the dynamics of open dsDNA close to thermal denaturation, based upon both 
the approximations summarized  above in this section and a Smoluchowski equation. Such a study lies   outside 
the scope of the present  paper, devoted to constrained macromolecular chains at equilibrium and, hence, will be omitted. 
See \cite{CAEDNA2}.

\chapter{Concluding comments}
\label{sec:RFAE.GFC.11}

We have concentrated on quantizing Fraenkel's model (very stiff flexible chain), by taking into account 
  \cite{Go1,Go2} and, 
mostly,  
  \cite{Ral}
: see   the last paragraph of subsection  
 \ref{subsec:RFAE.GFC.2.2}.  In a nutshell: upon quantization, does one obtain the physically expected 
large quantum zero-point energies  of  hard degrees of freedom? Do those zero-point energies depend  on  soft degrees 
of freedom? To the best of our knowledge, no exact or asymptotically exact  answer,  operating directly on the Schr\"{o}dinger 
equation and providing affirmative answers to those questions, exists at present. However,   a moderate step forward has been made in the last  years 
\cite{AEPR,AEM,AE,Calvo,RaCa,CAEDNA,CAEDNA2,CAEDNA1} for equilibrium quantum-mechanical partition functions of three-dimensional
 macromolecules, based always on the same underlying physical ideas.  Specifically, upon quantizing Fraenkel's  model and 
applying Peierls`   quantum-mechanical  variational  inequality  for large vibrational frequencies, 
the expected and physically correct large quantum zero-point energies  of the constrained (hard) degrees of freedom are 
obtained exactly. They are  proved  analytically  to be constant (and, hence, independent on the unconstrained soft variables) 
for several ss macromolecules 
(open, closed-ring, freely-jointed, freely-rotating, etc) and for ds open ones.   
We stress the crucial importance of choosing the variational trial wave functions   adequately: otherwise, 
one may obtain either contributions depending  on the unconstrained soft variables or results  which, even if independent 
on the  latter variables, do not coincide with the  physically expected  quantum zero-point energies (see the comment 
after \ (\ref{eq:HQANG00})). 
The proof of constancy  becomes  considerably complicated, as one proceeds from the open freely-jointed  chain 
  to  chains with constrained bond lengths and bond angles or closed-ring constraints: we have provided  in appendices A and D 
some (so far, unpublished) details which complement, in outline, various aspects  given  previously  
\cite{Calvo,RaCa,CAEDNA,CAEDNA1}. That  constancy leads to specific quantum partition functions  and hamiltonians 
($Z_{Q}$ and $H_{Q}$) for  the  soft angular  variables, neatly separated from the hard degrees of freedom. 
One always finds structures similar to those in Eqs. \ (\ref{eq:hfrec1}),  \ (\ref{eq:definham1})  and  
\ (\ref{eq:HQANG00}), with different $w_{ang}$ (associated to  all   constraints  except  those for the 
constrained bond lengths): see section \ref{sec:RFAE.GFC.6}.
\par
We are aware of the increasing complexities of the $Z_{Q}$'s  and $H_{Q}$'s derived for the successively  quantized chains, with 
increasingly complicated constraints, 
in sections \ref{sec:RFAE.GFC.4} through  \ref{sec:RFAE.GFC.7}. Our justification for them is twofold. First,
 Peierls`   inequality  yields, after  delicate exact  cancellations,  the correct  constant quantum zero-point energies. 
 Second, the resulting  $Z_{Q}$'s  and $H_{Q}$'s are rotationally invariant and the total quantum angular momentum is 
conserved,  which are very  important physical properties. In view of those nontrivial results, the quantum variation 
inequality appears to operate in the right direction even if, recognizedly, it may not be providing us, as yet, the definitive quantum-mechanical formulation for constrained macromolecules.
\par
What practical consequences do $Z_{Q}$'s  and $H_{Q}$'s  imply? How do their applications differ from and/or improve  those from cCHDa? For that purpose,  by taking  the classical  limit, the  $Z_{Q}$'s become the classical  partition functions $Z_{C}$'s, which are, respectively,  different from the classical partition functions found in cCHDa for similar  chains. The  $Z_{C}$'s for open chains yield, after certain large-distance approximations,  several quantities (bond-bond correlations,    squared end-to-end distance, probability distribution for the end-to-end vector, and others),  which agree consistently with those from the standard Gaussian model 
in Polymer Science. Since a similar consistency is met with cCHDa, the comparison with the standard Gaussian model does not distinguish, thus far, the consequences of our variational QMa from those from cCHDa. 
\par 
As another methodological application, we have also applied the  above  Peierls`inequality approach plus  classical and 
long-distance approximations to three-dimensional ds macromolecules, like  dsDNA. We have derived  the corresponding $Z_{C}$'s, 
 bearing structures which  generalize other  one-dimensional models of dsDNA \cite{Pey,Pey1}.   One could also develop similar 
quantum-mechanical approaches, starting from   Eq.\ (\ref{eq:H1DS}) with similar $\sum_{r=1}^{2} U^{(r)}_{b}$,  
$\sum_{r=1}^{2} U^{(r)}_{a}$   (so as to constrain bond lengths and angles)  but with some suitably chosen 
 $\sum_{r=1}^{2} V^{(r)}_{1} $+$V_{ds}$, so as to generate other kinds of (eventually weaker) angular constraints in 
each strand among neighbouring 
atoms (which be neither nearest-neighbours nor next-to-nearest-neighbours). Then, by extending the approaches in sections 
\ref{sec:RFAE.GFC.7} and \ref{sec:RFAE.GFC.8}, one could 
get eventually other models for dsDNA. 
\par
Thus, whether  our variational QMa has practical  applications  for macromolecules at thermal equilibrium which differ neatly from and/or improve  those from 
cCHDa remains open, thus far. 
\par
We have limited ourselves to static properties (equilibrium partition functions) 
upon quantizing very stiff flexible chains. We have not undertaken the far more difficult 
task of  quantizing very stiff flexible macromolecules off-equilibrium (that is, of analyzing dynamical properties {\em ab initio}, at the quantum level). Anyway, 
a model for the non-equilibrium evolution  of  a ds macromolecule, based upon the Smoluchowski equation and  the  approximate $Z$  obtained  in \cite{CAEDNA},  has been proposed in \cite{CAEDNA2}. In so doing, we have followed a  pragmatic procedure (not to be confused with a first-principles treatment): 
having obtained an approximate classical partition function ($Z$), we have constructed directly the classical  Smoluchowski equation which has  as 
equilibrium solution the distribution function characterizing uniquely $Z$. This model has been employed to study thermal denaturation of dsDNA: specifically, an approximate formula for the time duration required for thermal denaturation to occur (about the melting temperature) has been obtained. That pragmatic method   appears to be consistent,  at least  in spirit, with that followed in other approaches to the dynamics of constrained macromolecules: compare, for instance,  with \cite{Doi,SSTtiMaHa}. 
\par   
Biological macromolecules and the processes which occur in or involve them give rise to a fantastic variety of phenomena. A good number of the latter may well be described, to a sufficient degree of approximation, by classical approaches (cCHDa).   At certain stages in the analysis of  macromolecules, it may be unclear  whether  quantum-mechanical approaches  have enough practical consequences which differ from and/or improve those from  classical ones. But it is impossible to accept that classical formulations  will account for all the above phenomena, at all scales down to the nanometer one. Thus,  and as a matter of principle and  of scientific 
strategy, it seems that, if possible,  one should  pursue on disposing of both classical and quantum formulations. In fact, 
at some short spatial and/or temporal scales  (at some adequately large energy scale), one will have to deal with some 
genuinely quantum features and to analyze them. See the comments in section\ref{sec:RFAE.GFC.3}, in connection with Schr\"{o}dinger's book \cite{Schrod}.
 This has been a {\em leitmotif} to motivate the researches reported in this tutorial review.

\section*{Acknowledgments}
\label{sec:20} 
The authors of this review are grateful to Drs. G. Ciccotti and P. Echenique for their valuable comments.  They also thank Prof. A. 
Gonz\'{a}lez-L\'{o}pez for providing references \cite{Poly1,Poly2,Fink,Enc} and interesting discussions about them.  RFAE acknowledges the financial support from Project FIS2008-01323, Ministerio de Ciencia e Innovaci\'{o}n, Spain.  One of us (R. F. A.-E.) is an associate member of Instituto de Biocomputacion  y Fisica de  Sistemas Complejos, Universidad de Zaragoza, Zaragoza, Spain. GFC acknowledges financial support from Projects MTM2009-13832 (Ministerio de Ciencia e Innovaci\'on, Spain) and PEII11-0178-4092 (Junta de Comunidades de Castilla-La Mancha, Spain).  
  
\appendix


\chapter{}

\section*{Single-stranded   open freely-jointed   chain: proof  of ~(\ref{eq:hfrec})-~(\ref{eq:definham}) }
 \label{sec:A} 
    We shall use $[d{\bf y}]=[\prod_{i=1}^{N-1}y_{i}^{2}dy_{i}][{\bf d\Omega}]$ and  $[{\bf d\Omega}]$, $\Phi ({\bf y})$ and 
$\phi_{l} (y_{l})$  
( Eqs. ~(\ref{eq:37}),  (\ref{eq:tot}) and   (\ref{eq:osc}), respectively).  We shall consider   successively all  contributions from: 
 \begin{eqnarray}&&
\int  [\prod_{i=1}^{N-1}y_{i}^{2}dy_{i}][{\bf d\Omega}][\prod_{i=1}^{N-1}\phi_{l} (y_{l})\psi_{\sigma} (\theta , \varphi ) ]^{*}( 
H_{Q,in} +U({\bf y}))
[\prod_{i=1}^{N-1}\phi_{l} (y_{l})\psi_{\sigma} (\theta , \varphi ) ]\; ,\label{eq:ofjir0}
 \end{eqnarray}
  $ H_{Q,in} $ being given in  Eq. (\ref{eq:T1}). 
 First, we shall suppose that $U({\bf y})$ is the sum of    Morse potentials   in 
Eq. (\ref{eq:morsei}). We shall start with  the simplest structures. 
 The contributions from $2^{-1}B_{i}[{\bf a}_{i} \cdot {\bf a}_{i}/ y_{i}^{2}]$ and ${\bf a}_{i-1} \cdot {\bf a}_{i}/(M_{i}
(y_{i-1} y_{i}))$ follow immediately, by  letting 
  $\omega_{l} \rightarrow +\infty$ and using Eqs.  (\ref{eq:dirac}) and  (\ref{eq:Del2}). One gets: 
\begin{eqnarray}&&
\int   [{\bf d\Omega}]\psi_{\sigma} (\theta , \varphi ) ^{*}\frac{B_{i} }{2}  \frac{{\bf a}_{i} \cdot {\bf a}_{i}}{ d_{i}^{2}} 
 \psi_{\sigma} (\theta , \varphi )\; ,
\label{eq:ofjir01}\\&&\int   [{\bf d\Omega}]\psi_{\sigma} (\theta , \varphi ) ^{*}\frac{1}{M_{i} }  \frac{{\bf a}_{i-1} \cdot {\bf a}_{i}}
{d_{i-1} d_{i}} 
 \psi_{\sigma} (\theta , \varphi )\; .
\label{eq:ofjir01bis}
 \end{eqnarray}
Next, the contributions  from $M_{i}^{-1}[ \hbar^{2} {\bf u}_{i-1} \cdot {\bf u}_{i} (\partial^{2}/\partial 
y_{i-1}\partial y_{i}) - i\hbar (({\bf u}_{i-1} \cdot{\bf a}_{i} )/y_{i})(\partial/\partial y_{i-1})-  
 i\hbar (({\bf a}_{i-1} \cdot{\bf u}_{i} )/y_{i-1})(\partial/
\partial y_{i})]$ follow  by integrating  by parts over $y_{i}$ and $y_{i-1}$, letting 
  $\omega_{l} \rightarrow +\infty$ and using Eqs.  (\ref{eq:dirac}) and  (\ref{eq:Del2}).  One finds: 
\begin{eqnarray}&&
\int   [{\bf d\Omega}]\psi_{\sigma} (\theta , \varphi ) ^{*} 
\frac{1}{M_{i} }\left\{ \frac{ \hbar^{2} {\bf u}_{i-1} \cdot {\bf u}_{i}}{ d_{i-1}d_{i}
 } - \frac{i\hbar {\bf u}_{i-1} \cdot{\bf a}_{i} }{ d_{i-1}d_{i} } 
 -    \frac{ i\hbar {\bf a}_{i-1} \cdot{\bf u}_{i} }{ d_{i-1}d_{i} }\right\}
 \psi_{\sigma} (\theta , \varphi )\; .
\label{eq:ofjir02}
 \end{eqnarray}
  We shall now  deal with     somewhat more complicated structures.  We shall make use of the following identity for an 
arbitrary radial function $\phi (y)$:
\begin{eqnarray} &&
 \phi (y)\frac{\partial\phi (y)}{\partial y} = \frac{1}{2}\frac{\partial\phi (y)^{2}}{\partial y} \; .\label{eq:iip1}  
\end{eqnarray}
We consider: 
 \begin{eqnarray}&&
\int  [\prod_{i=1}^{N-1}y_{i}^{2}dy_{i}][{\bf d\Omega}][\prod_{i=1}^{N-1}\phi_{l} (y_{l})\psi_{\sigma} (\theta , \varphi ) ]^{*} 
\frac{B_{i}}{2}(-
 2\hbar^{2}) \frac{1}{y_{i}} \frac{\partial}{\partial y_{i}}[\prod_{i=1}^{N-1}\phi_{l} (y_{l})\psi_{\sigma} (\theta , \varphi ) ]\; .
\label{eq:ofjir1}
 \end{eqnarray}
We apply Eq. (\ref{eq:iip1}), integrate by parts over $y_{i}$, notice that the contributions from  $y_{i}=0$ and from 
$y_{i}\rightarrow +\infty$ vanish, let  $\omega_{l} \rightarrow +\infty$ and use Eqs.  (\ref{eq:dirac}), (\ref{eq:Del2}) and 
(\ref{eq:normaliz}). The result is:
\begin{eqnarray}&&
 \frac{B_{i}\hbar^{2}}{2d_{i}^{2}} \; .
\label{eq:ofjir2}
 \end{eqnarray}
We  treat, also as $\omega_{l} \rightarrow +\infty$, the remaining contributions: 
 \begin{eqnarray}&&
\int  [\prod_{i=1}^{N-1}y_{i}^{2}dy_{i}][{\bf d\Omega}][\prod_{i=1}^{N-1}\phi_{l} (y_{l})\psi_{\sigma} ]^{*} 
[\sum_{i=1}^{N-1}\frac{(-B_{i})\hbar^{2}}{2}   \frac{\partial^{2}}{\partial y_{i}^{2}}+U({\bf y})]\prod_{i=1}^{N-1}\phi_{l} (y_{l})
\psi_{\sigma} ]\; .
\label{eq:ofjir21}
 \end{eqnarray}
There are more than one way to treat them. See \cite{AE}. Possibly, the simplest one is: 
\begin{eqnarray}&&
 [\frac{(-B_{i})\hbar^{2}}{2}     \frac{\partial^{2}}{\partial y_{i}^{2}}  + V_{M}(y_{i})  ]\phi_{M,i,n=0} (y_{i}) 
\rightarrow  E_{M,i, 0}\phi_{M,i,n=0} (y_{i}) \; .
\label{eq:ofjir3} 
\end{eqnarray}
   $E_{M,i,n=0}$ is given by
 ~(\ref{eq:levels}), with the corresponding $D_{i}$, $\alpha_{i}$, $\omega_{i}$ (see the comment justifying  
~(\ref{eq:levels})).  By using  Eqs. (\ref{eq:ofjir3}),  (\ref{eq:dirac}) and  (\ref{eq:Del2}) in Eqs. (\ref{eq:ofjir21}),  by  recalling  the results in 
Eqs. (\ref{eq:ofjir01}), (\ref{eq:ofjir01bis}), (\ref{eq:ofjir02}),  (\ref{eq:ofjir1}),  (\ref{eq:ofjir2})  and by defining the new variable 
${\bf e}_{i}\equiv i\hbar {\bf u}_{i} -{\bf a}_{i}$, we arrive at Eqs. (\ref{eq:hfrec}),  (\ref{eq:definham}) and  (\ref{eq:HQANG}) 
for the single-stranded open  freely-jointed  quantum chain. In so doing, use is made of: 
${\bf u}_{i}  \cdot {\bf a}_{i}=0$ and of ${\bf a}_{i} \cdot{\bf u}_{i} =2i\hbar$. 

Next, we shall turn to the   case in which $U({\bf y})$ is the sum of  harmonic-oscillator-like  potentials given in Eq. (\ref{eq:Uy}). 
As $\omega_{i} \rightarrow +\infty$, one has: 
\begin{eqnarray}&&
 [\frac{(-B_{i})\hbar^{2}}{2}     \frac{\partial^{2}}{\partial y_{i}^{2}}  +\frac{\omega_{i}^{2}}{2B_{i}}(y_{i}-d_{i})^{2}  ]
\phi_{i} (y_{i})\rightarrow \frac{\hbar\omega_{i}}{2}\phi_{i} (y_{i})\; .
\label{eq:ofjir31} 
\end{eqnarray}
 The results  in     Eqs. 
(\ref{eq:ofjir01}),  (\ref{eq:ofjir01bis},    and  (\ref{eq:ofjir2}) continue to hold. The main difference now 
 is that $\sum_{l=1}^{N-1}E_{M,l,n=0}$ is replaced by  $\sum_{i=1}^{N-1}\hbar \omega_{i}/2$. All that yields 
~(\ref{eq:hfrec})-~(\ref{eq:definham}). 

We  remark  that neither $i\hbar {\bf u}_{l}$  nor ${\bf a}_{l}$ are Hermitean operators but, on the contrary, 
 ${\bf e}_{l}= i\hbar {\bf u}_{l} 
-{\bf a}_{l}$ is.   Using Cartesian components, we write 
${\bf e}_{l}=  ( e_{l,1},e_{l,2},e_{l,3})$ and ${\bf u}_{l}=( u_{l,1},u_{l,2},u_{l,3})$. Then, one can derive 
the following ``angular commutation relations'': $[u_{l,\alpha}, e_{k,\beta}]=i\hbar\delta_{lk}(\delta_{\alpha\beta}-
u_{l,\alpha}u_{l,\beta})$ \cite{AEM}. It suggests that Cartesian components of ${\bf u}_{l}$ could be regarded as  conjugate 
 variables (quantum-machanically) of  ${\bf e}_{l}$, in some extended sense. Those  (purely algebraic) 
``angular commutation relations'' ( together with the closed algebra formed by (\ref{eq:comrel}) and the standard commutation relations for orbital angular 
momentum in subsection \ref{subsec:RFAE.GFC.4.1}) hold  independently of the  $\omega_{l} \rightarrow +\infty$ limit, and do not look particularly 
interesting 
   before taking the latter limit,  because radial variables also matter. 
However, after    $\omega_{l} \rightarrow +\infty$,  all those algebraic relations  constitute a distinguishing 
feature of  the remaining 
( unconstrained) angular variables. Whether or up to what extent they constitute strict  quantum equivalents of  
classical radial constraints constitute   open questions.

\chapter{}

\section*{Weak next-to-nearest-neighbours interaction: $G({\bf q}) $ and $\langle({\bf x}_{N} - {\bf x}_{1})^{2}\rangle$}
\label{sec:B}

  For small $q$, the dominant contribution to $G({\bf q})$, as given by the ratio in Eq.\ (\ref{eq:Gq}), can be evaluated 
approximately as follows.   Studies  of $ [\Delta_{N-1}]^{-1/2}$   for large $N$ have been carried out, with successive  improvements, 
in  ~\cite{AE} (Subsection 5.4), ~\cite{RaCa} (Appendix B) and ~\cite{CAEDNA} (Appendix C): in short,  $ [\Delta_{N-1}]^{-1/2}$  takes 
on its dominant contributions when all $({\bf u}_{i}\cdot {\bf u}_{j})^{2}=1$, $i,j=1,..N-1$, $i\neq j$. Then, 
 in each integral over $\theta_{s}$, $s=2,\ldots,N-1$ in both the numerator and denominator in  Eq.\ (\ref{eq:Gq}), we keep   only 
the contributions over two adequately small intervals of size
$\delta\theta_{s}$ about  $\theta_{s-1}$ and  $\pi-\theta_{s-1}$.  So, we   
 extend the  arguments in \cite{AE,RaCa,CAEDNA}, with two modifications: i) we   impose the classical limit restrictions, by including a factor $\rho_{Q}$ for  $\theta_{s+1}\simeq \pi-\theta_{s}$, ii) the contributions of the values taken by
$U_{nnn}$  when $\theta_{s}$ is close to either  $\theta_{s-1}$ or   $\pi-\theta_{s-1}$.  The only  
contributing values of $v$ are $v(2d)$ (for   ${\bf u}_{i}\cdot {\bf u}_{i+1}\simeq+1$), and $v(\lambda_{th}/(2^{1/2} d))$  
(for ${\bf u}_{i}\cdot {\bf u}_{i+1}\simeq -1+\lambda_{th}^{2}/2 d^{2}$).  Thus,   having  integrated over all  
$\theta_{s}$, $s=2,\ldots,N-1$ (by retaining    only two  adequately small    regions in those   integrations),  
one gets  the following  approximate  representation: 
\begin{eqnarray}
G({\bf q}) & \simeq &
\frac{\int_{0}^{\pi}\sin\theta_{1}d\theta_{1}z(\theta_{1}; q)_{N}}{\int_{0}^{\pi}\sin\theta_{1}d\theta_{1}
z(\theta_{1}; q=0)_{N} } \; . \label{eq:Gq2} 
 \end{eqnarray}
  $z(\theta_{1};
q)_{N}$ turns out to be a sum of $2^{N-2}$  terms:
\begin{eqnarray}
z(\theta_{1};q)_{N} = \exp\left[(N-2)(k_{B}T)^{-1}v(2d)\right]
\left[\prod_{s=2}^{N-1}\delta\theta_{s}\right]
\sum_{l}\alpha_{l}^{(N)}(\rho)\exp\left[iqdl\cos\theta_{1} \right] \; . \label{eq:zI-l2}
 \end{eqnarray} 
  $\rho$ was given in  subsection  \ref{subsec:RFAE.GFC.4.3}.  
  $\alpha_{l}^{(N)}(\rho)$($>0$) are  certain polynomials in $\rho$ characterized below [with 
$\alpha_{-(N-1)}^{(N)}(\rho)=0$]. The summation in  Eq.  (\ref{eq:zI-l2})
should be understood as follows: for $N$ even (odd), $l=2k+1$ ($l=2k$), the
integer $k$ ranging from $-(N/2)+1$ up to $(N/2)-1$ (from  $-(N-1)/2$ up to
$(N-1)/2$). The   $\alpha_{l}^{(N)}(\rho)$'s,   following   
  from the above integrations,  are obtained
successively  from  the
following recurrence: 
$\alpha_{l}^{(N+1)}(\rho)=\alpha_{l-1}^{(N)}(\rho)+\rho\alpha_{-(l-1)}^{(N)}(\rho)$,
with $\alpha_{N}^{(N+1)}(\rho)=\alpha_{N-1}^{(N)}(\rho)$, and
$\alpha_{-(N-2)}^{(N+1)}(\rho)=\rho\alpha_{N-1}^{(N)}(\rho)$. For instance,
$\alpha_{3}^{(4)}(\rho)=1$, $\alpha_{1}^{(4)}(\rho)=\rho +\rho^{2}$,
$\alpha_{-1}^{(4)}(\rho)=\rho$.  Eqs. \ (\ref{eq:Gq2}) and (\ref{eq:zI-l2})  yield
\begin{eqnarray}
G({\bf q}) & \simeq & \sum_{j}\beta_{j}^{(N)}(\rho)\frac{\sin \left[(N-1-2j)qd\right]}{(N-1-2j)qd}
\; ,
\label{eq:Gq3} \\ 
\beta_{j}^{(N)}(\rho) & = & \frac{\alpha_{N-1-2j}^{(N)}(\rho) +
  \alpha_{-(N-1-2j)}^{(N)}(\rho)}{(1+\rho)^{N-2}} \; .
\label{eq:beta}
\end{eqnarray}
The summation in  Eq.  (\ref{eq:Gq3}) should be understood as follows: for $N$
even (odd), the integer $j$ ranges from $0$ up to $(N-2)/2$
(from  $0$ up to $(N-1)/2$). The following properties hold: $
\sum_{j}\beta_{j}^{(N)}(\rho) = 1$, $\beta_{j}^{(N)}(1) = {N-1 \choose j}/2^{N-2}$ if
$N$ is even, and $\beta_{j}^{(N)}(1) = {N-1 \choose j}/2^{N-2}$ if
$N$ is odd, except $\beta_{(N-1)/2}^{(N)}(1) = {N-1 \choose
  j}/(2^{N-1})$.  Notice that, upon performing the
inverse Fourier transform of Eq.  (\ref{eq:Gq3}),  $W({\bf r})\simeq (4\pi
d\mid {\bf r}\mid)^{-1}\sum_{j}j^{-1}\beta_{j}^{(N)}(\rho)\delta(\mid {\bf
  r}\mid-jd)$ ($\delta$ denoting Dirac's delta function). This representation
for  $W({\bf r})$ strongly reminds (although  it does not coincide with)  the end-to-end
distribution appearing in the random-flight approach to polymers \cite{McQ}. 
  Eq. \ (\ref{eq:Gq3}) yields:  
\begin{eqnarray}&&
\langle({\bf x}_{N} - {\bf x}_{1})^{2}\rangle =
\frac{d^{2}\eta}{(1+\rho)^{N-2}}  \; ,\quad \eta  \equiv  \sum_{j=0}^{N-2}c_{j}^{(N)}\rho^{j}\;  .\label{eq:endtoend} 
 \end{eqnarray}
 Using the properties of the $\beta_{j}^{(N)}(\rho)$, one   finds
the following   expressions for the case $N=$ even (the case $N=$ odd being analogous): 
$c_{j}^{(N)} = [(N-1-j)(N-2-j)(N-1)!][(j+2)j!(N-1-j)!]^{-1}$ if $j$ is odd, while 
$ c_{j}^{(N)} =[(N-1-j)^{2}(N-1)!][(j+1)j!(N-1-j)!]^{-1}$  if $j$ is  even. The computation of $\eta$, leading from 
(\ref{eq:endtoend})
 to   (\ref{eq:endtoendF}), is outlined in appendix C. 

We claim that the approximations leading from Eq. \ (\ref{eq:Gq}) to Eq. \ (\ref{eq:Gq3})  hold  for small $q$, say, up to and including order 
$q^{2}$ (and, so, they are  valid to compute $\langle({\bf x}_{N} - {\bf x}_{1})^{2}\rangle$). This statement is supported by the consistency 
obtained with the standard Gaussian model up to order $q^{2}$ when $U_{nnn}=0$ \cite{McQ}. 

\chapter{}

\section*{Additional computations for subsection ~\ref{subsec:RFAE.GFC.4.3} and Appendix B}
 \label{sec:C} 

Without loss of generality, we limit ourselves to the case $N$ even. Then, some direct algebra yields
\begin{eqnarray}
\eta&=&(N-1)^{2}[1+T_{1}]+(N-1)NT_{2}+T_{3}-(N-1)T_{4}\nonumber\\
&-&2^{-1}[2(N-1)+1]T_{5} \; .
\label{eq:C1}
\end{eqnarray}
The quantities $T_{i}$, $i=1,\ldots,5$ are certain sums involving ratios of factorials (combinatorial coefficients). We shall 
give below their initial and final expressions.  
\begin{eqnarray}
T_{1}&=&\sum_{i=1}^{(N-2)/2}\frac{1}{2i+1}\frac{(N-1)!}{(N-1-2i)!(2i)!}\rho^{2i}=\frac{(1+\rho)^{N}-(1-\rho)^{N}}{2N\rho}-1
\; ,
\label{eq:C2}\\
T_{2}&=&\sum_{i=1}^{(N-2)/2}\frac{1}{2i+1}\frac{(N-1)!}{(N-2i)!(2i-1)!}\rho^{2i-1}\nonumber \\ 
&= &  \frac{(1+\rho)^{N}+(1-\rho)^{N}}{2N\rho}- \frac{\rho^{N-1}}{N+1}
-  \frac{(1+\rho)^{N+1}-(1-\rho)^{N+1}}{2N(N+1)\rho^{2}} \; ,
\label{eq:C3}\\
T_{3}  &=& \sum_{i=1}^{(N-2)/2}\frac{4i^{2}}{2i+1}\left[\frac{(N-1)!}{(N-1-2i)!(2i)!}\rho^{2i}+\frac{(N-1)!}{(N-2i)!(2i-1)!}\rho^{2i-1}\right]
\nonumber \\ 
&= & T_{1}+T_{2}+(N-1)\rho\left[(1+\rho)^{N-2}-\rho^{N-2}\right] \nonumber\\
&-&\frac{(1+\rho)^{N-1}+(1-\rho)^{N-1}}{2}+1\; ,
\label{eq:C4}\\
T_{4}&=&\sum_{i=1}^{(N-2)/2}\frac{4i}{2i+1}\frac{(N-1)!\rho^{2i-1}}{(N-2i)!(2i-1)!} \nonumber\\
&=&[(1+\rho)^{N-1}+(1-\rho)^{N-1}]-2-2T_{1}\; ,
\label{eq:C5}\\
T_{5}&=&\sum_{i=1}^{(N-2)/2}\frac{4i}{2i+1}\frac{(N-1)!\rho^{2i}}{(N-1-2i)!(2i)!} \nonumber\\
&=&[(1+\rho)^{N-1}-(1-\rho)^{N-1}]-2\rho^{N-1}-2T_{2}\; .
\label{eq:C6}
\end{eqnarray}
When performing those sums, use has been made of the following formulas
\begin{eqnarray}&&
\sum_{k=0}^{m}\frac{1}{k+1}\frac{m!}{(m-k)!k!}\rho^{k} = 
\frac{(1+\rho)^{m+1}-1}{(m+1)\rho} \; ,
\label{eq:C7}\\&&
\sum_{k=0}^{m}\frac{(-1)^{k}}{k+1}\frac{m!}{(m-k)!k!}\rho^{k}  = 
\frac{1-(1-\rho)^{m+1}}{(m+1)\rho} \; ,
\label{eq:C8} 
\end{eqnarray}
and
\begin{eqnarray}&&
\sum_{i=1}^{(N-2)/2}\frac{1}{2i(2i+1)}\frac{(N-1)!}{(N-2i)!(2i-1)!}\rho^{2i-1}
= \nonumber \\&&
= \frac{1}{(N+1)N\rho^{2}}\left[\frac{(1+\rho)^{N+1}-(1-\rho)^{N+1}}{2} -
  (N+1)\rho-\rho^{N+1}\right] \; ,
\label{eq:C9}\\&&
\sum_{i=1}^{(N-2)/2}\left[
  2i\frac{(N-1)!}{(N-1-2i)!(2i)!}\rho^{2i}+(2i-1)\frac{(N-1)!}{(N-2i)!(2i-1)!}\rho^{2i-1}\right]
= \nonumber \\&&
= (N-1)\rho\left[(1+\rho)^{N-2}-\rho^{N-2}\right] \; .
\label{eq:C10}
\end{eqnarray}

Upon combining Eqs.\ (\ref{eq:C2}) through  Eq.\ (\ref{eq:C6}) and further
substitution in Eq. \ (\ref{eq:C1}), one finally arrives at Eq.\ (\ref{eq:endtoendF}).  

\chapter{}

\section*{Single-stranded  closed-ring  freely-jointed   chain: proof  of ~(\ref{eq:Hamiltoniancl}) }
\label{sec:D} 
 
In this appendix, 
we report some details and the results of the evaluation of $\langle \tilde{H}_{Q} \rangle = 
\langle H_{Q,in} \rangle + \langle U({\bf y}) \rangle$. Use will be made of harmonic oscillators (Eq. \ (\ref{eq:Uy})), of the unit vector 
${\bf u}_{N}$ defined through: ${\bf y}_{N} =-\sum_{i=1}^{N-1} {\bf y}_{i}= y_{N}{\bf u}_{N}  $ and of the methods in Appendix A.  
 The results given below  will come from   Eqs.\ (\ref{eq:T1}) and  \ (\ref{eq:Uy}). Using 
 Successive integrations by parts yield (as all frequencies $\omega_{i}$, 
$i=1,\ldots,N$, grow very large):
\begin{eqnarray}&&
\bigg\langle  - \frac{1}{2}\sum_{i=1}^{N-1} B_{i} \hbar^{2} \frac{\partial^{2}}{\partial y_{i}^{2}} \bigg\rangle = \sum_{i=1}^{N-1} 
\frac{\hbar \omega_{i}}{4}  - \sum_{i=1}^{N-1} \frac{\hbar^{2}B_{i}}{2d_{i}^{2}}  \nonumber \\&& 
 + \sum_{i=1}^{N-1} \frac{\hbar \omega_{N} B_{i}}{4 B_{N}} \left\{ \int [{\bf d\Omega}] |\psi_{\sigma} (\theta , \varphi )|^{2} 
({\bf u}_{i}\cdot{\bf u}_{N})^{2}\frac{\delta (y_{N} - d_{N})}{d_{N}^{2}} \right\} \nonumber \\
&&-  \sum_{i=1}^{N-1} \frac{\hbar^{2}B_{i}}{2d_{i}}  \left[ \frac{\partial}{\partial y_{i}} \left\{ \int [{\bf d\Omega}] 
|\psi_{\sigma} (\theta , \varphi )|^{2}  \frac{\delta (y_{N} (\theta , \varphi; y_{i})  - d_{N})}{d_{N}^{2}} \right\} \right]_{y_{i} = 
d_{i}} \nonumber \\ 
&& -  \sum_{i=1}^{N-1} \frac{\hbar^{2}B_{i}}{4} \left[ \frac{\partial^{2}}{\partial y_{i}^{2}} \left\{ \int [{\bf d\Omega}] |
\psi_{\sigma} (\theta , \varphi )|^{2}  \frac{\delta (y_{N} (\theta , \varphi; y_{i})  - d_{N})}{d_{N}^{2}} \right\} \right]_{y_{i} = 
d_{i}}\; ,  \label{eq:D1} 
\end{eqnarray}
\begin{eqnarray}&&
\bigg\langle  - \sum_{i=1}^{N-1} B_{i}\hbar^{2} \frac{1}{y_{i}} \frac{\partial}{\partial y_{i}} \bigg\rangle = \sum_{i=1}^{N-1} 
\frac{\hbar^{2}B_{i}}{2d_{i}^{2}}\; ,  \label{eq:D2} 
\end{eqnarray}
\begin{eqnarray}
\bigg\langle   \frac{1}{2} \sum_{i=1}^{N-1} B_{i} \frac{{\bf a}_{i} \cdot{\bf a}_{i}}{ y_{i}^{2}} \bigg\rangle &=& \nonumber \\
& +& \sum_{i=1}^{N-1} \frac{B_{i}}{2d_{i}^{2}} \left\{ \int [{\bf d\Omega}] \psi_{\sigma}^{*} (\theta , \varphi ) \left[({\bf a}_{i} 
\cdot {\bf a}_{i})\psi_{\sigma} (\theta , \varphi )\right] \frac{\delta (y_{N} - d_{N})}{d_{N}^{2}} \right\} \nonumber \\
&+&  \sum_{i=1}^{N-1} \frac{B_{i}}{2d_{i}^{2}} \left\{ \int [{\bf d\Omega}] \psi_{\sigma}^{*} (\theta , \varphi ) \left[ {\bf a}_{i}
\psi_{\sigma} (\theta , \varphi )\right] \cdot \left[{\bf a}_{i} \frac{\delta (y_{N} - d_{N})}{d_{N}^{2}} \right] \right\} \nonumber \\
&+&  \sum_{i=1}^{N-1} \frac{\hbar \omega_{N} B_{i}}{4 B_{N}} \left\{ \int [{\bf d\Omega }] |\psi_{\sigma} (\theta , \varphi )|^{2} 
({\bf u}_{\theta_{i}}\cdot {\bf u}_{N})^{2} \frac{\delta (y_{N} - d_{N})}{d_{N}^{2}} \right\} \nonumber \\
&+&  \sum_{i=1}^{N-1} \frac{\hbar \omega_{N} B_{i}}{4 B_{N}} \left\{ \int [{\bf d\Omega}] |\psi_{\sigma} (\theta , \varphi )|^{2} 
({\bf u}_{\varphi_{i}}\cdot{\bf u}_{N})^{2} \frac{\delta (y_{N} - d_{N})}{d_{N}^{2}} \right\} \nonumber \\
&+&  \sum_{i=1}^{N-1} \frac{B_{i}}{4d_{i}^{2}} \left\{ \int [{\bf d\Omega}] |\psi_{\sigma} (\theta , \varphi )|^{2} \left[ 
({\bf a}_{i} \cdot{\bf a}_{i}) \frac{\delta (y_{N} - d_{N})}{d_{N}^{2}} \right] \right\}\; ,  \label{eq:D3}
\end{eqnarray}
\begin{eqnarray}
&&\bigg\langle \sum_{i=2}^{N-1}\frac{\hbar^{2}}{M_{i}}\left\{ {\bf u}_{i-1} \cdot {\bf u}_{i} \frac{\partial^{2}}{\partial
 y_{i-1}\partial y_{i}} \right\} \bigg\rangle = \nonumber \\
&& - \sum_{i=2}^{N-1}\frac{\hbar \omega_{N}}{2M_{i}B_{N}}\left\{ \int [{\bf d\Omega}] |\psi_{\sigma} (\theta , \varphi )|^{2} 
({\bf u}_{i-1} \cdot{\bf u}_{i})({\bf u}_{i-1} \cdot{\bf u}_{N})({\bf u}_{i} \cdot{\bf u}_{N}) \frac{\delta (y_{N} - 
d_{N})}{d_{N}^{2}} \right\} \nonumber \\
&& +  \sum_{i=2}^{N-1}\frac{\hbar^{2}}{4M_{i}}\left[ \frac{\partial^{2}}{\partial y_{i-1}\partial y_{i}} \left\{ \int [{\bf d\Omega}] 
|\psi_{\sigma} (\theta , \varphi )|^{2} ({\bf u}_{i-1} \cdot{\bf u}_{i})\frac{\delta (
y_{N} (\theta , \varphi; y_{i-1}, y_{i})
 - d_{N})}{d_{N}^{2}}\right\} \right]_{y_{i-1} = d_{i-1}}^{y_{i} = d_{i}} 
\nonumber \\
 &&+  \sum_{i=2}^{N-1}\frac{\hbar^{2}}{M_{i}d_{i}d_{i-1}}
\left\{ \int [{\bf d\Omega}] |\psi_{\sigma} (\theta , \varphi )|^{2} ({\bf u}_{i-1} \cdot{\bf u}_{i})\frac{\delta (y_{N} - 
d_{N})}{d_{N}^{2}} \right\}\; , \label{eq:D4} 
\end{eqnarray}
\begin{eqnarray}
&&\bigg\langle -  \sum_{i=2}^{N-1}\frac{1}{M_{i}} \frac{{\bf a}_{i-1} \cdot{\bf a}_{i}}{y_{i-1} y_{i}} \bigg\rangle = \nonumber \\
&&- \sum_{i=2}^{N-1}\frac{\hbar \omega_{N}}{2M_{i}B_{N}}  \left\{ \int [{\bf d\Omega}] |\psi_{\sigma} (\theta , \varphi )|^{2} 
({\bf u}_{\theta_{i-1}} \cdot{\bf u}_{N})({\bf  u}_{\theta_{i-1}} \cdot{\bf u}_{\theta_{i}})({\bf u}_{\theta_{i}} \cdot
{\bf u}_{N})
\frac{\delta (y_{N} - d_{N})}{d_{N}^{2}} \right\} \nonumber \\ 
&& -  \sum_{i=2}^{N-1}\frac{\hbar \omega_{N}}{2M_{i}B_{N}} \left\{ \int [{\bf d\Omega}] |\psi_{\sigma} (\theta , \varphi )|^{2} 
({\bf u}_{\theta_{i-1}} \cdot{\bf  u}_{N})({\bf  u}_{\theta_{i-1}} \cdot{\bf  u}_{\varphi_{i}}) ({\bf u}_{\varphi_{i}} \cdot
{\bf  u}_{N})
\frac{\delta (y_{N} - d_{N})}{d_{N}^{2}} \right\} \nonumber \\ 
&& -  \sum_{i=2}^{N-1}\frac{\hbar \omega_{N}}{2M_{i}B_{N}}  \left\{ \int [{\bf d\Omega}] |\psi_{\sigma} (\theta , \varphi )|^{2} 
({\bf u}_{\varphi_{i-1}} \cdot{\bf  u}_{N})({\bf u}_{\varphi_{i-1}}\cdot{\bf  u}_{\theta_{i}})({\bf u}_{\theta_{i}} \cdot{\bf  u}
_{N})
\frac{\delta (y_{N} - d_{N})}{d_{N}^{2}} \right\} \nonumber \\ 
&&- \sum_{i=2}^{N-1}\frac{\hbar \omega_{N}}{2M_{i}B_{N}}  \left\{ \int [{\bf d\Omega}] |\psi_{\sigma} (\theta , \varphi )|^{2} 
({\bf u}_{\varphi_{i-1}} \cdot{\bf  u}_{N})({\bf u}_{\varphi_{i-1}}\cdot{\bf  u}_{\varphi_{i}}) ({\bf u}_{\varphi_{i}} \cdot{\bf  u}
_{N})
\frac{\delta (y_{N} - d_{N})}{d_{N}^{2}} \right\} \nonumber \\ 
&& - \sum_{i=2}^{N-1}\frac{1}{M_{i}d_{i}d_{i-1}} \left\{ \int [{\bf d\Omega}] \psi_{\sigma}^{*} (\theta , \varphi ) \left[ 
({\bf a}_{i-1}\cdot{\bf  a}_{i})\psi_{\sigma} (\theta , \varphi ) \right] \frac{\delta (y_{N} - d_{N})}{d_{N}^{2}} \right\} 
\nonumber \\
&& -  \sum_{i=2}^{N-1}\frac{1}{2M_{i}d_{i}d_{i-1}} \left\{ \int [{\bf d\Omega}] |\psi_{\sigma} (\theta , \varphi )|^{2} 
\left[ ({\bf a}_{i-1}\cdot{\bf  a}_{i}) \frac{\delta (y_{N} - d_{N})}{d_{N}^{2}} \right] \right\} \nonumber  \\
&&-  \sum_{i=2}^{N-1}\frac{1}{2M_{i}d_{i}d_{i-1}} \left\{ \int [{\bf d\Omega}] \psi_{\sigma}^{*} (\theta , \varphi ) 
\left[ {\bf a}_{i-1}\psi_{\sigma} (\theta , \varphi ) \right] \cdot \left[ {\bf a}_{i}\frac{\delta (y_{N} - d_{N})}{d_{N}^{2}} \right]
 \right\} \nonumber \\
&& -  \sum_{i=2}^{N-1}\frac{1}{2M_{i}d_{i}d_{i-1}} \left\{ \int [{\bf d\Omega}] \psi_{\sigma}^{*} (\theta , \varphi ) 
\left[ {\bf a}_{i}\psi_{\sigma} (\theta , \varphi ) \right] \cdot \left[ {\bf a}_{i-1}\frac{\delta (y_{N} - d_{N})}{d_{N}^{2}} 
\right] \right\}\; ,  \label{eq:D5} 
\end{eqnarray}
\begin{eqnarray}
&&\bigg\langle  -  \sum_{i=2}^{N-1}\frac{i\hbar}{M_{i}}\left\{ {\bf u}_{i-1} \cdot {\bf a}_{i} \frac{1}{y_{i}} 
\frac{\partial}{\partial y_{i-1}} + {\bf a}_{i-1} \cdot {\bf u}_{i} \frac{1}{y_{i-1}} \frac{\partial}{\partial y_{i}}\right\} 
\bigg\rangle = \nonumber \\
&& -  \sum_{i=2}^{N-1}\frac{\hbar \omega_{N}}{2M_{i}B_{N}}  \left\{ \int [{\bf d\Omega}] |\psi_{\sigma} (\theta , \varphi )|^{2} 
({\bf u}_{i-1} \cdot{\bf u}_{N})({\bf u}_{i-1} \cdot{\bf u}_{\theta_{i}})({\bf u}_{\theta_{i}} \cdot{\bf u}_{N})\frac{\delta (y_{N} - 
d_{N})}{d_{N}^{2}} \right\} \nonumber \\
&&- \sum_{i=2}^{N-1}\frac{\hbar \omega_{N}}{2M_{i}B_{N}} \left\{ \int [{\bf d\Omega}] |\psi_{\sigma} (\theta , \varphi )|^{2} ({\bf u}_{i-1} \cdot u_{N})({\bf u}_{i-1} 
\cdot{\bf u}_{\varphi_{i}})({\bf u}_{\varphi_{i}} \cdot{\bf u}_{N})\frac{\delta (y_{N} - d_{N})}{d_{N}^{2}} \right\} \nonumber \\
&& -  \sum_{i=2}^{N-1}\frac{\hbar \omega_{N}}{2M_{i}B_{N}}  \left\{ \int [{\bf d\Omega}] |\psi_{\sigma} (\theta , \varphi )|^{2} 
({\bf u}_{i} \cdot{\bf  u}_{N})({\bf u}_{i} \cdot{\bf  u}_{\theta_{i-1}})({\bf u}_{\theta_{i-1}} \cdot{\bf  u}_{N}) 
\frac{\delta (y_{N} - d_{N})}{d_{N}^{2}} \right\} \nonumber \\
&& -\sum_{i=2}^{N1}\frac{\hbar \omega_{N}}{2M_{i}B_{N}} \left\{ \int [{\bf d\Omega}] |\psi_{\sigma} (\theta , \varphi )|^{2} ({\bf u}_{i} \cdot u_{N})({\bf u}_{i} 
\cdot{\bf u}_{\varphi_{i-1}})({\bf u}_{\varphi_{i-1}} \cdot{\bf u}_{N}) \frac{\delta (y_{N} - d_{N})}{d_{N}^{2}} \right\} \nonumber \\
&&-  \sum_{i=2}^{N-1}\frac{i\hbar}{4M_{i}}  
\left[\frac{\partial}{\partial y_{i-1}} 
\left\{ \int [{\bf d\Omega}] |\psi_{\sigma} (\theta , \varphi )|^{2}  \frac{({\bf u}_{i-1}\cdot{\bf a}_{i})}{d_{i}} 
\frac{\delta (y_{N}(\theta, \varphi; y_{i-1}) - d_{N})}{d_{N}^{2}} \right\} \right]_{y_{i-1} = d_{i-1}} 
\nonumber \\
&& -  \sum_{i=2}^{N-1}\frac{i\hbar}{4M_{i}}  
\left[\frac{\partial}{\partial y_{i}} 
\left\{ \int [{\bf d\Omega}] |\psi_{\sigma} (\theta , \varphi )|^{2}  \frac{({\bf u}_{i}\cdot{\bf a}_{i-1})}{d_{i-1}} 
\frac{\delta (y_{N}(\theta, \varphi; y_{i}) - d_{N})}{d_{N}^{2}} \right\} \right]_{y_{i} = d_{i}} 
\nonumber \\
&& +  \sum_{i=2}^{N-1}\frac{i\hbar}{M_{i}d_{i}d_{i-1}} \left\{ \int [{\bf d\Omega}] \psi_{\sigma}^{*} (\theta , \varphi ) 
({\bf u}_{i-1}\cdot{\bf a}_{i}\psi_{\sigma} (\theta , \varphi )) \frac{\delta (y_{N} - d_{N})}{d_{N}^{2}} \right\} \nonumber \\ 
&& + \sum_{i=2}^{N-1}\frac{i\hbar}{M_{i}d_{i}d_{i-1}} \left\{ \int [{\bf d\Omega}] \psi_{\sigma}^{*} (\theta , \varphi )
({\bf u}_{i}\cdot{\bf a}_{i-1}\psi_{\sigma} (\theta , \varphi )) \frac{\delta (y_{N} - d_{N})}{d_{N}^{2}} \right\}\nonumber \\ 
&&+  \sum_{i=2}^{N-1}\frac{i\hbar}{2M_{i}}  \left\{ \int [{\bf d\Omega}] |\psi_{\sigma} (\theta , \varphi )|^{2} \left[ \left[ 
\frac{({\bf u}_{i-1}\cdot{\bf a}_{i})}{d_{i}d_{i-1}} + \frac{({\bf u}_{i}\cdot{\bf a}_{i-1})}{d_{i}d_{i-1}} \right] 
\frac{\delta (y_{N} 
- d_{N})}{d_{N}^{2}} \right] \right\}\; ,   \label{eq:D6} 
\end{eqnarray}
\begin{eqnarray}
&&\bigg \langle \sum_{i=1}^{N}\frac{\omega_{i}^{2}}{2B_{i}}(y_{i}-d_{i})^{2} \bigg \rangle =   
\sum_{i=1}^{N}\frac{\hbar \omega_{i}}{4}\; . \label{eq:D7}
\end{eqnarray}
Besides $y_{N}$ (the meaning of which is given in Eq. (\ref{eq:constraint3})), the above equations contain two new functions, namely, $y_{N} (\theta , \varphi; y_{i})$ and $y_{N} (\theta , \varphi; y_{i-1}, y_{i})$, which depend on the variables displayed respectively. In turn, those  two functions are defined as follows: 
 
a) $y_{N} (\theta , \varphi; y_{i} )  \equiv \left[ \sum_{l,j=1}^{N-1} {\bf y}_{l}\cdot{\bf y}_{j}\right]^{1/2}$, for  all $y_{h}=d_{h}$, $h=1,.., i-1,i+1,..,N-1$, that is,  $h=i$ is excluded, 

b) $y_{N} (\theta , \varphi; y_{i-1}, y_{i} )\equiv \left[ \sum_{l,j=1}^{N-1} {\bf y}_{l}\cdot{\bf y}_{j}\right]^{1/2}$, for 
 all $y_{h}=d_{h}$, $h=1,.., i-2,i+1,..,N-1$, that is, both  $h=i-1$ and  $h=i$ are excluded. 

The meaning of, say, $\left[ {\bf a}_{i}\psi_{\sigma}(\theta , \varphi )\right] \left[ {\bf a}_{i} \delta (y_{N} - d_{N})\right]$ in  
Eq.\ (\ref{eq:D3}) is the following. First,  the differential operator ${\bf a}_{i}$ acts upon  $\psi_{\sigma}$  by  regarding  
all angular variables $\theta_{l}$, $\varphi_{l}$, $l=1,..,N-1$, as if they were independent on one another (that is, as if  the 
constraint $\delta (y_{N} - d_{N})$ were not operative). On the other hand, the interpretation of ${\bf a}_{i} \delta (y_{N} - d_{N})$ 
is similar to  that of the first derivative of $\delta(f(x))$ with respect to $x$ ($f(x)$ being a given function of $x$): 
$d\delta(f(x))/dx= (df(x)/dx)d\delta(f)/df$, $d\delta(f)/df$ being the   first derivative of $\delta(f)$ \cite{MesI,GalPas}. 
Then, ${\bf a}_{i} \delta (y_{N} - d_{N})$ also   embodies  the essentials of the  closed-ring constraint.  After the operator  ${\bf a}_{i}$ has acted upon $\psi_{\sigma}$ with the  understanding explained above, then   
${\bf a}_{i} \delta (y_{N} - d_{N})$ acts and implies that there is one relationship among those $2(N-1)$ angles. A similar 
interpretation applies for other therms containing differential operators with respect to angles acting upon $ \delta (y_{N} - d_{N})$ 
and upon $\psi_{\sigma}$.   

Eqs. \ (\ref{eq:D1}) and \ (\ref{eq:D7}) display the simplifying feature  that all frequencies $\omega_{i}$, $i=1,\ldots ,N-1$ go 
multiplied by constant (angle-independent) factors. However, in Eqs. \ (\ref{eq:D1}), \ (\ref{eq:D3})-\ (\ref{eq:D6}) one sees terms linear in  $\omega_{N}$, which  go multiplied by complicated,  in principle angle-dependent, expressions. 
We shall prove that there are two crucial  cancellations, when we add all those matrix elements which  contain
 the frequency  
$\omega_{N}$ times various functions (displaying angular dependences). 
The first cancellation is obtained if we add the corresponding terms of Eq. \ (\ref{eq:D1}) and Eq. \ (\ref{eq:D3}),  which contain  $\omega_{N}$ times  
the following angle-dependent factors:
\begin{eqnarray}&&
 \,  \sum_{i=1}^{N-1} \frac{\hbar \omega_{N} B_{i}}{4 B_{N}} \nonumber \\&&
\times  \left\{ \int [{\bf d\Omega}]  |\psi_{\sigma} (\theta , \varphi )|^{2} \left[ ({\bf u}_{i}\cdot{\bf u}_{N})^{2} + 
({\bf u}_{\theta_{i}}\cdot{\bf u}_{N})^{2} + ({\bf u}_{\varphi_{i}}\cdot{\bf u}_{N})^{2} \right] \frac{\delta (y_{N} - 
d_{N})}{d_{N}^{2}} 
 \right\} \nonumber \\&&
 =  \sum_{i=1}^{N-1} \frac{\hbar \omega_{N} B_{i}}{4 B_{N}}\left\{ \int [{\bf d\Omega}]  |\psi_{\sigma} (\theta , \varphi )|^{2} 
\frac{\delta (y_{N} - d_{N})}{d_{N}^{2}}  \right\} 
 = \sum_{i=1}^{N-1} \frac{\hbar \omega_{N} B_{i}}{4 B_{N}}\; . \label{eq:D8}
\end{eqnarray}
as ${\bf u}_{N}^{2}=1$. The second (and rather non-trivial!) cancellation arises if we add the corresponding terms of Eq. \ (\ref{eq:D4}), Eq. \ (\ref{eq:D5}) 
 and Eq. \ (\ref{eq:D6}), 
which  contain   $\omega_{N}$ times the following functions containing angular dependences:
\begin{eqnarray}&&
 -  \sum_{i=2}^{N-1}\frac{\hbar \omega_{N}}{2M_{i}B_{N}}\left\{ \int [{\bf d\Omega}] |\psi_{\sigma} (\theta , \varphi )|^{2} 
\left[ ({\bf u}_{i-1} \cdot{\bf u}_{i})({\bf u}_{i-1} \cdot{\bf u}_{N})({\bf u}_{i} \cdot{\bf u}_{N}) 
\right. \right. \nonumber \\&&
 +  ({\bf u}_{\theta_{i-1}} \cdot{\bf u}_{N})({\bf  u}_{\theta_{i-1}} \cdot{\bf u}_{\theta_{i}})({\bf u}_{\theta_{i}} 
\cdot{\bf u}_{N}) + 
({\bf u}_{\theta_{i-1}} \cdot{\bf u}_{N})({\bf  u}_{\theta_{i-1}} \cdot{\bf u}_{\varphi_{i}} ) ({\bf u}_{\varphi_{i}} \cdot{\bf u}
_{N}) 
\nonumber \\&& 
+  ({\bf u}_{\varphi_{i-1}} \cdot{\bf u}_{N})({\bf u}_{\varphi_{i-1}}\cdot{\bf u}_{\theta_{i}})({\bf u}_{\theta_{i}} \cdot
{\bf u}_{N}) + 
({\bf u}_{\varphi_{i-1}} \cdot{\bf u}_{N})({\bf u}_{\varphi_{i-1}}\cdot{\bf u}_{\varphi_{i}}) ({\bf u}_{\varphi_{i}} \cdot
{\bf u}_{N}) 
\nonumber \\ &&
 +  ({\bf u}_{i-1} \cdot{\bf u}_{N})({\bf u}_{i-1} \cdot{\bf u}_{\theta_{i}})({\bf u}_{\theta_{i}} \cdot{\bf u}_{N}) + 
({\bf u}_{i-1} \cdot{\bf u}_{N})({\bf u}_{i-1} \cdot{\bf u}_{\varphi_{i}})({\bf u}_{\varphi_{i}} \cdot{\bf u}_{N}) \nonumber \\ &&
 +  \left. \left. ({\bf u}_{i} \cdot{\bf u}_{N})({\bf u}_{i} \cdot{\bf u}_{\theta_{i-1}})({\bf u}_{\theta_{i-1}} \cdot{\bf u}_{N}) + 
({\bf u}_{i} \cdot{\bf u}_{N})({\bf u}_{i} \cdot{\bf u}_{\varphi_{i-1}})({\bf u}_{\varphi_{i-1}} \cdot{\bf u}_{N}) \right] 
\frac{\delta (y_{N} - d_{N})}{d_{N}^{2}} \right\} \nonumber \\ &&
 =  - \sum_{i=2}^{N-1}\frac{\hbar \omega_{N}}{2M_{i}B_{N}}\left\{ \int [{\bf d\Omega}] |\psi_{\sigma} (\theta , \varphi )|^{2} 
\frac{\delta (y_{N} - d_{N})}{d_{N}^{2}} \right\}   
 =  - \sum_{i=2}^{N-1}\frac{\hbar \omega_{N}}{2M_{i}B_{N}}\; .  \label{eq:D9}
\end{eqnarray}
Both cancellations in  Eqs. \ (\ref{eq:D8}) and \ (\ref{eq:D9}) are due to the following properties: i) the   vectors ${\bf u}_{i}, 
{\bf u}_{\varphi_{i}}, 
{\bf u}_{\theta_{i}}$ are orthonormalized (as are  ${\bf u}_{i-1}, {\bf u}_{\varphi_{i-1}}, 
{\bf u}_{\theta_{i-1}}$  ), ii) ${\bf u}_{N}^{2}=1$. 

Notice that if we add the first term of the right hand side of Eq. \ (\ref{eq:D1})  plus those in the right hand sides of 
Eq. \ (\ref{eq:D7}), Eq. \ (\ref{eq:D8}) and Eq. \ (\ref{eq:D9}), we find: 
\begin{eqnarray}
&&  \sum_{i=1}^{N-1}\frac{\hbar \omega_{i}}{4} + \sum_{i=1}^{N}\frac{\hbar \omega_{i}}{4} + 
\sum_{i=1}^{N-1} \frac{\hbar \omega_{N} B_{i}}{4 B_{N}} - \sum_{i=2}^{N-1}\frac{\hbar \omega_{N}}{2M_{i}B_{N}} 
 =  \sum_{i=1}^{N}\frac{\hbar \omega_{i}}{2}\; ,\label{eq:D10}
\end{eqnarray}
which is the  total vibrational zero point energy  of the closed chain in its ground state. Use has been made of the expressions for 
$ B_{i}$, $i=1...N$.  
Thus, all the above 
cancellations have enabled to transform  the initial very  lengthy expressions (containing angle-dependent   expressions, multiplying the 
frequencies) 
 into somewhat shorter ones, in which all coefficients multiplying the  frequencies are constant. The resulting expressions 
are still somewhat lengthy. The final result is  collected 
in appendix A in \cite{Calvo}. Eq.  (\ref{eq:Hamiltoniancl}) is nothing but a  simplified way of presenting the final expression, 
which displays explicitly only 
the most relevant  terms: $\bigg \langle  {\cal O}_{\text{ang}}^{(\text{C})}(\hbar) \bigg \rangle $ in (\ref{eq:Hamiltoniancl}) 
denotes 
the remainder, which follows directly from   appendix A in \cite{Calvo}.

If Morse potentials are employed, instead of harmonic-oscillator-like ones, the  computations and cancellations in this appendix 
continue to hold, with $\hbar \omega_{i}/2$ replaced by $E_{M,i,n=0}$, $i=1...N$.


\begin{thebibliography}{}
  \bibitem{SSTtiEche} P. Echenique, C. N. Cavasotto, C. N. and P. Garc\'{\i}a-Risue\~{n}o, 
  Eur. Phys. J. Special Topics (2011). 

\bibitem{Leh} A. L. Lehninger, D. L. Nelson   and M. M. Cox,   \textit{Principles of Biochemistry} 2nd ed. (Worth Publishers, New York 1993).

\bibitem{Volk} M. V. Volkenshtein,    \textit{  Biophysics}  (Mir Publishers, Moskow 1983).

   
\bibitem{Flo} P. J.   Flory,    \textit{ Statistical Mechanics of Chain Molecules}, 2nd ed. (Wiley-Interscience, 
New York 1975).  

\bibitem{Gros} A. Y. Grossberg   and A. R. Khokhlov, \textit{Statistical Physics of Macromolecules}  
  (AIP Press, AIP Series in Polymers and Complex Materials, New York 1994).
 
\bibitem{McQ} D. A. McQuarrie,   \textit{ Statistical Thermodynamics} (Harper and Row, New York 1964).

 \bibitem{Doi} M.  Doi    and S. F.  Edwards,    \textit{ The Theory of Polymer Dynamics} (Oxford University Press, Oxford 1986).
\bibitem{Freed} K. F.  Freed,   \textit{ Renormalization Group Theory of Macromolecules} (John Wiley and Sons, New York 1987).
\bibitem{Proh} E.  Prohofsky,     \textit{ Statistical Mechanics and Stability of Macromolecules}  (Cambridge University Press, 
  Cambridge  1995). 
\bibitem{Frank} M. D. Frank-Kamenetskii,     Phys. Reports  \textbf{288}, (1997)   13.
\bibitem{Elias} H.-G. Elias,   \textit{ An introduction to Polymer Science} (VCH, John Wiley and Sons, New York 1997).
\bibitem{Gotoh} O. Gotoh,     Adv. Biophys. \textbf{ 16}, (1983) 1.

\bibitem{deGen} P. G. de Gennes,   \textit{  Scaling Concepts in  Polymer Physics} (Cornell University Press, Ithaca 1979).

\bibitem{desClois} J. des Cloiseaux and G. Jannink,  \textit{   Polymeres en Solution } (Les  Editions de Physique, Paris 1987).
\bibitem{Pol2} R. M.  Wartell    and A. S. Benight,     Phys. Reports \textbf{ 126}, (1985) 67.
 
\bibitem{SSTtiSkRe} R. D. Skeel and  S. and Reich, Eur. Phys. J. Special Topics  (2011).
\bibitem{SSTtiHaCi} C. Hartmann and G. Ciccotti, Eur. Phys. J. Special Topics  (2011). 
\bibitem{SSTtiMaHa} J. Maddocks and C. Hartmann,  Eur. Phys. J. Special Topics  (2011). 
\bibitem{SSTtiElHe} R. Elber  and B. Hess,  Eur. Phys. J. Special Topics  (2011).
\bibitem{SSTtiHunRe} T. Hundertmark  and S. Reich, Eur. Phys. J. Special Topics  (2011).
\bibitem{Brillou} L. Brillouin,   \textit{ Tensors in Mechanics and Elasticity}, p. 231 (Academic Press, New York  1964).
\bibitem{JenKo} H. Jensen and H. Koppe,  Ann.   Phys.   \textbf{ 63}, (1971) 596.
\bibitem{DaCos1} R. C. T. da Costa,   Phys. Rev. A \textbf{ 23}, (1981) 1982.
\bibitem{DaCos3} R. C. T. da Costa,  Eur. J.  Phys.  \textbf{7}, (1986) 269.
\bibitem{DaCos2} R. C. T. da Costa,   Phys. Rev. A \textbf{25}, (1982) 2893.
 \bibitem{AE} R. F.  Alvarez-Estrada,      Macromol. Theory Simul. \textbf{ 9}, (2000) 83.
     
\bibitem{Rubi} J. M. Rubi, D. Bedeaux and S. Kjelstrup,  J. Phys. Chem. B \textbf{110}, (2006) 12733.
\bibitem{Ritort} F. Ritort,  J. Phys.: Condens. Matter \textbf{ 18}, (2006) R531.


\bibitem{HuMcC} Y. Huang and W. F. McColl,  J.  Phys. A: Math. Gen.  \textbf{ 30}, (1997) 7919.
\bibitem{Kram} H. A. Kramers,    J. Chem. Phys. \textbf{ 14}, (1946) 415.
\bibitem{KirkRise} J. G. Kirwood and J. Riseman,      J. Chem. Phys. \textbf{ 16}, (1948) 565.


\bibitem{Hass} O. Hassager,       J. Chem. Phys. \textbf{ 60}, (1974) 2111, 4001.

\bibitem{Hass2} C. F. Curtiss, R. B. Bird and   O. Hassager,    Adv. Chem. Phys. \textbf{ 35}, (1976) 31.
 



\bibitem{Fix}  M. Fixman,     Proc. Natl. Acad.  Sci. U.S.A. \textbf{71 }, (1974) 3050.
\bibitem{EdGood}  S. F. Edwards and A. G. Goodyear,   J.  Phys. A  \textbf{ 5}, (1972) 965 and 1188.
 
\bibitem{Ryck} J.-P. Ryckaert,  Mol. Phys. \textbf{55}, (1985) 549.
\bibitem{Ciccotti} G. Ciccotti, in \textit{Liquides, Cristallisation, Transition Vitreuse} (Proc. of the Les Houches Summer School of Theoretical Physics, 
Session LI, 1989).  Editors:J. P. Hansen, 
D. Levesque and J. Zinn-Justin (Elsevier, Amsterdam 1991). 
 

\bibitem{Shake} J.-P. Ryckaert, G. Ciccotti and H. J. C. Berendsen,  J. Comp. Phys.  \textbf{23}, (1977) 327.
 
\bibitem{Mazars1} M. Mazars, Phys.Rev. E \textbf{ 53}, (1996) 6297.

\bibitem{Mazars2} M. Mazars, J.  Phys. A: Math. Gen.  \textbf{ 31}, (1997) 1949;   \textbf{ 32}, (1999) 1841.

 
\bibitem{Mazars3} M. Mazars, J.  Phys. A: Math. Theor.  \textbf{ 43}, (2010) 425002.
 



\bibitem{Fraenk} G. K. Fraenkel, J. Chem. Phys. \textbf{ 20}, (1952) 642.

\bibitem{FixKo} M. Fixman and J. Kovac,   J. Chem. Phys. \textbf{ 61}, (1974) 4939 and 4950.
\bibitem{FixEv} M. Fixman and  G. T.Evans,  J. Chem. Phys. \textbf{ 64}, (1976) 3474.
\bibitem{Tit1} U. M. Titulaer,  J. Chem. Phys. \textbf{ 66},  (1977) 1631.
\bibitem{Tit2} U. M. Titulaer and J. M. Deutch,   J. Chem. Phys. \textbf{ 63}, (1975) 4505.
\bibitem{Rou} P. E. Rouse, J. Chem. Phys. \textbf{ 21}, (1953) 1272.

\bibitem{Zimm} B. H. Zimm,  J. Chem. Phys. \textbf{ 24}, (1956) 269.
\bibitem{Lod} A. S. Lodge and Y. Yu,   Rheol. Acta \textbf{ 10}, (1971) 539.
\bibitem{Hinc} E. J. Hinch,  J. Fluid Mechs. \textbf{ 75}, (1976) 765.
 
\bibitem{Bird} R. B. Bird, M. S. Johnson and C. F. Curtiss,  J. Chem. Phys. \textbf{ 51}, (1969) 3023.
\bibitem{Ed} S. F. Edwards and  K. F. Freed,    J. Chem. Phys. \textbf{ 61}, (1974)  3626.
\bibitem{Freed1} K. F Freed and S. F. Edwards,  J. Chem. Phys. \textbf{ 61}, (1974)  1189.
\bibitem{Go1} N. Go  and  H. A. Scheraga,    J. Chem. Phys. \textbf{  51}, (1969) 4751.  
 
 
  
\bibitem{Gottl} M. Gottlieb  and R. B. Bird,   J. Chem. Phys. \textbf{ 65 }, (1976) 2467.

\bibitem{Ral} J. M.  Rallison,     J. Fluid. Mech. \textbf{ 93}, (1979) 251.
\bibitem{vanKampen} N. G. Van Kampen,    Appl.  Sci. Res.  \textbf{ 3}, (1981) 67.
 \bibitem{Pear} M. Pear and  J. W. Weiner,   J. Chem. Phys. \textbf{ 71 }, (1979) 212.
 
 \bibitem{ErKirk} J. J. Erpenbeck and J. G. Kirkwood,  J. Chem. Phys. \textbf{ 29}, (1958) 909. 
ibid   J. Chem. Phys. \textit{ 38}, (1963) 1023.
\bibitem{Helfand} E. Helfand,   J. Chem. Phys. \textbf{ 71}, (1979) 5000.


 
\bibitem{Go2}N.  Go  and  H. A. Scheraga,     Macromolecules \textbf{ 9}, (1976) 535.
 
 
  
 
 \bibitem{Eche} P. Echenique, I. Calvo  and J. L. Alonso,     J.  of Comput. Chem.  \textbf{ 27}, (2006)  1733.
 
  
\bibitem{Podolsk} B. Podolsky,    Phys. Rev. \textbf{ 32}, (1928) 812.  
\bibitem{deWitt} B. S.  De Witt,   Rev. Mod. Phys.  \textbf{ 29}, (1957) 377.
 \bibitem{MesI} A.  Messiah,    \textit{ Quantum Mechanics}, Vol. I, (North Holland, Amsterdam 1961). 
 \bibitem{GalPas} A. Galindo  and P. Pascual,      \textit{ Quantum Mechanics} Vol. I (Springer-Verlag, Berlin 1990). 
\bibitem{AEPR} R. F. Alvarez-Estrada, Phys. Rev. A \textbf{ 46}, (1992 ) 3206. 
 
 
 
 \bibitem{MesII} A. Messiah,    \textit{ Quantum Mechanics}, Vol. II (North Holland, Amsterdam 1962). 
 \bibitem{Schrod} E.  Schroedinger,   \textit{ What is Life? The Physical Aspect of the Living Cell} 
(Cambridge University Press, Cambridge  1967 ). 
\bibitem{Morse} P. M. Morse,      Phys. Rev. \textbf{34}, (1929) 57.
 
\bibitem{Calvo} G. F.  Calvo    and  R. F. Alvarez-Estrada,     Macromol. Theory Simul.    \textbf{ 9}, (2000) 585.
\bibitem{RaCa} R. F. Alvarez-Estrada   and G. F.  Calvo,    Molecular Phys.  \textbf{100}, (2002) 2957.
 \bibitem{CAEDNA} G. F. Calvo   and R. F. Alvarez-Estrada,     J. Phys.: Condens. Matter \textbf{ 17}, (2005) 7755.
\bibitem{CAEDNA2} G. F.  Calvo   and R. F. Alvarez-Estrada,      J. Phys.: Condens. Matter \textbf{ 20}, (2008) 035101.
\bibitem{CAEDNA1} R. F. Alvarez-Estrada   and G. F.  Calvo,      J. Phys.: Condens. Matter \textbf{ 16}, (2004) S2037.
  
\bibitem{Hua} K.  Huang,    \textit{ Statistical Mechanics} 2nd ed. (John Wiley \& Sons, New York 1987).
\bibitem{Peie} R. E.   Peierls,       Phys. Rev.  \textbf{ 54}, (1938) 918. 

 
  
  \bibitem{Schweb} S. S.  Schweber,     \textit{ An Introduction to Relativistic Quantum Field Theory} (Harper and Row, 
  New YorK 1966) 
\bibitem{Deut1} J. M. Deutsch, Phys. Rev.  E \textbf{77}, (2008) 051804.
\bibitem{Deut2} J. M. Deutsch, Phys. Rev. Lett.  \textbf{99}, (2007) 238301.
\bibitem{Simon} B. Simon, Ann. Inst. Henri Poincare, Sect. A  \textbf{38}, (1983) 295.
 \bibitem{Poly1} A.  Polychronakos, Phys. Rev. Lett. \textbf{70}, (1993) 2329. 
\bibitem{Poly2} A.  Polychronakos, Nucl. Phys. B  \textbf{543}, (1999) 553.
\bibitem{Fink} F. Finkel and A. Gonzalez-Lopez, Phys. Rev.  B \textbf{72}, (2005) 174411(6). 
 \bibitem{Enc} A. Enciso, F. Finkel,  A. Gonzalez-Lopez and M. A. Rodriguez, Nucl. Phys. B  \textbf{707}, (2005) 553.
\bibitem{AEM} R. F.  Alvarez-Estrada,      Macromol. Theory Simul. \textbf{ 7}, (1998) 457.
 
\bibitem{Topol1} S. F. Edwards,   J.  Phys. A  \textbf{ 1}, (1968) 15.
\bibitem{Topol2} E. Orlandini and S. G. Whittington, Rev. Mod. Phys. \textbf{ 79}, (2007) 611.
 
 
\bibitem{Pey} M.  Peyrard   and  A. R. Bishop,     Phys. Rev. Lett.  \textbf{62}, (1989) 2755.
 
 
\bibitem{Pey04} M. Peyrard,  Nonlinearity \textbf{ 17}, (2004) R1.

   

\bibitem{Pey1} T. Dauxois, M.   Peyrard  and  A. R. Bishop,  Phys. Rev.  E \textbf{47}, (1993) 684 and R44.

 

 

\bibitem{Ares} S. Ares, N. K. Voulgarakis, K. O.  Rasmussen   and  A. R. Bishop,  Phys. Rev. Lett. \textbf{94}, (2005) 035504.
\bibitem{Yaku} L.  V.  Yakushevich,     \textit{ Non-Linear Physics of DNA}.  2nd rev. ed. (Wiley-VCH,Weinheim  2004).


\bibitem{Strick} T. R. Strick et al, Rep. Prog. Phys. \textbf{66}, (2003) 1.
 \bibitem{Kumar} S. Kumar and M. S. Li, Phys. Rep. \textbf{486}, (2010) 1.
\bibitem{Marko} J. F. Marko and E. D. Siggia, Macromolecules \textbf{28}, (1995) 209.
\bibitem{Bou}C. Bouchiat  et al., Biophys. J. \textbf{76}, (1999) 409.
 
\end{thebibliography}
\end{document}